\documentclass[]{aa}
\usepackage{natbib} 
\usepackage[varg]{txfonts} 
\usepackage{xcolor}             
\usepackage{verbatim}     
\usepackage{xtab,afterpage} 
\usepackage{graphicx}
\usepackage{tikz}
\usepackage{epsfig}
\usepackage{subfig}
\usepackage{longtable}

\newcommand{\multi}{{\tt MULTI}}
\newcommand{\tlusty}{{\tt TLUSTY}}
\newcommand{\synspec}{{\tt SYNSPEC}}
\newcommand{\Na}[1]{\ion{Na}{#1}}
\newcommand{\Mg}[1]{\ion{Mg}{#1}}
\newcommand{\K}[1]{\ion{K}{#1}}
\newcommand{\Ca}[1]{\ion{Ca}{#1}}
\newcommand{\teff}{$T_\mathrm{eff}$}
\newcommand{\logg}{$\log\,g$} 
\newcommand{\feh}{\mbox{[Fe/H]}}
\newcommand{\fig}[1]{Fig. \ref{#1}}

\newcommand{\NLTEs}{\mbox{NLTE-s}}
\newcommand{\NLTEm}{\mbox{NLTE-m}}

\begin{document}
\title{NLTE for APOGEE: Simultaneous Multi-Element NLTE Radiative Transfer.}
\author{
Y. Osorio\inst{1,2} \and C. Allende Prieto\inst{1,2} \and I. Hubeny\inst{3} \and Sz. M\'esz\'aros\inst{4,5} \and M. Shetrone\inst{6}}
\institute{Instituto de Astrof\'isica de Canarias, E-38205 La Laguna, Tenerife, Spain \and Departamento de Astrof\'isica, Universidad de La Laguna (ULL), E-38206 La Laguna, Tenerife, Spain \and Steward Observatory, University of Arizona, Tucson, USA \and ELTE E\"otv\"os Lor\'and University, Gothard Astrophysical Observatory, 9700 Szombathely, Szent Imre h. st. 112, Hungary \and MTA-ELTE Exoplanet Research Group, 9700 Szombathely, Szent Imre h. st. 112, Hungary \and University of Texas at Austin, McDonald Observatory, McDonald Observatory, TX 79734, USA }
\offprints{Yeisson Osorio}
\mail{yeisson.osorio@iac.es}
\titlerunning{multi-element NLTE.}
\authorrunning{Osorio Y. et. al.}
\date{}
\abstract
{ 
    Relaxing the assumption of Local Thermodynamic Equilibrium (LTE) in modelling stellar spectra is a necessary step to determine chemical abundances better than about 10 \% in late-type stars.
} 
{ 
    In this paper we describe our efforts gearing up to perform multi-element (Na, Mg, K, and Ca) Non-LTE (NLTE) calculations that can be applied to the APOGEE survey.
}
{ 
    The new version of \tlusty\  allows for the calculation of restricted NLTE in cool stars using pre-calculated opacity tables. We demonstrate that \tlusty\ gives consistent results with \multi, a well-tested code for NLTE in cool stars. We use \tlusty\ to perform LTE and a series of NLTE calculations using all combinations of 1, 2, 3 and the 4 elements mentioned above simultaneously in NLTE.
}
{ 
    In this work we take into account how departures from LTE in one element can affect others through changes in the opacities of Na, Mg, K, and Ca. We find that atomic Mg, which provides strong UV opacity, and exhibits significant departures from LTE in the low-energy states, can impact the NLTE populations of Ca, leading to abundance corrections as large as 0.07~dex. The differences in the derived abundances between the single-element and the multi-element cases can exceed those between the single-element NLTE determinations and an LTE analysis, warning that this is not always a second-order effect. 
    By means of detailed tests for three stars with reliable atmospheric parameters (Arcturus, Procyon and the Sun) we conclude that our NLTE calculations provide abundance corrections that can amount in the optical up to 0.1, 0.2 and 0.7 dex for Ca, Na and K, but LTE is a good approximation for Mg. In the H-band, NLTE corrections are much smaller, and always under to 0.1 dex. The derived NLTE abundances in the optical and in the IR are consistent. For all four elements, in all three stars, NLTE line profiles fit better the observations than the LTE counterparts.
}
{ 
    The atomic elements in ionisation stages where over-ionisation is an important NLTE mechanism are likely affected by departures from LTE in \Mg. Special care must be taken with the collisions adopted for high-lying levels when calculating NLTE profiles of lines in the H-band. The derived NLTE corrections in the optical and in the H-band differ, but the derived NLTE abundances are consistent between the two spectral regions.
}
\maketitle

\section{Introduction}

The determination of chemical abundances in stars relies on the comparison with physical models of stellar atmospheres involving a number of approximations. Among these, the assumption that the gas is in local thermodynamic equilibrium (LTE), i.e. finding the atomic populations using the Saha and Boltzmann equations from the local temperature and electron density, is known to be one, if not the main, factor limiting the determination of accurate abundances. One of the difficulties in relaxing the LTE assumption, performing Non-LTE (NLTE) calculations, is the need for detailed data on the radiative and collisional processes that affect the atoms of interest. Over the last decade, much progress has been made in this direction, and a substantial effort investment has been made to implement NLTE in the analysis of stellar spectra  \citep{2016LRSP...13....1A,Asplund:2005bp}.

Implementing NLTE is particularly hard for very large spectroscopic samples with high-resolution and wide spectral coverage, since it requires collecting all the necessary data for many ions and performing  time-consuming calculations. Surveys such as GALAH  have started to provide NLTE abundances for a few elements \citep{2018MNRAS.478.4513B}, and we are gearing up to do the same for APOGEE  \citep{Majewski_2017}. In this paper we address the important question of whether NLTE calculations can be performed for a chemical element at a time, or NLTE effects from one element affect others significantly.

In contrast to NLTE radiative transfer calculations for hot stars,  where H and other other elements must be considered in NLTE in order to obtain reliable atmospheric structure and spectrum \citep{1969ApJ...156..157A, 1969ApJ...156..681A, 1970ApJ...160..233A}, NLTE radiative transfer calculations in cool stars are performed using the trace element approach, where the effects of the NLTE populations on the atmospheric structure and background opacities are neglected. The justification for this approach lies in the argument that the abundance of the element under study, and the differences between the LTE and NLTE populations, are small enough that they do not affect other elements or the stellar atmospheric structure.

In the calculations described in this paper we do not allow the atmospheric structure to change, i.e., the temperature, electron and density distributions are kept fixed. In this case interactions between the NLTE populations for different elements in our calculations take place trough the opacity. In the specific case of Mg and Ca, the broad resonance lines of their first ions provide an important contribution to the opacities between  $\sim$2\,700 and 4\,000~\AA\ in solar-type stars, and NLTE populations of the excited levels of those ions can also affect the opacity "seen" by other elements, and thus modify their NLTE populations. Elements with small contributions to the opacity will then have a negligible impact on other elements calculated simultaneously in NLTE.    

Section 2 describes the model atoms and the reference observations we will be testing our models against. Section 3 provides an overview of the basics of our NLTE calculations. Section 4 compares calculations for a single element with two different NLTE codes: MULTI and TLUSTY. Section 5 describes our multi-element calculations and in Section 6 we confront them with observations. We focus on the most relevant results for the APOGEE survey in Section 7, and give our conclusions in section 8.

\section{Model atoms and reference stars}

For this work, we used 4 different atomic elements: \Na{i}, \Mg{i \& ii}, \K{i}\ and \Ca{i \& ii}. The atomic data were drawn from different sources. The \Mg\ and \Ca\ model atoms were basically the same as in \cite{2015A&amp;A...579A..53O,CaPaperI} respectively.  while the model atoms for \Na\ and \K\ were created from scratch for this work. Three reference stars with observational data of exceptional quality and well-determined stellar parameters were adopted: Procyon, the Sun and Arcturus, The observational data and the synthetic model atmospheres were the same as in \cite{CaPaperI}. A description of these data will be given in \S~\ref{sec:observations}.  

\subsection{Mg and Ca}

Energy levels were taken mostly from the NIST database \citep{NIST2015}, line data were taken from VALD \citep{2000BaltA...9..590K,2015PhyS...90e4005R} , NIST, and in the case of \Ca{i}, from \cite{YU2018263}. Continuum data were taken from the TOPBASE database \citep{TOPBASE}. Hydrogen collisions  \citep{2012PhRvA..85c2704B,2016PhRvA..93d2705B,2017PhRvA..95f9906B} and electron collisions \citep{2007A&A...469.1203M,PhysRevA.99.012706,2017A&A...606A..11B} were included in the NLTE calculations, when no data were available, approximation formulas were used. Our \Mg\ model atoms have 96 levels of \Mg{i}, 29 of \Mg{ii} and the ground level of \Mg{iii}. The \Ca\ atoms we used have 66 levels of \Ca{i}, 24 of \Ca{ii} and the ground level of \Ca{iii}.  A detailed description of the atomic data used for \Mg\ and \Ca\ is given in \cite{2015A&amp;A...579A..53O,CaPaperI}, respectively. There are only two differences between the Mg and Ca data from the works mentioned above and the present study: 1) given that Van der Waals broadening parameters accepted by \tlusty\ is $\Gamma_6$, we have computed these parameters from the Anstee-Barklem-O'mara (ABO) theory \citep{1998PASA...15..336B}. 2) the photo-ionisation cross sections in this work where resonance averaged photo-ionisation cross sections \citep{2003ApJS..147..363A} based on TOPBASE data, while the same data without smoothing were used for Mg and Ca in \cite{2015A&amp;A...579A..53O,CaPaperI} respectively.

\subsection{Na and K}

\begin{figure}[t]
  \begin{tabular}{c}
    \begin{tikzpicture} 
      \node[anchor=south east, inner sep=0] (image) at (0,0) {
        \subfloat{\includegraphics[width=0.47\textwidth]{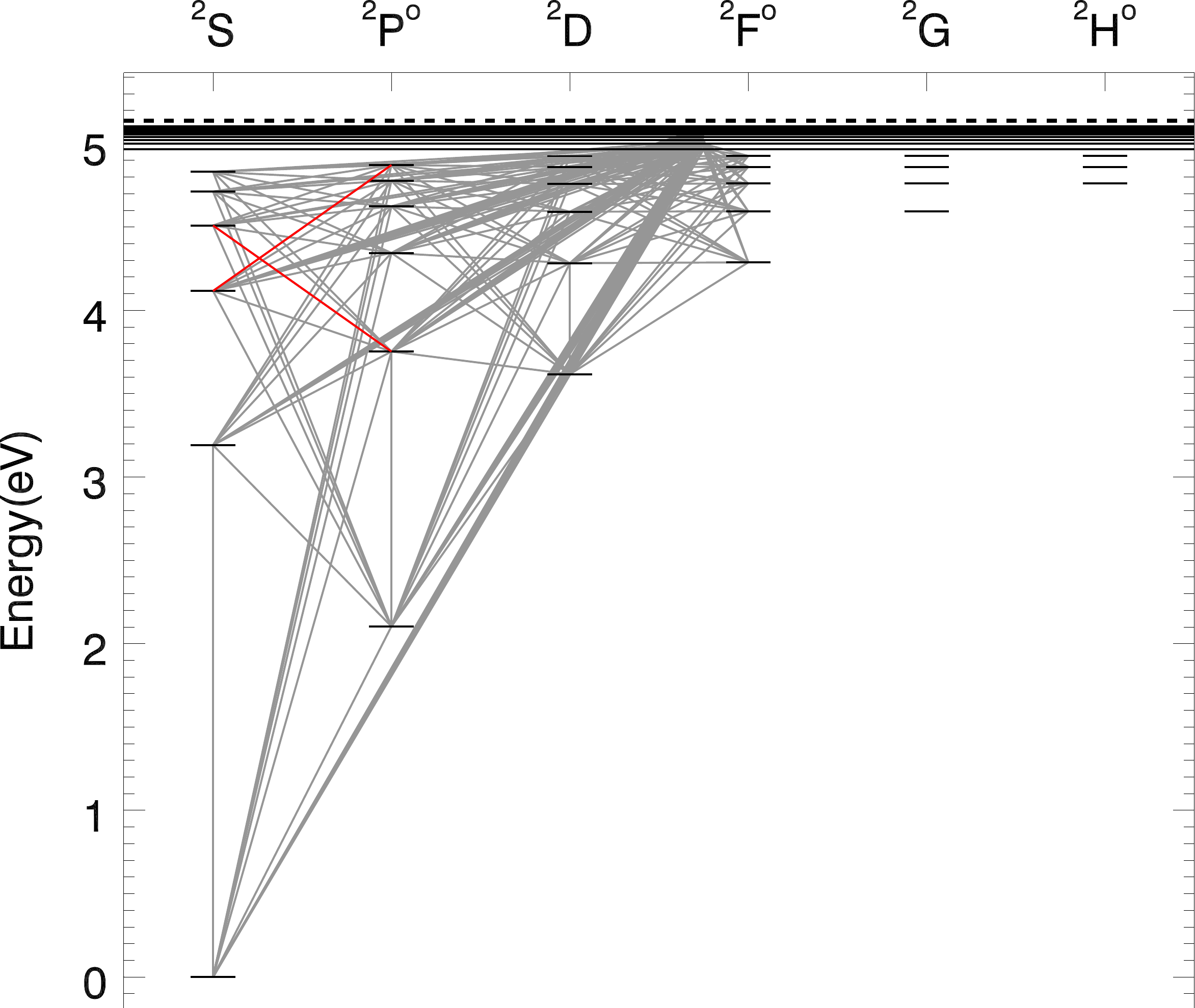}}
      };
      \node at (-0.05\textwidth,0.08\textwidth) {\scalebox{1.50}{\Na{i}}};
    \end{tikzpicture}
    \\[-0.01\textwidth]
    \begin{tikzpicture} 
      \node[anchor=south east, inner sep=0] (image) at (0,0) {
        \subfloat{\includegraphics[width=0.47\textwidth]{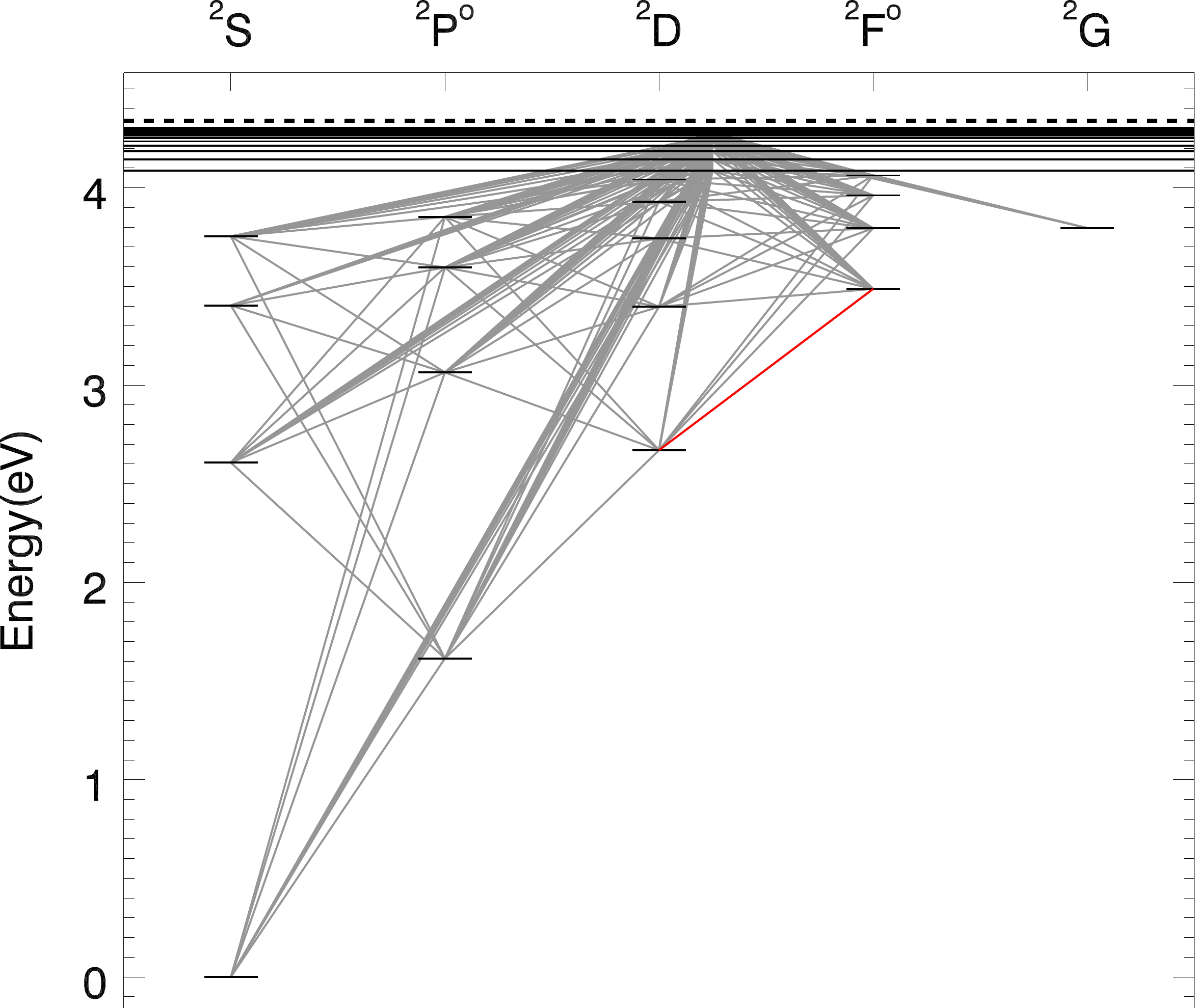}}
      };
      \node at (-0.05\textwidth,0.08\textwidth) {\scalebox{1.50}{\K{i}}};
    \end{tikzpicture}
  \end{tabular} 
  \caption{Grotrian diagram of \Na\ (upper panel) and \K\ (lower panel) model atoms used in this work. The red lines correspond to transitions in the $H$-band. }\label{fig:grotrian}
\end{figure}

The final model atom of \Na\ has 42 levels of \Na i and the ground level of \Na{ii}, and a total of 190 radiative transitions. 
while for \K\ we used 31 levels of \K i and the ground level of \K{ii} and a total of 219 radiative transitions. Figure \ref{fig:grotrian} shows the Grotrian diagram with the levels and the transitions of the \K\ and \Na\ model atoms used in this work. 

\subsubsection{Energy levels and radiative data}

Energy levels for \Na{i}\  and \K{i}\ were adopted from NIST.  Fine structure components were ignored while the allowed angular momentum components lower than $l=6$ for a given principal quantum number $n$ were taken as separated levels up to $n=8$ for \Na\ and $n=6$ for \K. $n$ super levels were included up to $n=20$ for \Na\ and \K\ in order to assure coupling between the highest level of the neutral species and the ground level of their first ion. 

Photo-ionisation cross-sections of \Na{i} were taken from the TOPBASE data base. \K{i} photo-ionisation data were taken from \cite{PhysRevA.81.043423}. When bound-free data were not available, photo-ionisation cross-sections were calculated using the hydrogenic formula implemented in \tlusty\ \citep[][
for recent upgrades, detailed description and users' manual, see \cite{tlusty1,tlusty2,tlusty3}]{1995ApJ...439..875H}.  
 
Oscillator strengths (f-values) come from various sources collected in the NIST and VALD data bases. TOPBASE \citep{TOPBASE}, provide the atomic data (f-values, photo-ionisation cross sections) from the Opacity Project calculations \citep{1996Obs...116..177S}. For \Na, VALD has the experimental f-values provided by NIST so the \Na\ f-values from VALD have priority over NIST. For the rest of bound-bound transitions TOPBASE data were used when available. In the case of \K, data from VALD come from \cite{K12} while data from NIST is mostly from \cite{doi:10.1063/1.2789451}. When for the same transition, f-values from different sources differ, preference was given to NIST over VALD in this case. VALD also provides collisional broadening parameters from \cite{K12} and \cite{BPM} where available. Line broadening data are taken from \cite{BPM}, via VALD, when available. Otherwise we adopted the formula from \cite{1955psmb.book.....U} with an enhancement factor of 2, as suggested by \cite{2000ARep...44..790M} in their Na NLTE studies on the Sun. Radiative bound-bound data for the most important transitions of \Na\ and \K\ (including those in the APOGEE window) are presented in Table \ref{tab:bbNaK}.

\begin{table}[t]
    \centering
    \small
    \caption{Data adopted for the NLTE Calculations for characteristic lines of \Na\ and \K.}
    \label{tab:bbNaK}
    \begin{tabular}{l r r r  c  c}\hline
    low (J) &   Upp (J)  &  $\lambda$[\AA] &     loggf  & log $\Gamma_4$ & log $\Gamma_6$  \\\hline
        \multicolumn{6}{c}{\Na{i}}   \\\hline
    3s(1/2) & 3p(1/2) &  5\,895.9 &  -0.194 &  -5.64 & -7.53* \\
    3s(1/2) & 3p(3/2) &  5\,889.9 &   0.108 &  -5.64 & -7.53* \\
    3p(3/2) & 3d(1/2) &  8\,183.3 &   0.237 &  -5.38 & -7.23* \\
    3p(3/2) & 3d(1/2) &  8\,194.8 &   0.492 &  -5.38 & -7.23* \\
    3p(1/2) & 4d(1/2) &  5\,682.6 &  -0.706 &  -4.52 & -6.85* \\
    3p(3/2) & 4d(3/2) &  5\,688.2 &  -0.452 &  -4.52 & -6.85* \\
    3p(1/2) & 5s(1/2) &  6\,154.2 &  -1.547 &  -4.39 & -6.99 \\
    3p(3/2) & 5s(3/2) &  6\,160.7 &  -1.246 &  -4.39 & -6.99 \\
    3p(1/2) & 6s(1/2) &  5\,148.8 &  -2.044 &  -5.59 & -6.81 \\
    3p(3/2) & 7s(3/2) &  4\,751.8 &  -2.078 &  -5.63 & -6.67 \\ 
    4s(1/2) & 5p(3/2) & 10\,746.4 &  -1.294 &  -5.97 & -6.91 \\
    4p(1/2) & 6s(1/2) & 16\,373.9 &  -1.328 &  -4.23 & -6.84 \\
    4p(3/2) & 6s(1/2) & 16\,388.8 &  -1.027 &  -4.23 & -6.84 \\
    5s(1/2) & 8p(3/2) & 16\,393.9 &  -2.149 &  -3.14 & -6.52 \\      
    5s(1/2) & 8p(1/2) & 16\,395.2 &  -2.456 &  -3.14 & -6.52 \\\hline
    \\[-0.3cm]
        \multicolumn{6}{c}{\K{i}}   \\\hline
    4p(1/2) &  7s(1/2) &  5\,782.4 &  -1.909 &  -4.11 &  -6.79 \\
    4p(3/2) &  7s(1/2) &  5\,801.8 &  -1.605 &  -4.11 &  -6.79 \\
    4p(1/2) &  6s(1/2) &  6\,911.1 &  -1.409 &  -4.52 &  -6.91* \\
    4p(3/2) &  6s(1/2) &  6\,938.8 &  -1.252 &  -4.52 &  -6.90* \\
    4s(1/2) &  4p(3/2) &  7\,664.9 &   0.127 &  -5.64 &  -7.44* \\
    4s(1/2) &  4p(1/2) &  7\,699.0 &  -0.177 &  -5.64 &  -7.44* \\   
    4p(1/2) &  5s(1/2) & 12\,432.3 &  -0.444 &  -5.06 &  -7.34* \\ 
    4p(3/2) &  5s(1/2) & 12\,522.2 &  -0.135 &  -5.06 &  -7.34* \\ 
    3d(5/2) &  4f(7/2) & 15\,163.1 &   0.617 &  -4.79 &  -6.82 \\
    3d(3/2) &  4f(5/2) & 15\,168.4 &   0.494 &  -4.79 &  -6.96 \\\hline
    \end{tabular}
    \tablefoot{* The data were computed from the ABO theory \citep{1998PASA...15..336B}. }
\end{table}


\subsubsection{Collisional data}

\emph{Electron collisions.} The formula from \cite{1971JQSRT..11....7P}, which is an empirical adjustment of the Born approximation applicable only to neutral alkali atoms, has shown to be a good approximation for NLTE calculations of Li and Na in cool stars when compared with more sophisticated data \citep{2011A&A...529A..31O,2011A&A...528A.103L}. For \Na, the convergent close coupling (CCC) calculations from \cite{2008ADNDT..94..981I} cover all transitions for the levels up to $4f$ and those where included in our model atom. Regarding \K, very recent calculations from \cite{2019A&A...627A.177R} suggest that the Park formula tends to overestimate the \K\ collisional rates. They perform CCC and B-spline R-matrix (BSR) calculations (both agree extremely well) for electron collisional excitation between all levels up to $6d$; we decided to adopt their BSR data. 

We filled the missing transitions of Na in \cite{2011A&A...528A.103L} and of K in \cite{2019A&A...627A.177R} with the Park formula. It covers all the transitions between levels up to n=7 for \Na\ and n=6 for \K. For higher levels, it provides data for transitions between the 16 closest levels with n$_{upper}\leq20$. For the remaining transitions, the formula from \cite{1962amp..conf..375S} was implemented. 

Electron collisional ionisation was described via the hydrogenic approximation as presented in \citeauthor{2000asqu.book.....C} (2000, Sect 3.6.1) for low-lying levels, and the formula from \cite{Vrinceanu:2005em} for levels with principal quantum number $n\ge5$ for \Na\ and $n\ge6$ for~\K.

 \emph{Hydrogen collisions.}
For collisional excitation of \Na\ and \K\ we adopted data from \cite{2012A&A...541A..80B} and \cite{2016PhRvA..93d2705B,2017PhRvA..95f9906B} respectively. Those works cover the transitions between all levels from the ground state up to 5p for \Na\ and 4f for \K. For the missing one-electron transitions in these works, hydrogen collisional excitation was calculated using the code from \cite{2017ascl.soft01005B} based on the free electron model of \cite{1985JPhB...18L.167K,Kaulakys:1986tl}.

Charge transfer processes like
$\rm{A}+\rm{H} \leftrightarrow \rm{A}^{+}+\rm{H}^-$ 
were calculated by \cite{2012A&A...541A..80B} and \cite{2016PhRvA..93d2705B,2017PhRvA..95f9906B} for \Na\ and \K\ respectively and are included in this work.

\begin{figure*}[t]
  \begin{tabular}{c c}
    \begin{tikzpicture} 
      \node[anchor=south east, inner sep=0] (image) at (0,0) {
        \subfloat{\includegraphics[width=0.47\textwidth]{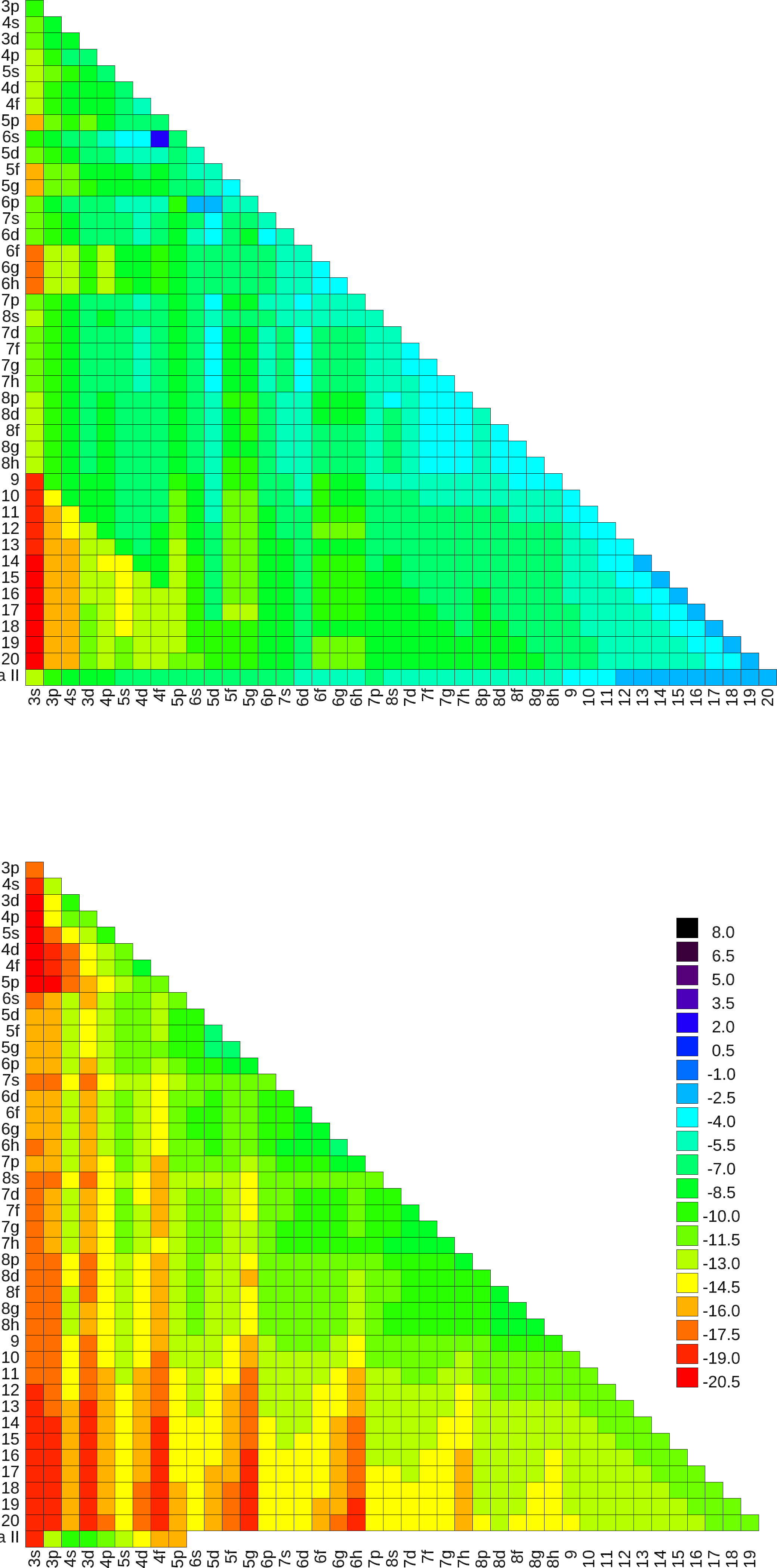}}
      };
      \node[right] at (-0.3\textwidth,0.90\textwidth) {\scalebox{1.2}{\Na{i}. electron collisions}};
      \node[right] at (-0.3\textwidth,0.40\textwidth) {\scalebox{1.2}{\Na{i}. H collisions}};
      \node[rotate=90] at (-0.08\textwidth,0.25\textwidth) {\scalebox{0.80}{Log(Na rate coeff. at 5\,000 K) [cm$^3$/s]}};
      \node[right] at (-0.369\textwidth,0.354\textwidth) {\scalebox{1.5}{$\Leftarrow$}};
    \end{tikzpicture}
    &
    \begin{tikzpicture} 
      \node[anchor=south east, inner sep=0] (image) at (0,0) {
        \subfloat{\includegraphics[width=0.47\textwidth]{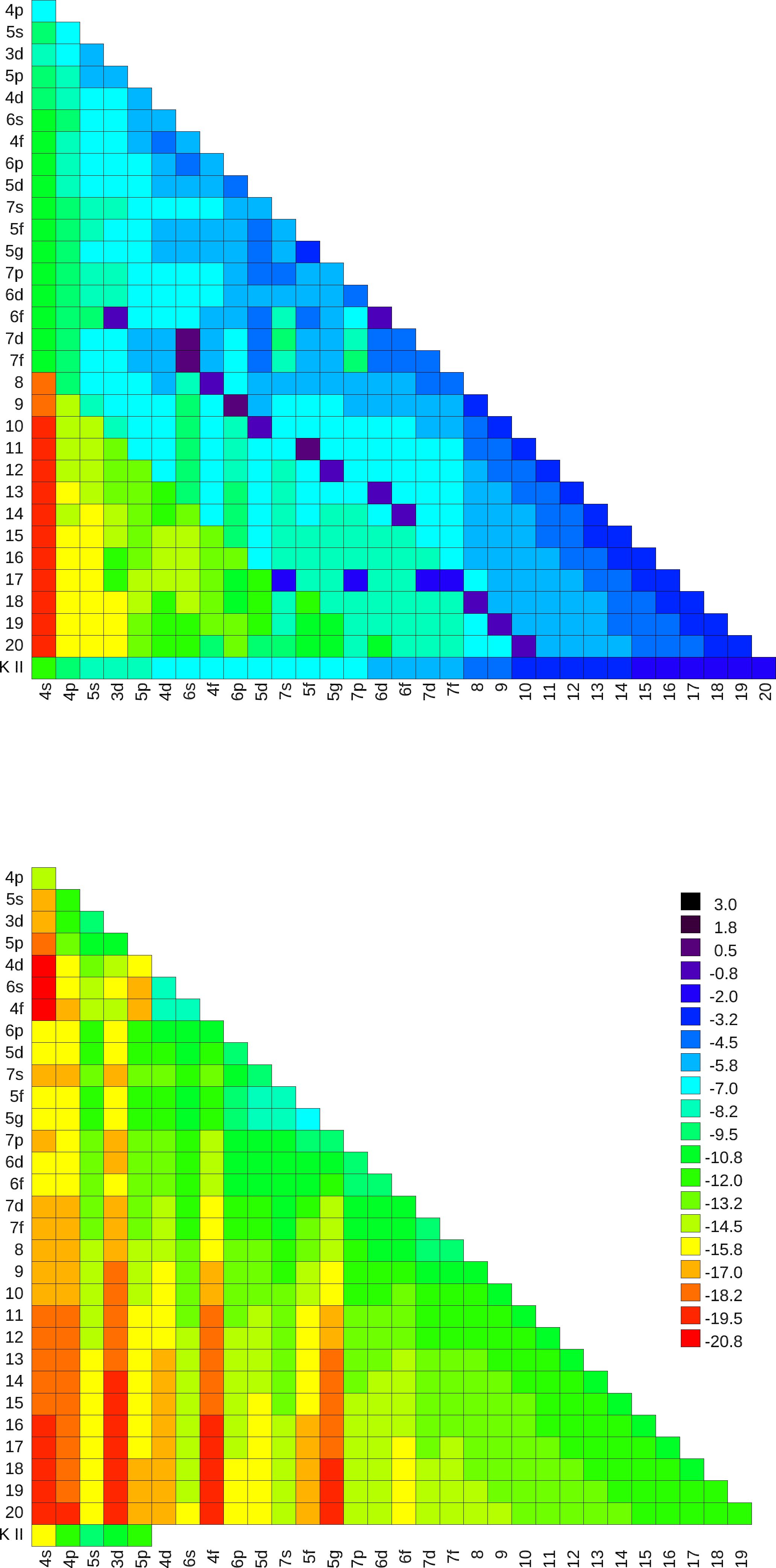}}
      };
      \node[right] at (-0.3\textwidth,0.90\textwidth) {\scalebox{1.0}{\K{i}. electron collisions}};
      \node[right] at (-0.3\textwidth,0.40\textwidth) {\scalebox{1.0}{\K{i}. H collisions}};
      \node[rotate=90] at (-0.08\textwidth,0.27\textwidth) {\scalebox{0.80}{Log(K rate coeff at 5\,000 K) [cm$^3$/s]}};
      \node[right] at (-0.352\textwidth,0.337\textwidth) {\scalebox{1.5}{$\Leftarrow$}};
    \end{tikzpicture}
  \end{tabular}
\caption{Rate coefficients of the electron related collisions (excitation and ionisation, top ) and hydrogen related collisions (excitation and charge-exchange, bottom) for \Na\ (left) and \K\ (right). Initial levels of the transitions are labelled in rows and final levels are labelled in columns. The arrows point towards the transition that separates the hydrogen collisional excitation data from  \citeauthor{2012A&A...541A..80B} and the data calculated using the formula from \cite{1985JPhB...18L.167K}. The colour scale at the left is for the \Na\ rates and the colour scale at the right is for the \K\ rates.}\label{fig:colls}
\end{figure*}
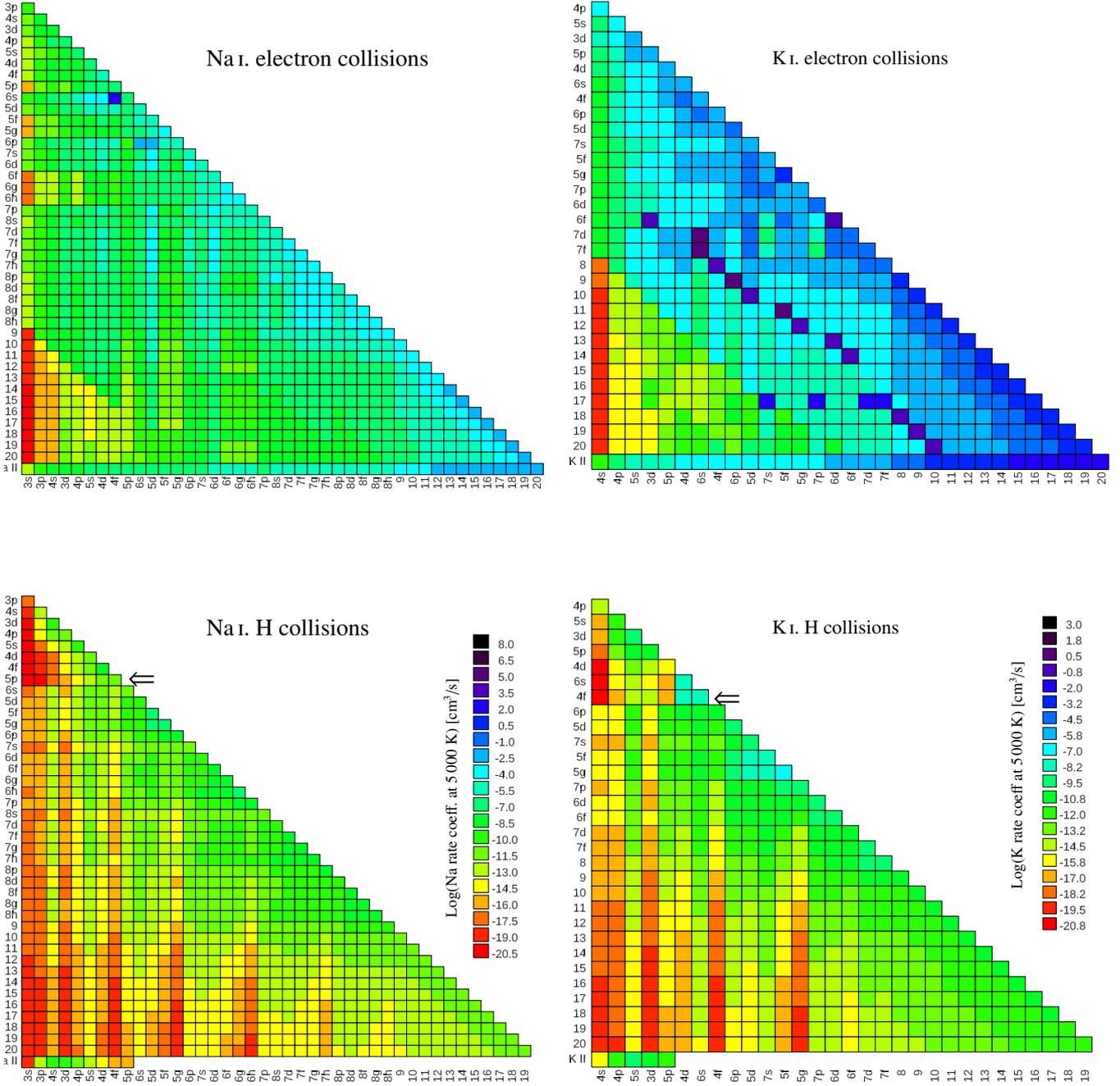

Figure \ref{fig:colls} shows the rate coefficients obtained at 5\,000 K for the different collisional processes described above for the \Na\ and \K\ atoms. It is important to remember that the rate coefficients are an estimation of the efficiency of the transition, and the actual rate of a transition is obtained multiplied multiplying by the population of the projectile. For solar metallicities the population of hydrogen is around four orders of magnitude larger than the population of free electrons, and therefore even if the rate coefficients for electron collisional excitation is one or two orders of magnitude stronger than their hydrogen counterpart, the contribution to the total collisional excitation rate from hydrogen collisions can be more important than the contribution from electron collisions. At lower metallicities, fewer free electrons are available, weakening even more their contribution to the total collisional rate and thus making the hydrogen contribution even more important.

\subsection{The APOGEE spectral window}

NLTE effects are different for different lines, and when the atomic lines to be observed are from transitions involving higher levels, having high quality collisional data for the low lying levels (which are the most populated) is important. However, one also must pay special attention to the collisional data for transitions associated with the lines of interest. For that reason, the study of the $H$-band (where most of the atomic transitions are between high excitation levels) requires the model atoms of the elements under study to have enough high-lying levels and the collisional data for transitions involving those levels to be as good as possible. Unfortunately, modern calculations of collisional excitation and ionisation usually do not cover such transitions.

We have built our model atoms with those concerns in mind. The levels used in the final model atom are such that the lines in the $H$-band do not involve transitions between super levels, required to keep the size of the model atom manageable, and at the same time have a realistic coupling with the continuum  \citep{2010EAS....43..189M,Asplund:2005bp}. Most of the collisional \mbox{(de-)} excitation data for the transitions in the $H$-band, and the collisional ionisation data for levels involved in those transitions, are outside the cover of modern, detailed, quantum mechanical calculations. Recent studies (and some not-too recent but not commonly used) give us better alternatives to the old, traditional approximations methods.   The above considerations demonstrate that we put considerable effort in the adoption of the best data available for transitions involving high-lying levels, as it is the case for many of the lines in the $H$-band.

\subsection{Reference stars}\label{sec:observations}

We decided to use the Sun, Arcturus and Procyon for the comparison between observations and our synthetic spectra. The observational data used in this work are the same as the ones described in \cite{CaPaperI}. Here, we can mention that our selection was based on the excellent quality of the data and the reliability of the atmospheric parameters, together with the wide wavelength coverage of the observations.

For Procyon, we used the observations from PEPSI \citep{2018A&A...612A..45S}. Its wavelength coverage is (3\,800 - 9\,100~\AA) at R$\sim220\,000$. The solar observations we used were the 2005 version of the flux atlas from \cite{kursun84}, it covers from 3\.000 to 10\,000~\AA, with a full width half maximum reducing power R$\sim400\,000$. For Arcturus we adopted the atlas from \cite{2000vnia.book.....H} that spans from 3\,727 to 9\,300~\AA\ and has R$\sim150\,000$. A more detailed description of the observations and a comparison with other atlases can be found in \cite{CaPaperI}.  The atmospheric parameters adopted are given in Table \ref{tab:parameters}. For more details regarding the observational data please refer to \cite{CaPaperI}.

\begin{table}[htp]
\caption{Parameters of the model atmospheres used for the construction of the synthetic spectra.}\label{tab:parameters}
\centering                           
\begin{tabular}{l@{\hskip 0.17cm}c@{\hskip 0.17cm}c@{\hskip 0.17cm}c@{\hskip 0.17cm}c}\hline\hline\\[-9pt]
Name     & \teff[K] & \logg\,[cm\,s$^{-2}$] & \feh       & v$_{mic}$[km\,s$^{-1}$]  \\\hline
Procyon$^{\dagger}$  & $6\,530$ & $4.00$ & +0.0  & 2.00     \\
Sun$^{\dagger\dagger}$ & $5\,772$ & $4.44$ & +0.0 & 1.10    \\
Arcturus$^{\dagger\dagger\dagger}$ & $4\,247$ & $1.59$ & $-$0.5 & 1.63   \\
\hline
\end{tabular}
\tablefoot{ Data adopted from: \\
$^{\dagger}$ \cite{2002ApJ...567..544A} \\
$^{\dagger\dagger}$ \cite{Asplund:2005bp} \\
$^{\dagger\dagger\dagger}$ \cite{2011ApJ...743..135R}
}
\end{table}%

\section{Computations}

There are two sets of computations done in this work: in the first one we used the results of the well-established NLTE radiative transfer code for cool stars \multi\ \citep{1986UppOR..33.....C,1992ASPC...26..499C}, and compare its results with the last version of \tlusty\ \citep{1995ApJ...439..875H,tlusty1,tlusty2,tlusty3} that now allows for the calculation of NLTE populations in cool stellar atmospheres. Given that \tlusty\ was created for the study of early-type stars and accretion discs, it allows for the treatment of multiple species in NLTE at the same time; a property long known to be necessary for NLTE studies in hot stars \citep{1969ApJ...156..157A}. We obtained Mg and Ca LTE/NLTE populations with the two codes using the same input atomic data and a configuration as similar as possible to each other. The results and analysis of these calculations are given in \S \ref{sec:comparison}. 

The second set of computations are performed exclusively with \tlusty\ (to calculate the LTE/NLTE populations) and \synspec\ (to calculate the detailed spectra). We compare observations against the results of two different NLTE calculations: the traditional (for cool stars) trace-element, single-species NLTE calculation, referred to as \NLTEs, and calculations where multiple species are treated in NLTE simultaneously, still within the trace-element framework, referred to as \NLTEm. We used Arcturus, the Sun and Procyon to calculate the LTE, \NLTEs, and \NLTEm\ populations and spectra of Na, Mg, K and Ca on these stars. From \S \ref{sec:nlte-m} on, this paper is dedicated to the second set of computations.

The \NLTEm\ calculations are a better approximation than the \NLTEs\ ones from the physical point of view: the effects of the NLTE radiation of one atomic species on the NLTE populations of other species is included in the \NLTEm\ and ignored in the \NLTEs\ calculations. The relevance of such effects in cool stars is studied for the first time in this work (see \S \ref{sec:inter-element}).

In order to obtain the NLTE populations we ran \tlusty\ v.207 in the opacity table mode. The opacity tables were constructed using \synspec\ v.53\footnote{The code can be found in:\\ \url{https://www.as.arizona.edu/~hubeny/pub/tlusty207.tar.gz}. A description of the code and an operation manual are found in \cite{tlusty1,tlusty2,tlusty3}.}. After experimentation with different numbers of wavelengths, temperatures and densities, the adopted table for the second set of calculations has 100\,000 wavelength points equally distributed in a logarithmic scale from 900 to 100\,000 \AA; 10 temperature grid points and 10 density grid points. The element abundances used in the the opacity tables are the appropriate for each star's metallicity. In each NLTE calculation, the opacity table used excludes the  contribution from the elements treated in NLTE. 

The detailed synthetic spectra (in LTE and NLTE) were calculated using the same code as the opacity tables \synspec\ v.53, adopting the same atomic and molecular line-list as in \cite{2012AJ....144..120M} but with $\log gf$ and broadening parameters of the \Na{i}, \Mg{i \& ii}, \K{i}\ and \mbox{\Ca{i \& ii}}\ lines replaced by the ones used in the NLTE calculations done with \tlusty.

The stellar model atmospheres are the same as in \cite{CaPaperI}: Kurucz model atmospheres computed with ATLAS9 \citep{1993PhST...47..110K}, using the setup described in \cite{2012AJ....144..120M}.

\section{Comparison between \tlusty\ and \multi}\label{sec:comparison}

The last updates of \tlusty\ bring the possibility to use the code for NLTE calculations for a subset of species, while the rest of atomic, as well as molecular, species are treated through pre-calculated opacity tables. We can thus compare with other NLTE radiative transfer codes used in the study of cool stars. 

Our first goal was to reproduce the results obtained in previous works, more precisely the solar departure coefficients obtained for Mg \citep{2015A&amp;A...579A..53O} and Ca \citep{CaPaperI}. In both cases the calculations were performed with the NLTE radiative transfer code \multi\ \citep{1986UppOR..33.....C} version 2.3; and now we repeat the calculations with \tlusty. The results of such comparison are shown in this section.

The same solar model atmosphere used in \cite{CaPaperI} is adopted here for both Mg and Ca. The input data for Mg and Ca used in both \multi\ and \tlusty\ were adapted for each code and this adaptation led to the differences described in \S \ref{sec:cmp-bf}, \ref{sec:cmp-vdw} and  \ref{sec:cmp-colls}.

\subsection{Background opacity}

In this work we use the term \emph{background opacity} to refer to all the sources of opacity that do not involve the elements under study. In \tlusty, those are stored in  pre-calculated opacity tables. In \multi, the continuous opacities are calculated  on-the-fly,  while only line opacities are stored in a read-in file (line-opacity table).  

For \multi, we used a line-opacity table based on the background opacities presented in \cite{Gustafsson:2008df} re-sampled to 10\,300 frequency points between 900 and 200\,000~\AA. The bound-free (b-f) and free-free (f-f) contribution from the most important atomic elements and also the b-f and f-f contribution from several molecules are calculated by the code.

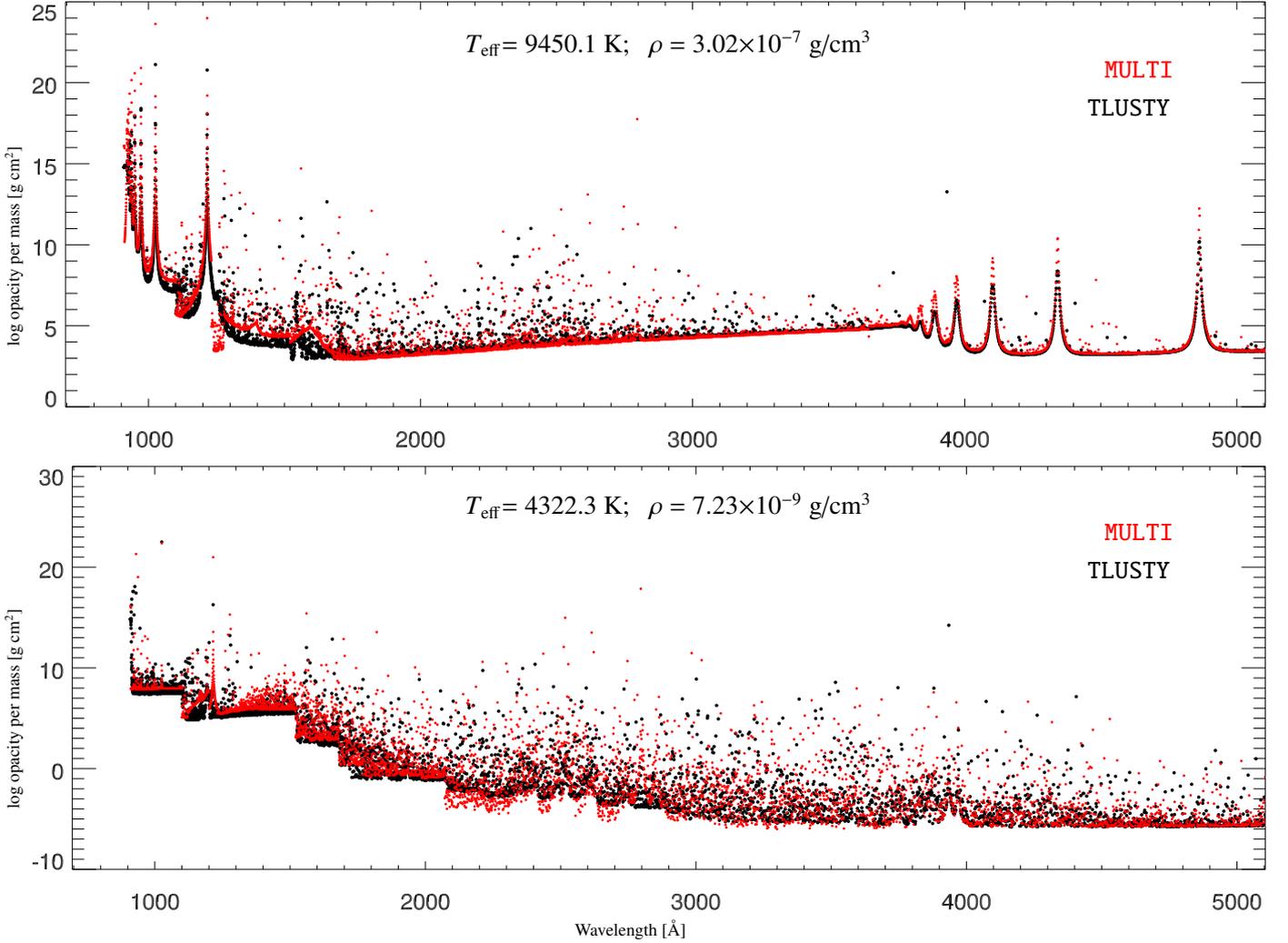
\begin{figure*}[t]
  \hspace{-0.02\textwidth}
  \begin{tabular}{c}
    \begin{tikzpicture} 
      \node[anchor=south east, inner sep=0] (image) at (0,0) {
        \subfloat{\includegraphics[width=0.97\textwidth]{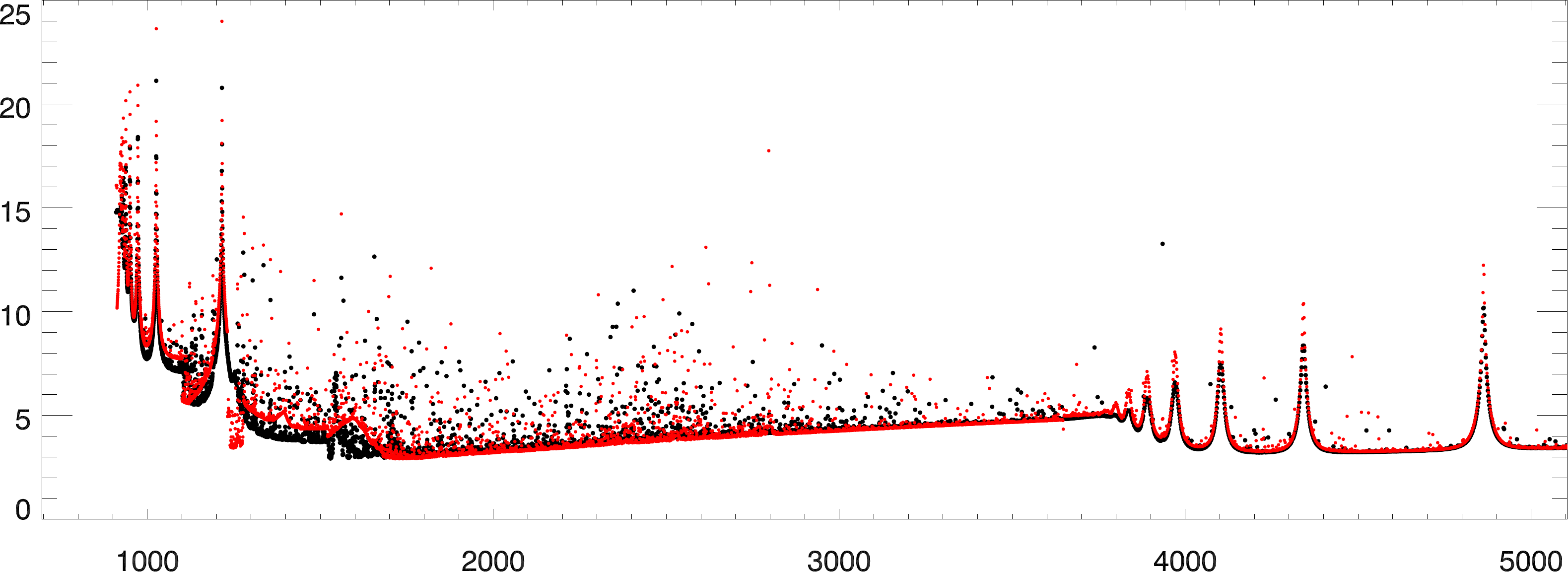}}
      };
      \node           at (-0.47\textwidth,0.32\textwidth) {\scalebox{1.2}{\teff = 9450.1 K;\,\,\, $\rho$ = 3.02$\times10^{-7}$ g/cm$^3$}};
      \node[color=red,align=right] at (-0.10\textwidth,0.30\textwidth) {\scalebox{1.2}{\multi}};
      \node[align=right] at (-0.108\textwidth,0.27\textwidth)            {\scalebox{1.2}{\tlusty}};
      \node[rotate=90] at (-0.985\textwidth,0.16\textwidth) {\scalebox{0.80}{log opacity per mass [g cm$^2$]}};
    \end{tikzpicture}
    \\[-0.01\textwidth]
    \begin{tikzpicture} 
      \node[anchor=south east, inner sep=0] (image) at (0,0) {
        \subfloat{\includegraphics[width=0.97\textwidth]{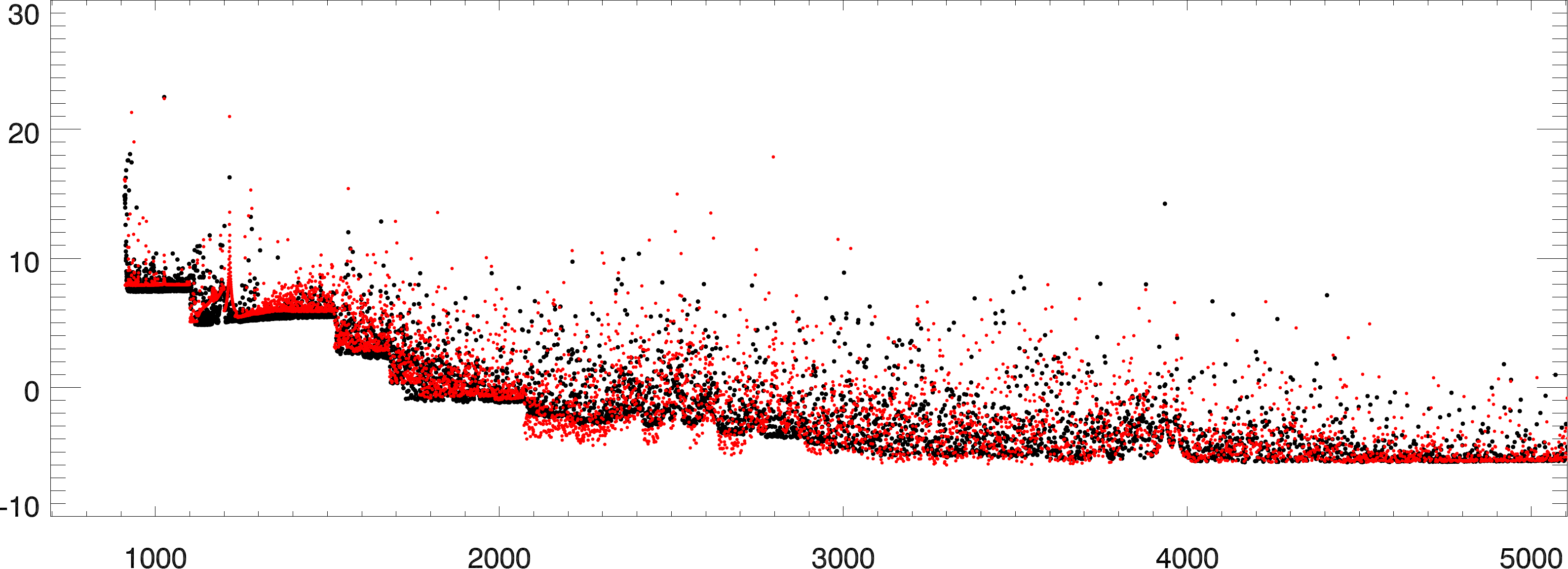}}
      };
      \node           at (-0.47\textwidth,0.32\textwidth) {\scalebox{1.2}{\teff = 4322.3 K;\,\,\, $\rho$ = 7.23$\times10^{-9}$ g/cm$^3$}};
      \node[color=red,align=right] at (-0.10\textwidth,0.30\textwidth) {\scalebox{1.2}{\multi}};
      \node[align=right] at (-0.108\textwidth,0.27\textwidth)            {\scalebox{1.2}{\tlusty}};
      \node[rotate=90] at (-0.985\textwidth,0.16\textwidth) {\scalebox{0.80}{log opacity per mass [g cm$^2$]}};
       \node at (-0.50\textwidth,-0.015\textwidth) {\scalebox{0.80}{Wavelength [\AA]}};
    \end{tikzpicture}

  \end{tabular}      
    \caption{Sample of the background opacities from the opacity tables in the UV region used in the NLTE calculation of Mg (i.e., removing the contribution of \Mg.) by \multi\ (red) and \tlusty\ (black) at two different points of the solar atmosphere.}\label{fig:opa_cmp}
\end{figure*}

Unlike previous versions, the latest version of \tlusty\ allows the use of opacity tables. These tables include also the bound-bound (b-b) contribution from atoms and some molecules. In order to make the comparison easier, the opacity tables used in \tlusty\ for this test have the same number of frequency points and wavelength coverage than those adopted for the \multi calculations. The atoms and molecules that contribute to the background opacities are listed in Table \ref{tab:cmpopt}. A sample of the background opacities used for the Mg NLTE calculations in \multi\ and \tlusty\ is presented in \fig{fig:opa_cmp}.

\begin{table}[h]
  \caption{Components of the background opacities used for the comparison calculations between \multi$^*$ and \tlusty.}\label{tab:cmpopt}
  \hspace{0.0\textwidth}
  \begin{tabular}{l c c c}\hline\hline
    Species                  &  \multicolumn{3}{c}{Process}  \\\hline
                             & \multi & \phantom{xx} &   \tlusty    \\\hline
 \ion{H}{$^-$}               &     bf, ff    & &     bf, ff   \\ 
    \ion{H}{i}               & bb, bf, ff    & & bb, bf$^{**}$, ff   \\ 
   \ion{He}{i}               & bb, bf, ff    & & bb, bf, ff   \\
\ion{He}{$^-$}               &         ff    & &     bf, ff   \\ 
  \ion{C}{i} \& \ion{}{ii}   & bb, bf        & & bb, bf, ff   \\
  \ion{C}{$^-$}              &         ff    & &              \\
  \ion{N}{i} \& \ion{}{ii}   & bb, bf        & & bb           \\
  \ion{N}{$^-$}              &         ff    & &              \\
  \ion{O}{i} \& \ion{}{ii}   & bb, bf        & & bb, bf, ff   \\
  \ion{O}{$^-$}              &         ff    & &              \\
  \ion{Na}{i} \& \ion{}{ii}  & bb            & & bb, bf, ff   \\
  \ion{Mg}{i}$^{\dagger}$    & bb, bf, ff    & & bb, bf, ff   \\
  \ion{Mg}{ii}$^{\dagger}$   & bb, bf        & & bb, bf, ff   \\
  \ion{Al}{i} \& \ion{}{ii}  & bb            & & bb, bf, ff   \\
  \ion{Si}{i}                & bb, bf, ff    & & bb, bf, ff   \\
  \ion{Si}{ii}               & bb, bf        & & bb, bf, ff   \\
  \ion{K}{i}                 &  bb           & & bb, bf, ff  \\
  \ion{Ca}{i} \& \ion{}{ii}$^{\dagger\dagger}$ bb  & bb, bf        & & bb, bf, ff   \\
  \ion{Fe}{i} \& \ion{}{ii}    &    bb, bf        & & bb, bf, ff   \\
 All other metals                &       bb, ff    & & bb           \\
  \ion{H}{$^+_2$}            &         ff    & &     ff   \\
  \ion{H}{$^-_2$}            &         ff    & &              \\
  \ion{CH}                   &     bf        & & bb, bf       \\
  \ion{OH}                   &     bf        & & bb, bf       \\
  \ion{CO}                   &     bf        & & bb           \\
  Other diatomic molecules$^{\dagger\dagger\dagger}$ & bb & & bb          \\
  \ion{H}{}\,\, Rayleigh Scattering       & yes           & & yes          \\
  \ion{He}{} Rayleigh Scattering      & yes           & & yes          \\
  \ion{H}{$_2$} Rayleigh Scattering   & yes           & & yes          \\
    e$^-$ (Thomson) Scattering         & yes           & & yes          \\\hline
  \end{tabular}
  \tablefoot{``bb'',``bf'' and ``ff'' stands for bound-bound, bound-free and free-free transitions respectively.\\
  $^*$ The information is taken from \cite{Gustafsson:2008df}, from which the opacity table used in \multi\ was based on.\\
  $^{**}$ The continua for the first three levels are extended short ward of the corresponding  edges to describe a "pseudo-continuum" - see \cite{1994A&A...282..151H}.\\
    $^{\dagger}$\,\, Removed for the \Mg\ NLTE calculations.\\
    $^{\dagger\dagger}$ Removed for the \Ca\ NLTE calculations.\\
    $^{\dagger\dagger\dagger}$ For \tlusty\ these are: H$_2$, NH, MgH, SiH, C$_2$, CN and SiO.}
\end{table}

\subsection{Photo-ionisation}\label{sec:cmp-bf}

\begin{figure}[t]
  \hspace{-0.02\textwidth}
    \begin{tikzpicture} 
      \node[anchor=south east, inner sep=0] (image) at (0,0) {
        \subfloat{\includegraphics[width=0.47\textwidth]{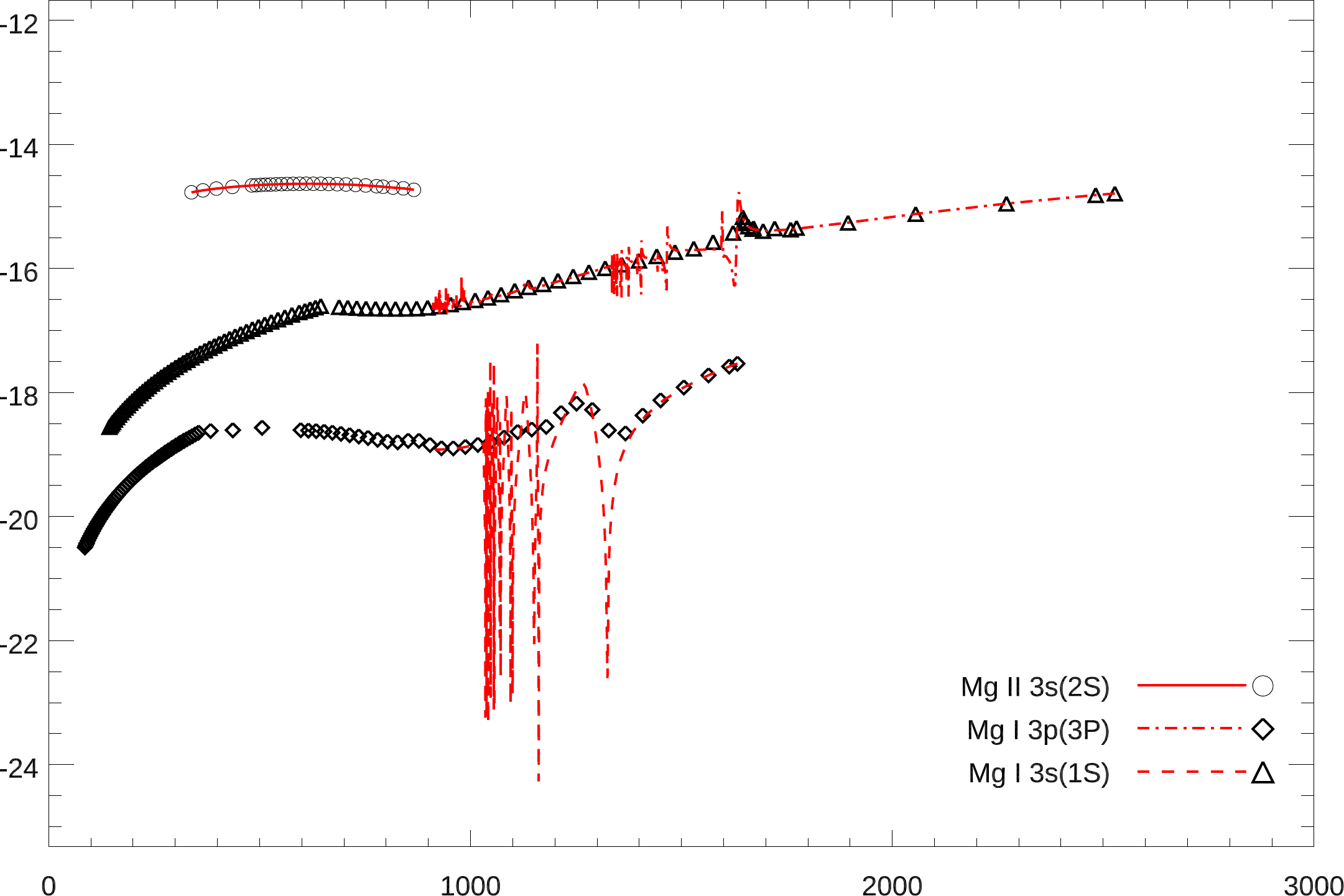}}
      };
      \node[rotate=90] at (-0.48\textwidth,0.15\textwidth) {\scalebox{0.75}{Photo-ionisation cross section ($\log$ cm$^2$)}};
      \node at (-0.22\textwidth,-0.015\textwidth) {\scalebox{0.75}{Photon wavelength (\AA)}};
    \end{tikzpicture}
  \caption{ Photo-ionisation cross section data used in the \multi\ (red lines) and \tlusty\ (black symbols) calculations for two b-f transitions of \Mg{i} and the one from the ground level of \Mg{ii}. The red lines are the values taken from the TOPBASE database with the cross sections bellow 911~\AA\ removed (except for the \Mg{ii} b-f transition). The symbols are the resonance averaged photo-ionisation cross sections of the TOPBASE data \cite[see][]{2003ApJS..147..363A}. The b-f cross section of the \Mg{ii} 3s(2S) and the \Mg{i} 3p(3P) levels are vertically shifted 4 and 2 log units respectively in order to ease visualisation.}\label{fig:bfMgI}
\end{figure}

The source of the data for Mg and Ca is the TOPBASE database \citep{TOPBASE}. 
For the \multi\ calculations, the data were used directly with the exclusion of some values. A maximum of 500 points per b-f transition  (starting from threshold) were used. If, for a given level, the cross section extended to energies higher than 13.6 eV (corresponding to 911~\AA), those data points were removed, since the lowest wavelength in the line-opacity table used by \multi\ is 900~\AA. This applies to all photo-ionisation cross sections, except the one of the ground level of \Mg{ii} because its threshold is at 14.3 eV (870~\AA, see \fig{fig:bfMgI}) and according to the above criteria, the whole cross section for this level would have been removed in the \multi\ calculations. Having the b-f taken directly makes it possible to have the resonances present in the data fully resolved for photon energies near the threshold. 

For \tlusty, the original TOPBASE data were smoothed and re-sampled using the resonance averaged photo-ionisation cross sections method presented by \cite{1998ApJS..118..259B}. This smoothing is physically justified given the uncertain in the atomic structure calculations, but it also brings a benefit due to the reduction in the number of frequencies required to describe the cross-sections. For these calculations, we used the full wavelength range provided by the sources for all the b-f transitions. Figure \ref{fig:bfMgI} shows the photo-ionisation data used for two levels of \Mg{i}\ in the \tlusty\ (symbols in black) and \multi\ (red lines) codes.

\subsection{Van der Waals broadening}\label{sec:cmp-vdw}

\begin{figure}  
    \begin{tikzpicture}
     \node[anchor=south east, inner sep=0] (image) at (0,0) {
        \subfloat{\includegraphics[width=0.45\textwidth]{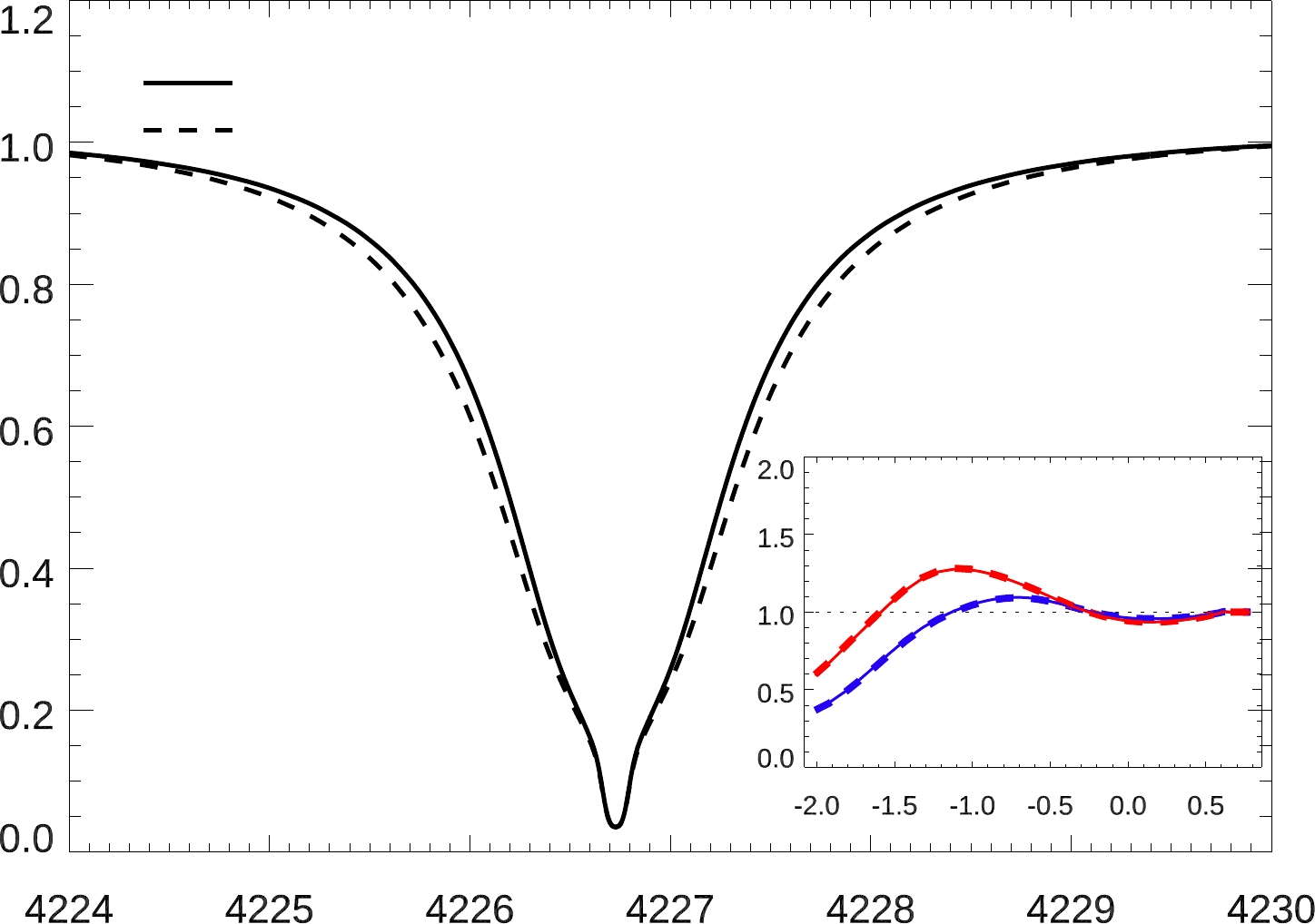}}
      };
      \node at (-0.347\textwidth,0.29\textwidth) {\scalebox{0.85}{ABO}};
      \node at (-0.342\textwidth,0.272\textwidth) {\scalebox{0.85}{Uns\"old}};
      \node[rotate=90] at (-0.47\textwidth,0.17\textwidth) {\scalebox{0.90}{Relative Flux}};
      \node at (-0.23\textwidth,-0.02\textwidth) {\scalebox{0.90}{$\lambda$ (\AA)}};
      \node[fill=white] at (-0.10\textwidth,0.025\textwidth) {\scalebox{0.60}{log(column mass)}};
      \node at (-0.20\textwidth,0.105\textwidth) {\scalebox{0.60}{$b$}};
      \node[color=red] at (-0.04\textwidth,0.13\textwidth) {\scalebox{0.90}{4s}};
      \node[color=blue] at (-0.04\textwidth,0.145\textwidth) {\scalebox{0.90}{4p}};
      \end{tikzpicture}
\caption{Comparison between \multi\ NLTE solar line profile and departure coefficients (inner figure) of the levels involved for the \Ca{i}\ transition 4p$^1$P - 4s$^1$S (4227~\AA); when the Van der Waals broadening is treated using the ABO theory (solid lines) and when the formulation from \cite{1955psmb.book.....U}, with a $\Gamma_6$ value derived from the ABO theory, is used (dashed lines).}\label{abovsgamma6}   
\end{figure}

When available, \multi\ uses the ABO theory directly by having the Van der Walls broadening input data in the  "$\sigma.\alpha$" format. \tlusty, on the other hand, uses the Van der Waals broadening coefficient $\Gamma_6$ at 10\,000~K. When available, the values adopted for \tlusty\ are the equivalent $\Gamma_6$ at 10\,000~K obtained from the ABO theory. This difference in the treatment of the spectral lines does not affect the derived NLTE populations, but it affects the line profiles. We have compared two \multi\ calculations: one using the ABO theory and the other treating van der Waals damping with the Uns\"old method \citep{1955psmb.book.....U} with an$\Gamma_6$ at T=10\,000~K derived from the ABO theory, and found no differences in the NLTE populations, although the wings of line profiles are clearly different (see \fig{abovsgamma6}). Therefore, the differences between the line profiles of \multi\ and \tlusty/\synspec\ for lines with ABO format data lie in the line-profile calculations between \multi\ and \synspec, and not in the NLTE populations derived by \multi\ and \tlusty. This is easy to understand because possible departures form LTE a driven by radiative rates, which in turn are dominated by the line cores.

\subsection{Collisional data}\label{sec:cmp-colls}  
\multi\ requires \emph{upward} rate coefficients for transitions resulting in ionisation and \emph{downward} rate coefficients for transitions resulting in excitation, while \tlusty\ demands \emph{upward} collisional rate coefficients in all cases. We used detailed balance relations in order to convert upward~$\leftrightarrow$~downward rate coefficients \citep[see ][\S 9.3]{HubenyMihalas2014}.

\begin{figure}[t]
  \hspace{-0.02\textwidth}
  \begin{tabular}{c}
    \begin{tikzpicture} 
      \node[anchor=south east, inner sep=0] (image) at (0,0) {
        \subfloat{\includegraphics[width=0.47\textwidth]{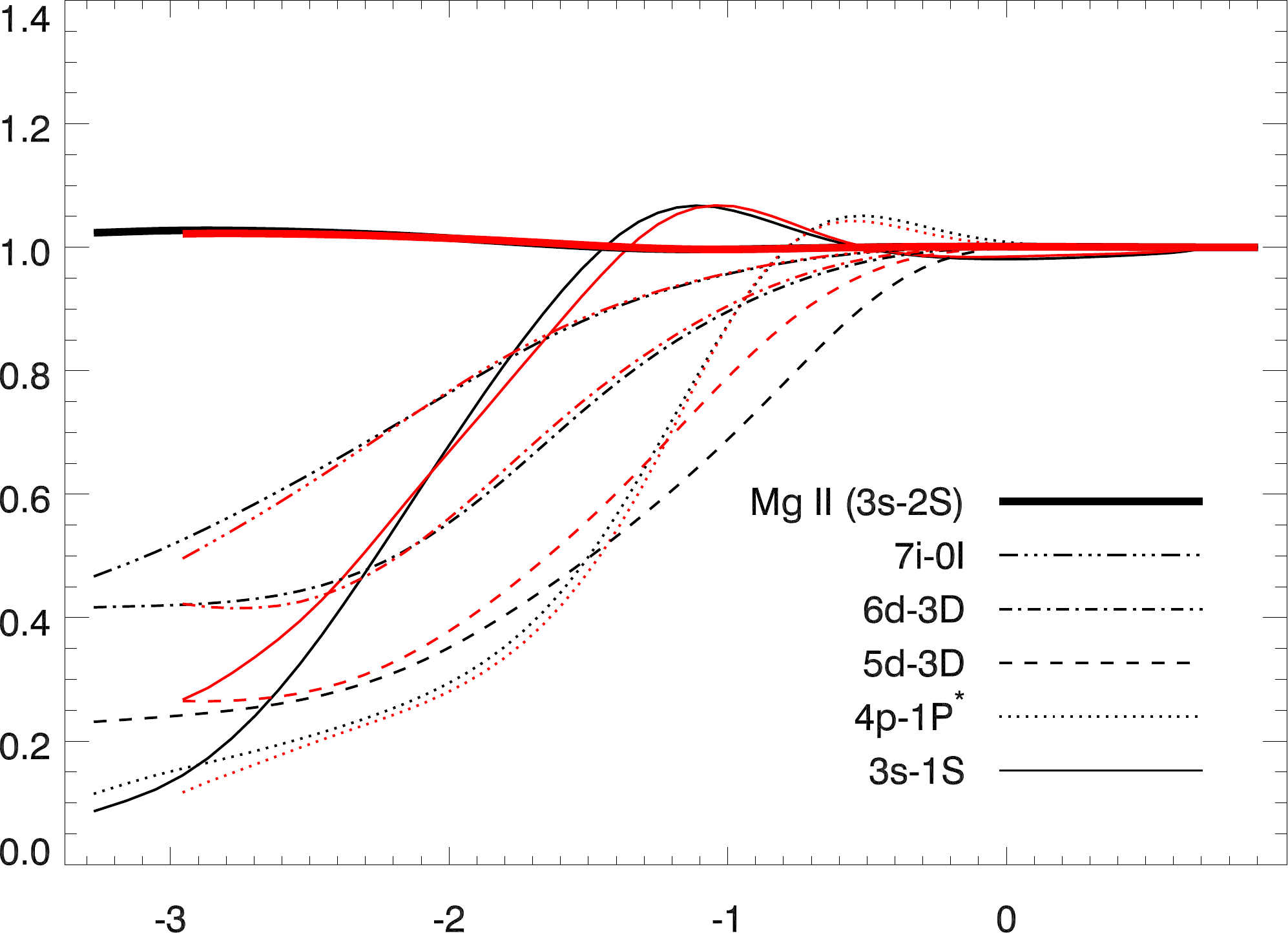}}
      };
      \node[color=red] at (-0.40\textwidth,0.32\textwidth) {\scalebox{0.90}{\tlusty}};
      \node at (-0.40\textwidth,0.30\textwidth) {\scalebox{0.90}{\multi}};
      \node at (-0.48\textwidth,0.18\textwidth) {\scalebox{1.00}{$b$}};
      \draw[color=gray,thick] (-0.04\textwidth,0.30\textwidth) -- (-0.04\textwidth,0.20\textwidth);
      \draw[color=gray,thick] (-0.137\textwidth,0.30\textwidth) -- (-0.137\textwidth,0.20\textwidth);
    \end{tikzpicture}
    \\[-0.01\textwidth]
    \begin{tikzpicture} 
      \node[anchor=south east, inner sep=0] (image) at (0,0) {
        \subfloat{\includegraphics[width=0.47\textwidth]{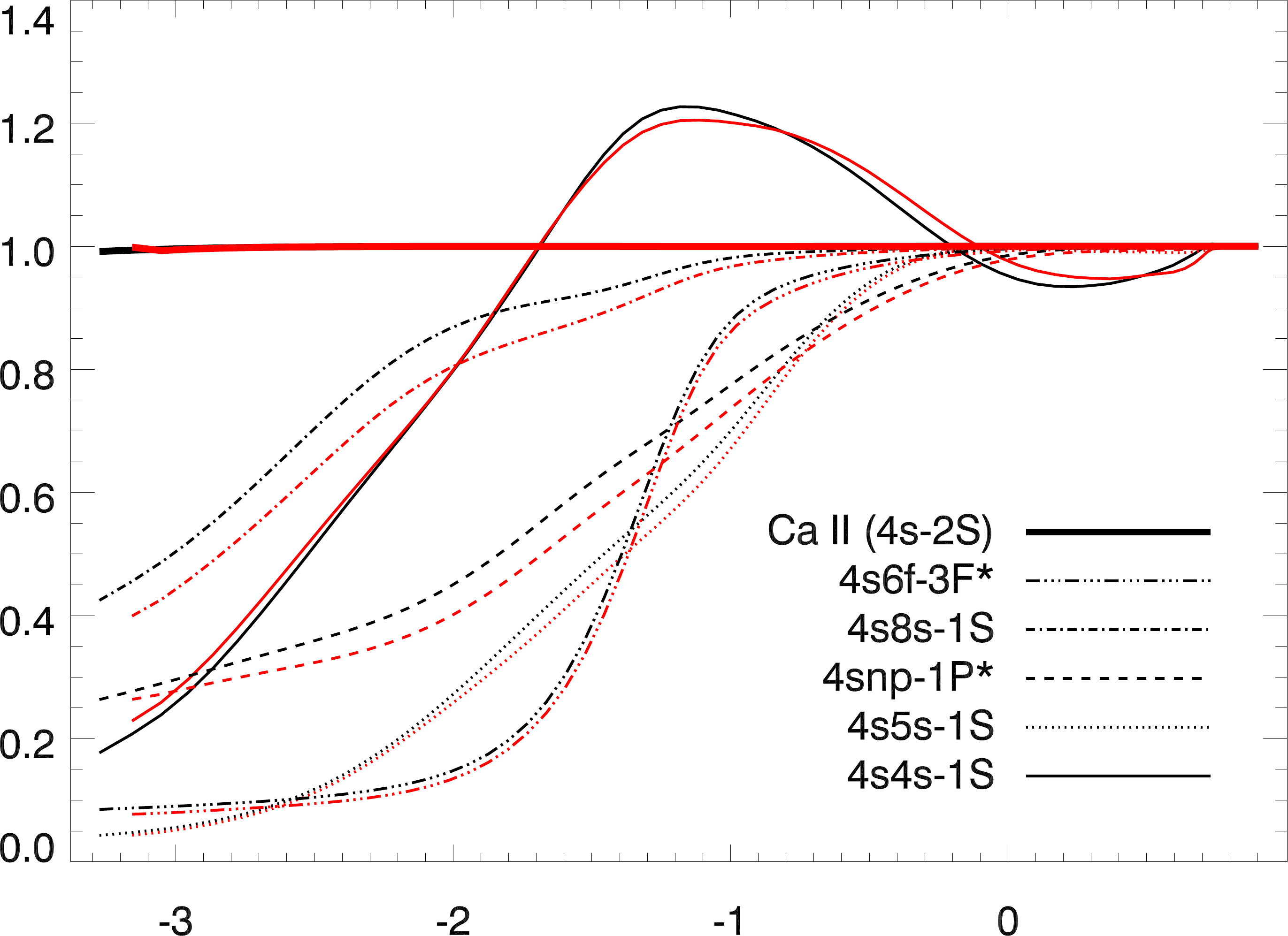}}
      };
      \node[color=red] at (-0.40\textwidth,0.32\textwidth) {\scalebox{0.90}{\tlusty}};
      \node at (-0.40\textwidth,0.30\textwidth) {\scalebox{0.90}{\multi}};
      \node at (-0.48\textwidth,0.18\textwidth) {\scalebox{1.00}{$b$}};
      \draw[color=gray,thick] (-0.04\textwidth,0.30\textwidth) -- (-0.04\textwidth,0.20\textwidth);
      \draw[color=gray,thick] (-0.137\textwidth,0.30\textwidth) -- (-0.137\textwidth,0.20\textwidth);
      \node at (-0.22\textwidth,-0.015\textwidth) {\scalebox{0.90}{log(column mass)}};
    \end{tikzpicture}
  \end{tabular}
  \caption{ Departure coefficients for some levels of \Mg{i}\ (upper panel) and \Ca{i}\ (lower panel) obtained with \tlusty\ (red) and \multi\ (black) for the same solar model atmosphere and the same atomic data. The grey vertical lines denote the depths at $\tau_{ross}=0.01$ (left) and $\tau_{ross}=1.0$ (right).}\label{fig:multl_cmp}
\end{figure}

\subsection{Results of the comparison}

The most direct comparison between these two codes is done by comparing the NLTE populations through the departure $b$ coefficients\footnote{The departure coefficient $b$  of level $i$ is defined as \[ b_{i}=n_{i}/n_{i}^*, \] where $n_i$ is the number density of level $i$ and the asterisk ($^*$) means LTE. so $n_i^*$ is the \emph{absolute} LTE population of level $i$, i.e. a population obtained by solving the set of LTE kinetic equations for all levels of all ionisation stages considered and \emph{not} by the usual definition $n_i^*=n_1^+ n_e \Phi_i(T)$ where $n_1^+$ is the actual population of the ground state of the next higher ion, $n_e$ the electron density and $\Phi_i$ the Saha-Boltzmann factor. see \cite[][Chapter 9, section 9.1]{HubenyMihalas2014}.} as function of the column mass. Traditionally, departure coefficients as shown as a function of optical depth $\tau$, but given that $\tau$ depends more on the background opacities (which is not exactly the same in \tlusty\ and \multi) than the column mass, and because the calculations in both codes are performed in the column mass scale, a more direct comparison between the two codes can be made using column mass as abscissae. Figure \ref{fig:multl_cmp} compares the obtained departure coefficients of some levels of \Mg{}\ and \Ca{}\ against column mass for a solar model. The behaviour is very similar for the two codes (see top panel of \fig{fig:multl_cmp}). The small discrepancies are due mainly to the differences in the background opacities. Figure \ref{fig:multl_cmp} gives us confidence on the ability of the new version of \tlusty\ to perform NLTE calculations in late-type stars.

Our experiments show that certain level of detail in the background opacities is required for calculating reliable NLTE populations. It is also important to ensure that the adopted background opacities cover the wavelengths associated to the most important b-b and b-f transitions of the species computed in NLTE. A more detailed comparison between \tlusty\ and other NLTE-trace-element radiative transfer codes will be presented in a future paper.

\section{Multi-element NLTE radiative transfer}\label{sec:nlte-m} 

In what follows we will label as \NLTEs\ the traditional single-element NLTE calculations (i.e. where only one atomic element was calculated in NLTE while all the other elements were kept in LTE) and \NLTEm\ to the simultaneous multi-element NLTE calculations. 
    
\begin{figure*}[!ht]
  \hspace{-0.03\textwidth}
  \begin{tabular}{c@{\hskip -1cm}c@{\hskip -1cm}c}
    \begin{tikzpicture} 
      \node[anchor=south east, inner sep=0] (image) at (0,0) {
        \subfloat{\includegraphics[width=0.38\textwidth]{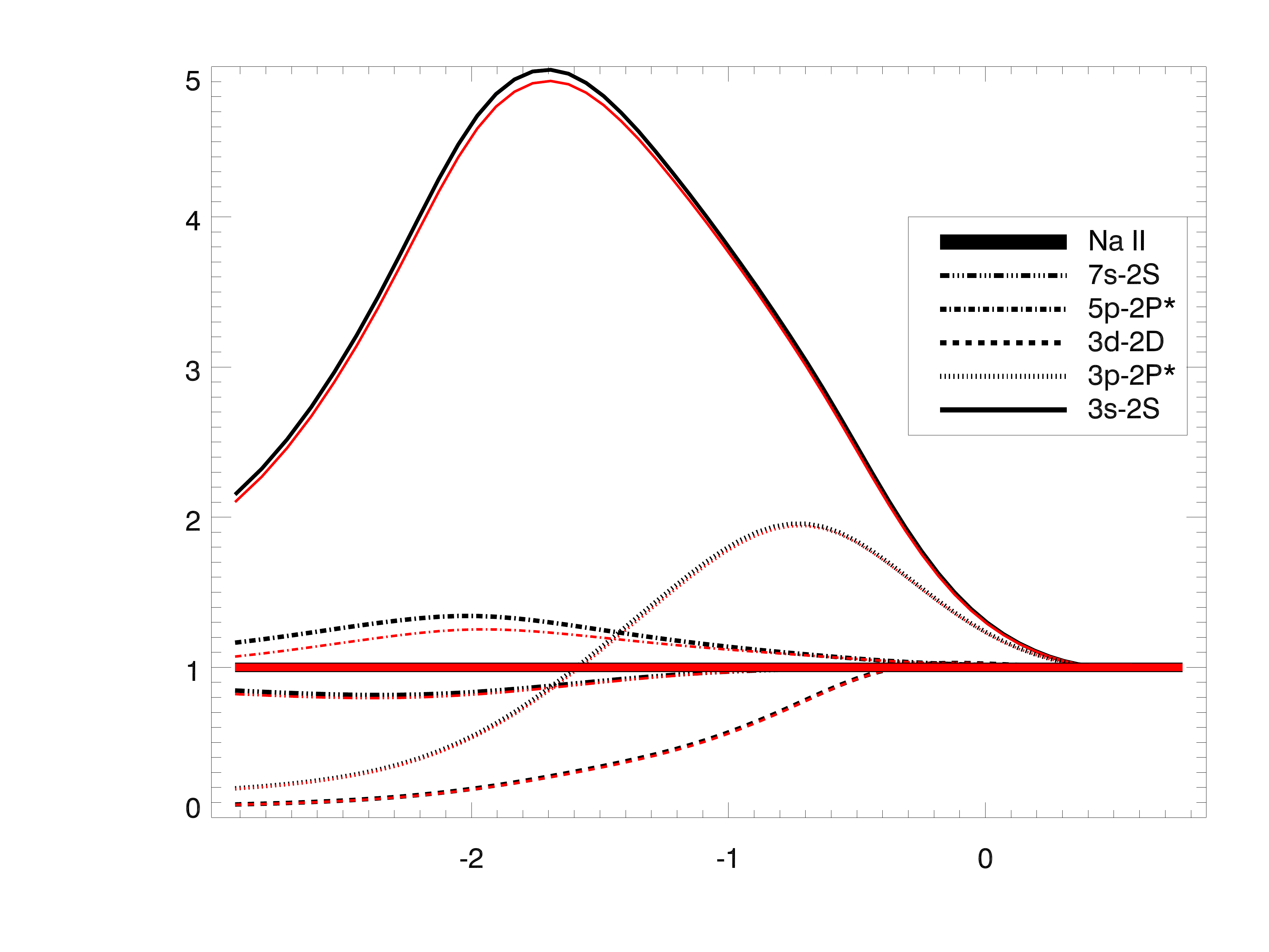}}
      };
      \node at (-0.267\textwidth,0.25\textwidth) {\scalebox{0.90}{Na \NLTEm}};
      \node[color=red] at (-0.27\textwidth,0.23\textwidth) {\scalebox{0.90}{Na \NLTEs}};
      \node at (-0.10\textwidth,0.25\textwidth) {\scalebox{1.00}{Procyon}};
       \draw[color=gray,thick] (-0.04\textwidth,0.12\textwidth) -- (-0.04\textwidth,0.04\textwidth);
       \draw[color=gray,thick] (-0.11\textwidth,0.12\textwidth) -- (-0.11\textwidth,0.04\textwidth);
      \node[rotate=0] at (-0.35\textwidth,0.15\textwidth) {\scalebox{0.80}{$b=\frac{n}{\,\,\,\,n_{_{lte}}}$}};
    \end{tikzpicture}
    & 
    \begin{tikzpicture} 
      \node[anchor=south east, inner sep=0] (image) at (0,0) {
        \subfloat{\includegraphics[width=0.38\textwidth]{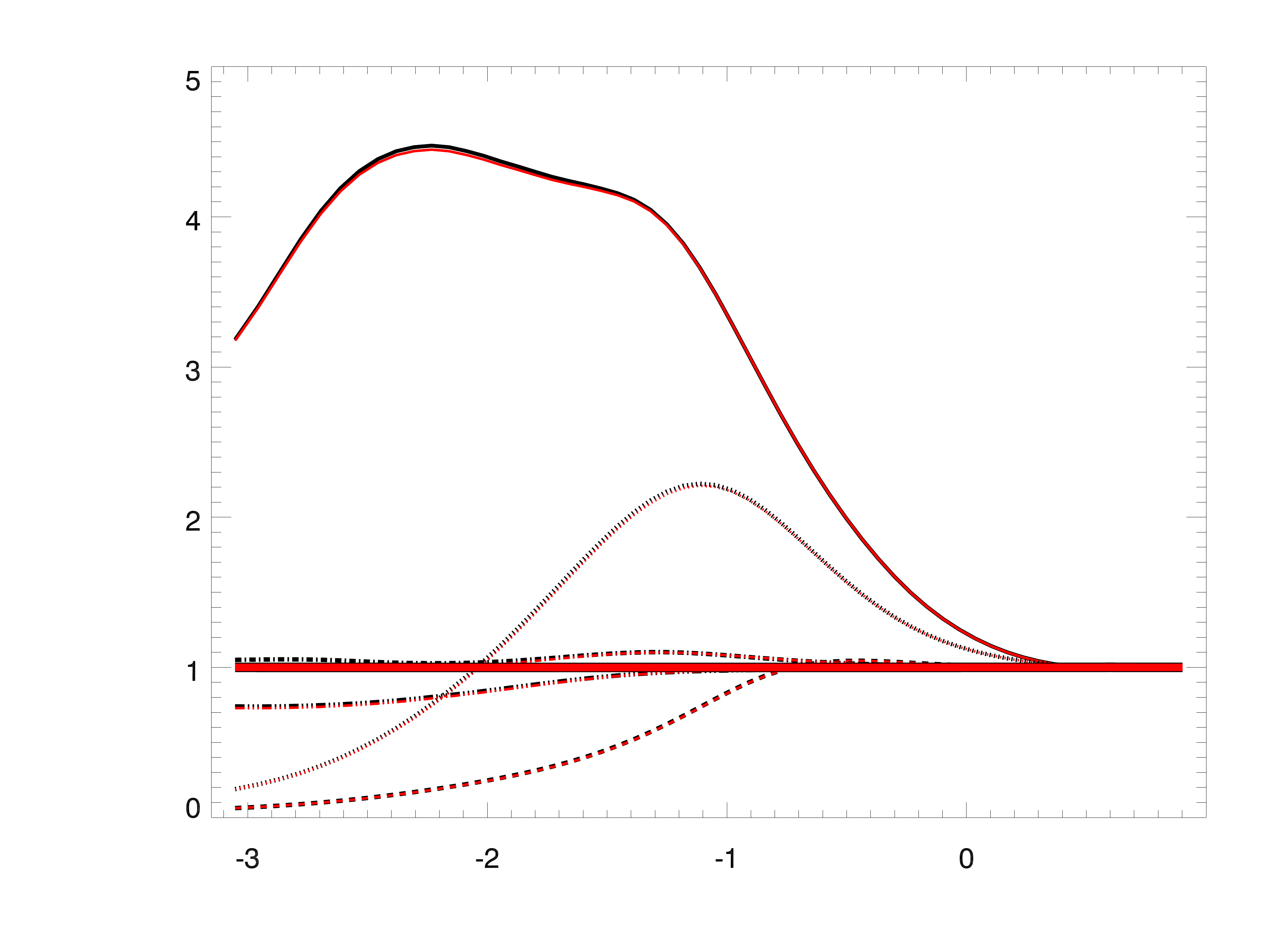}}
      };
      \node at (-0.10\textwidth,0.25\textwidth) {\scalebox{1.00}{Sun}};
       \draw[color=gray,thick] (-0.045\textwidth,0.12\textwidth) -- (-0.045\textwidth,0.04\textwidth);
       \draw[color=gray,thick] (-0.117\textwidth,0.12\textwidth) -- (-0.117\textwidth,0.04\textwidth);
    \end{tikzpicture}
    & 
        \begin{tikzpicture} 
      \node[anchor=south east, inner sep=0] (image) at (0,0) {
        \subfloat{\includegraphics[width=0.38\textwidth]{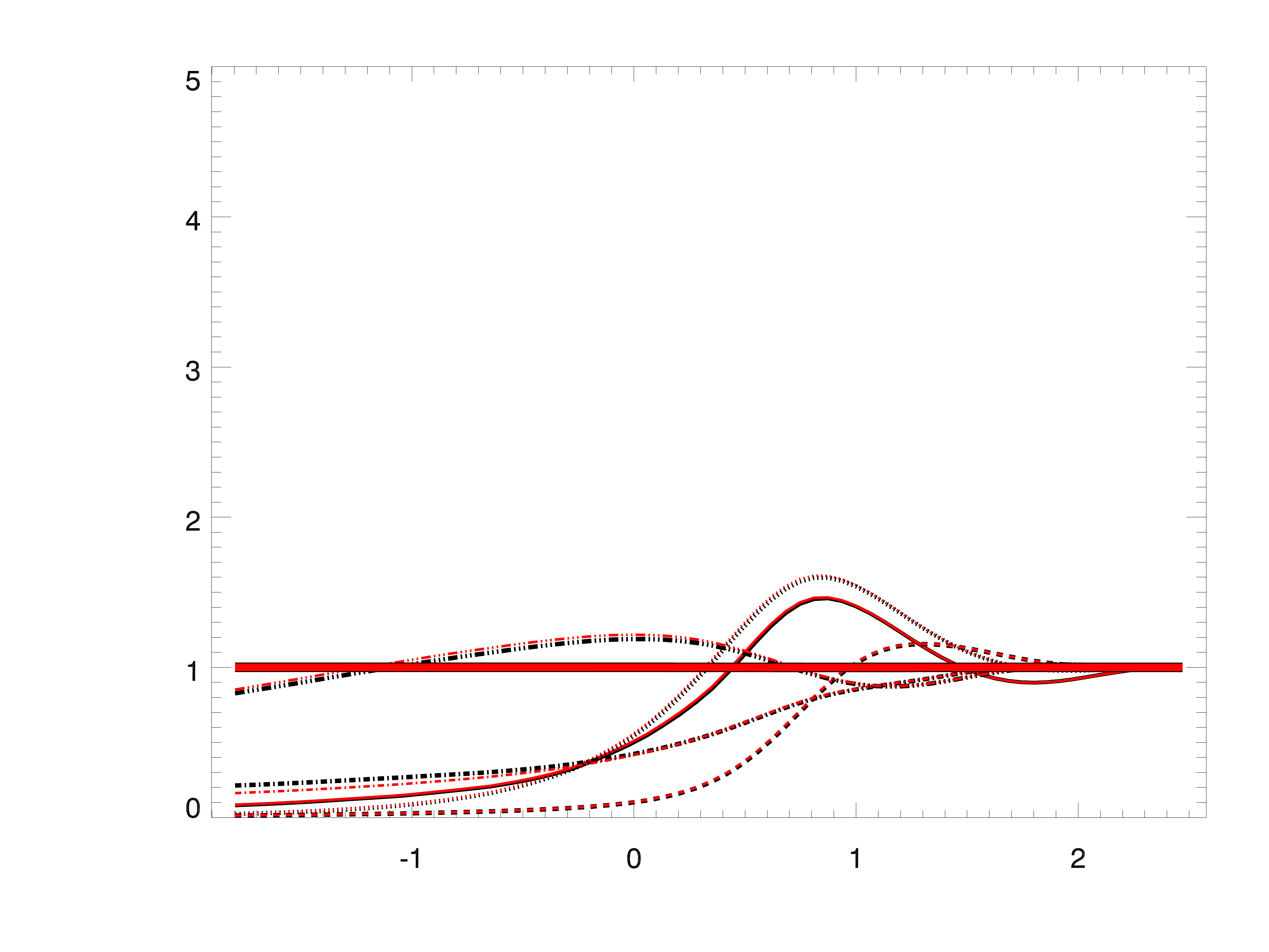}}
      };
      \node at (-0.10\textwidth,0.25\textwidth) {\scalebox{1.00}{Arcturus}};
       \draw[color=gray,thick] (-0.045\textwidth,0.12\textwidth) -- (-0.045\textwidth,0.04\textwidth);
       \draw[color=gray,thick] (-0.107\textwidth,0.12\textwidth) -- (-0.107\textwidth,0.04\textwidth);
    \end{tikzpicture}
        \\[-0.05\textwidth]
    \begin{tikzpicture} 
      \node[anchor=south east, inner sep=0] (image) at (0,0) {
        \subfloat{\includegraphics[width=0.38\textwidth]{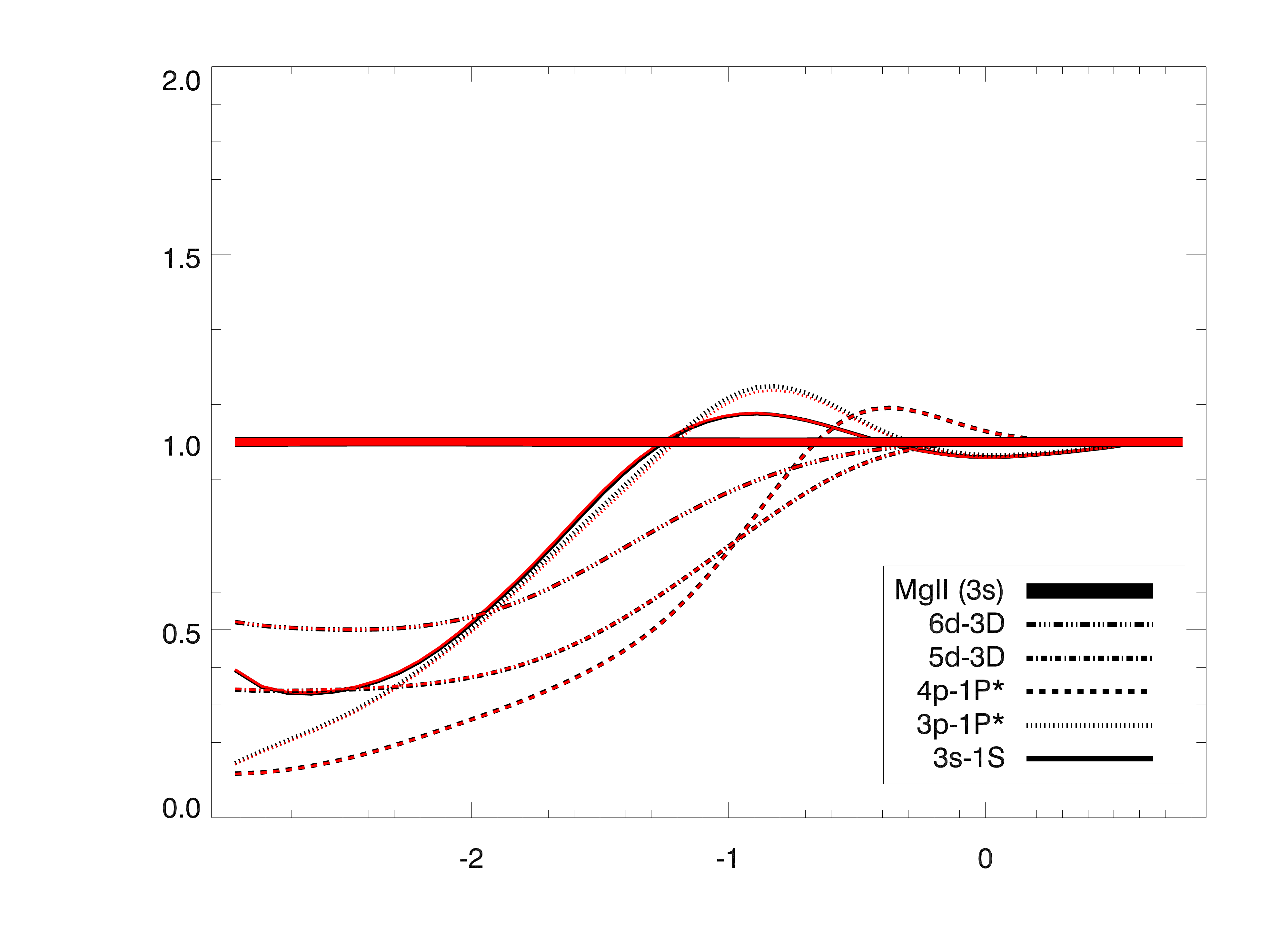}}
      };
      \node at (-0.267\textwidth,0.25\textwidth) {\scalebox{0.90}{Mg \NLTEm}};
      \node[color=red] at (-0.27\textwidth,0.23\textwidth) {\scalebox{0.90}{Mg \NLTEs}};
      \node at (-0.10\textwidth,0.25\textwidth) {\scalebox{1.00}{Procyon}};
       \draw[color=gray,thick] (-0.04\textwidth,0.12\textwidth) -- (-0.04\textwidth,0.20\textwidth);
       \draw[color=gray,thick] (-0.11\textwidth,0.12\textwidth) -- (-0.11\textwidth,0.20\textwidth);
      \node[rotate=0] at (-0.35\textwidth,0.14\textwidth) {\scalebox{0.80}{$b=\frac{n}{\,\,\,\,n_{_{lte}}}$}};
    \end{tikzpicture}
    & 
    \begin{tikzpicture} 
      \node[anchor=south east, inner sep=0] (image) at (0,0) {
        \subfloat{\includegraphics[width=0.38\textwidth]{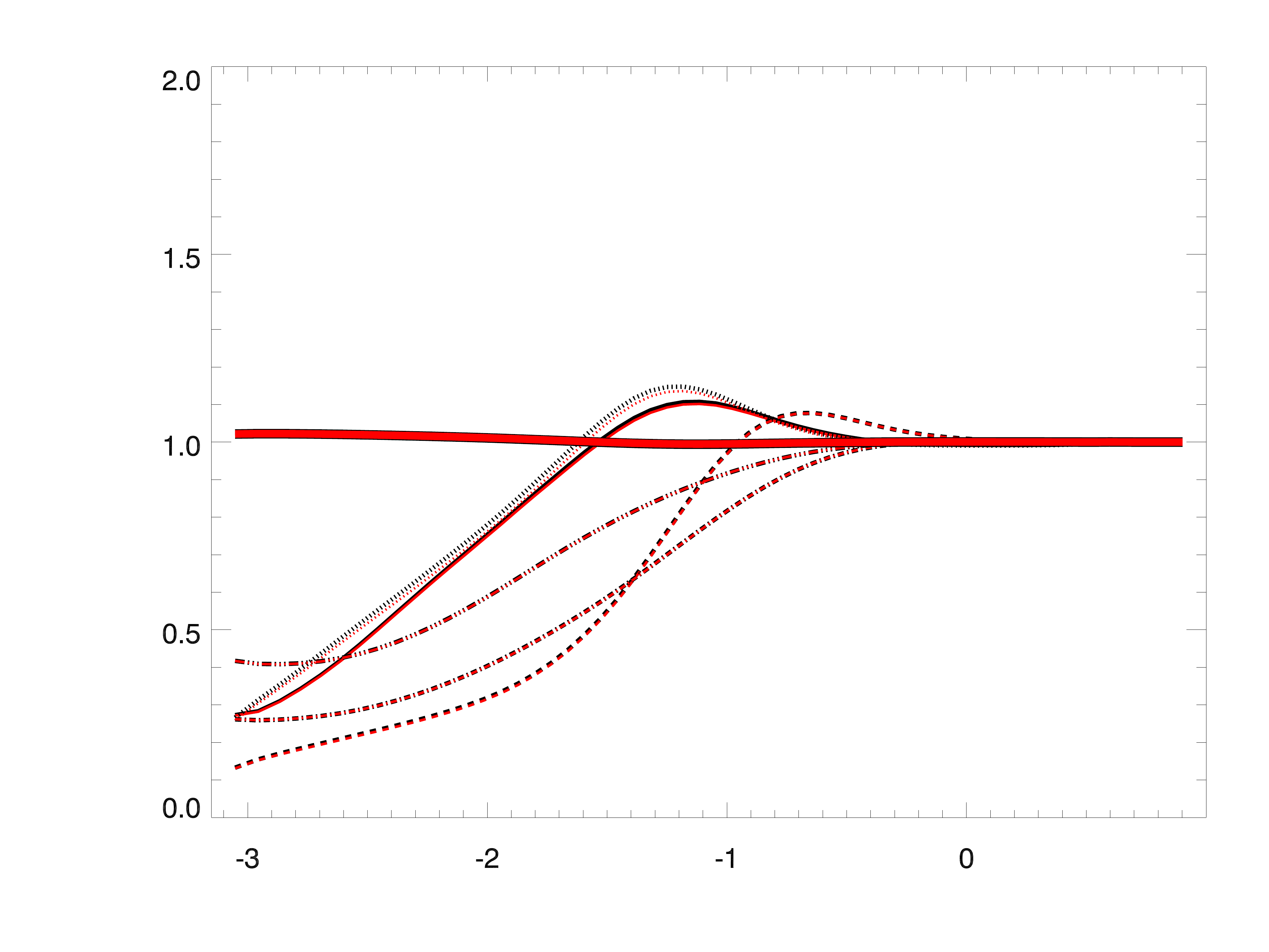}}
      };
      \node at (-0.10\textwidth,0.25\textwidth) {\scalebox{1.00}{Sun}};
       \draw[color=gray,thick] (-0.045\textwidth,0.20\textwidth) -- (-0.045\textwidth,0.12\textwidth);
       \draw[color=gray,thick] (-0.117\textwidth,0.20\textwidth) -- (-0.117\textwidth,0.12\textwidth);
    \end{tikzpicture}
    & 
        \begin{tikzpicture} 
      \node[anchor=south east, inner sep=0] (image) at (0,0) {
        \subfloat{\includegraphics[width=0.38\textwidth]{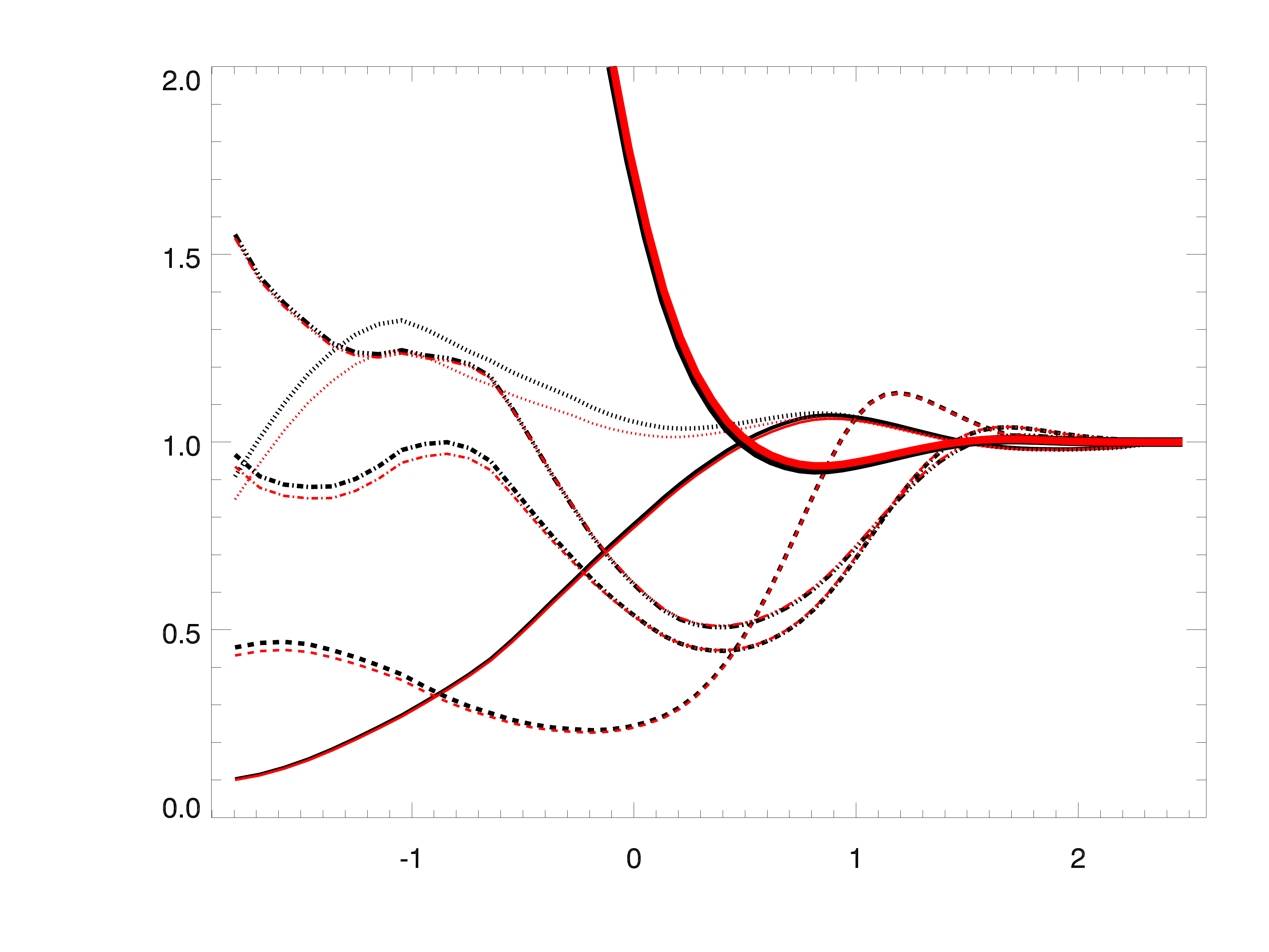}}
      };
      \node at (-0.10\textwidth,0.25\textwidth) {\scalebox{1.00}{Arcturus}};
       \draw[color=gray,thick] (-0.045\textwidth,0.20\textwidth) -- (-0.045\textwidth,0.12\textwidth);
       \draw[color=gray,thick] (-0.107\textwidth,0.20\textwidth) -- (-0.107\textwidth,0.12\textwidth);
    \end{tikzpicture}        
        \\[-0.05\textwidth]
    \begin{tikzpicture} 
      \node[anchor=south east, inner sep=0] (image) at (0,0) {
        \subfloat{\includegraphics[width=0.38\textwidth]{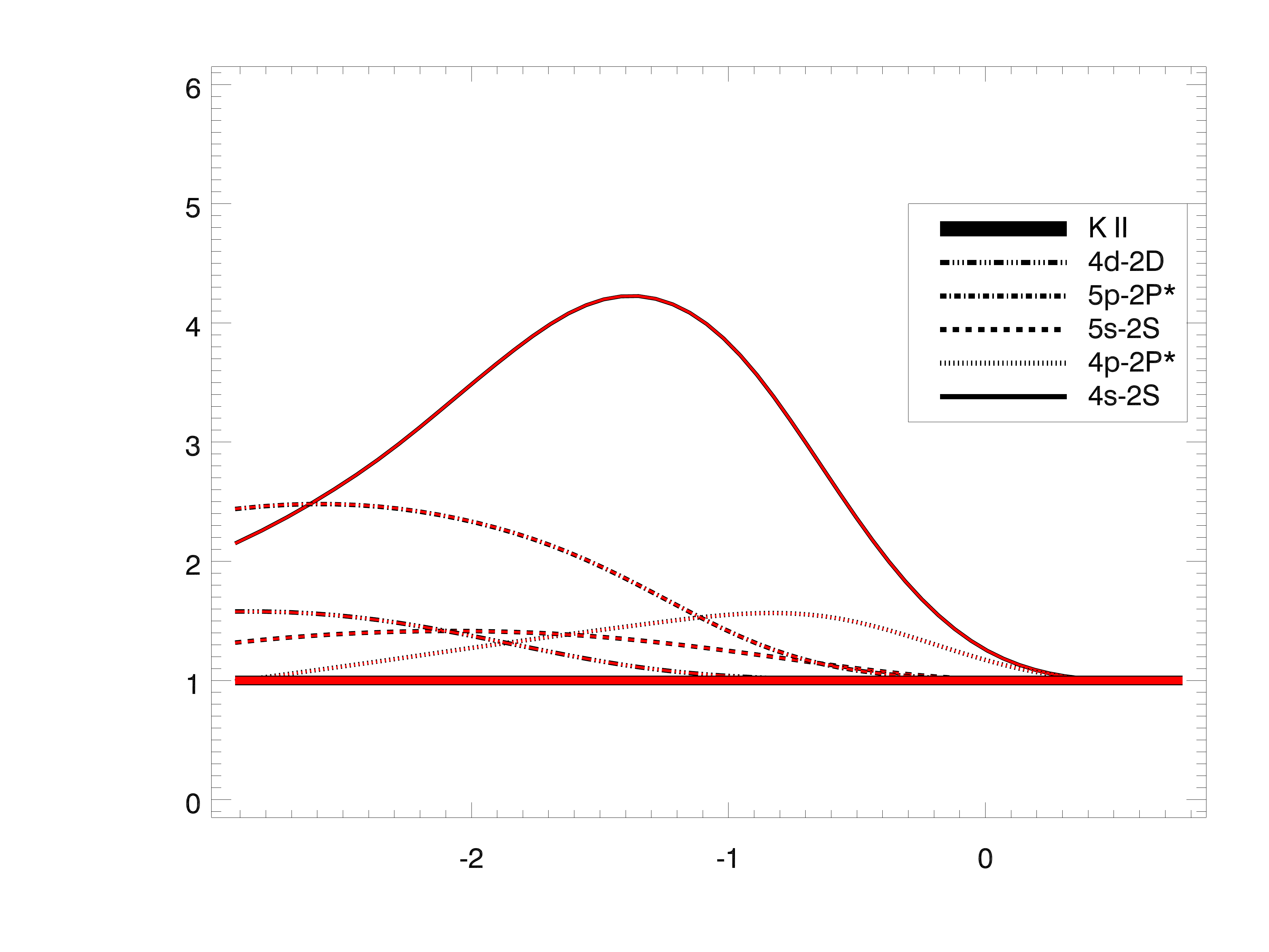}}
      };
      \node at (-0.267\textwidth,0.25\textwidth) {\scalebox{0.90}{K \NLTEm}};
      \node[color=red] at (-0.27\textwidth,0.23\textwidth) {\scalebox{0.90}{K \NLTEs}};
      \node at (-0.10\textwidth,0.25\textwidth) {\scalebox{1.00}{Procyon}};
       \draw[color=gray,thick] (-0.04\textwidth,0.12\textwidth) -- (-0.04\textwidth,0.04\textwidth);
       \draw[color=gray,thick] (-0.11\textwidth,0.12\textwidth) -- (-0.11\textwidth,0.04\textwidth);
      \node[rotate=0] at (-0.35\textwidth,0.14\textwidth) {\scalebox{0.80}{$b=\frac{n}{\,\,\,\,n_{_{lte}}}$}};
    \end{tikzpicture}
    & 
    \begin{tikzpicture} 
      \node[anchor=south east, inner sep=0] (image) at (0,0) {
        \subfloat{\includegraphics[width=0.38\textwidth]{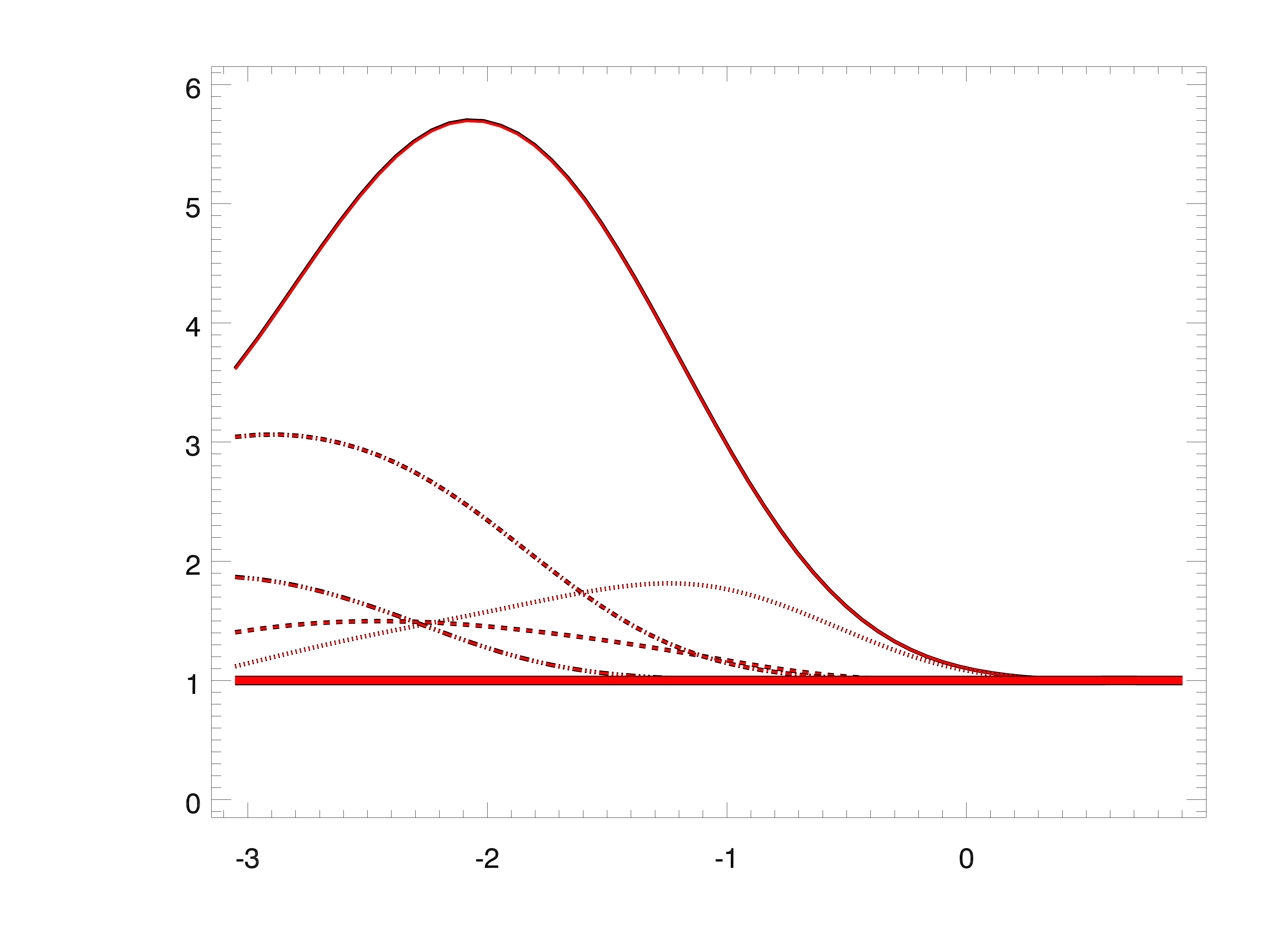}}
      };
      \node at (-0.10\textwidth,0.25\textwidth) {\scalebox{1.00}{Sun}};
       \draw[color=gray,thick] (-0.045\textwidth,0.12\textwidth) -- (-0.045\textwidth,0.04\textwidth);
       \draw[color=gray,thick] (-0.117\textwidth,0.12\textwidth) -- (-0.117\textwidth,0.04\textwidth);
    \end{tikzpicture}
    & 
        \begin{tikzpicture} 
      \node[anchor=south east, inner sep=0] (image) at (0,0) {
        \subfloat{\includegraphics[width=0.38\textwidth]{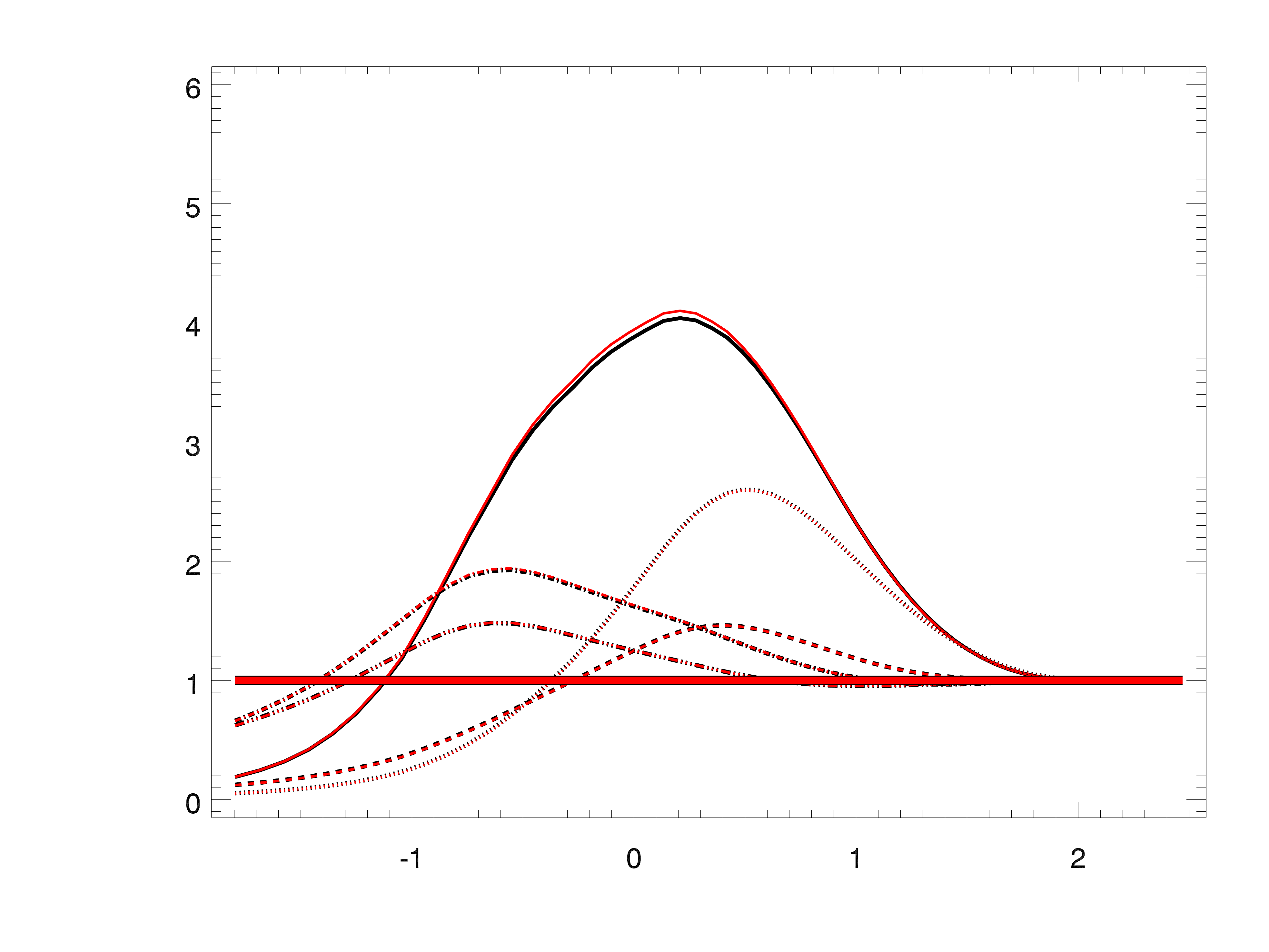}}
      };
      \node at (-0.10\textwidth,0.25\textwidth) {\scalebox{1.00}{Arcturus}};
       \draw[color=gray,thick] (-0.045\textwidth,0.12\textwidth) -- (-0.045\textwidth,0.04\textwidth);
       \draw[color=gray,thick] (-0.107\textwidth,0.12\textwidth) -- (-0.107\textwidth,0.04\textwidth);
    \end{tikzpicture}        
        \\[-0.05\textwidth]
    \begin{tikzpicture} 
      \node[anchor=south east, inner sep=0] (image) at (0,0) {
        \subfloat{\includegraphics[width=0.38\textwidth]{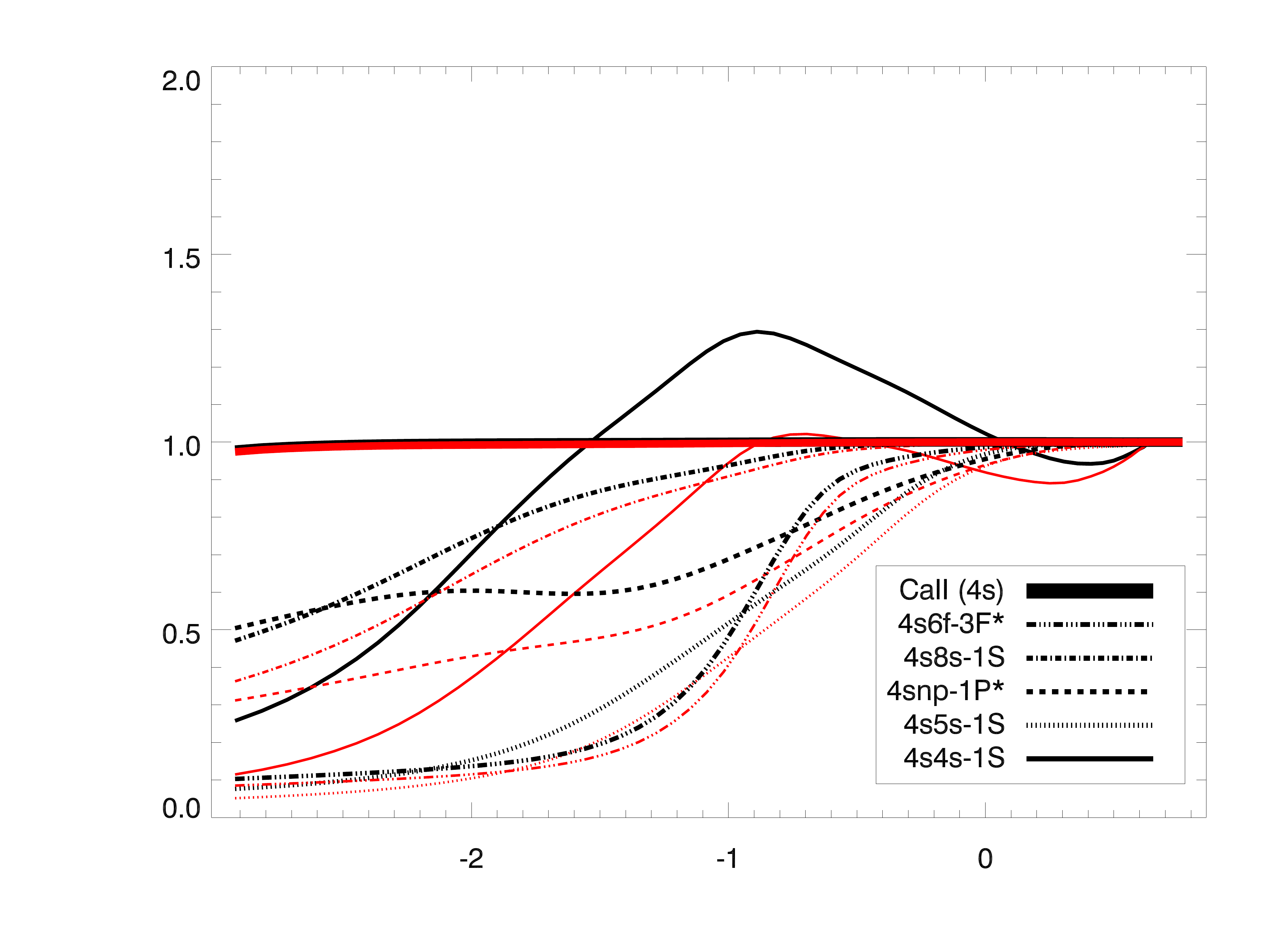}}
      };
      \node at (-0.267\textwidth,0.25\textwidth) {\scalebox{0.90}{Ca \NLTEm}};
      \node[color=red] at (-0.27\textwidth,0.23\textwidth) {\scalebox{0.90}{Ca \NLTEs}};
      \node at (-0.10\textwidth,0.25\textwidth) {\scalebox{1.00}{Procyon}};
       \draw[color=gray,thick] (-0.04\textwidth,0.12\textwidth) -- (-0.04\textwidth,0.20\textwidth);
       \draw[color=gray,thick] (-0.11\textwidth,0.12\textwidth) -- (-0.11\textwidth,0.20\textwidth);
      \node[rotate=0] at (-0.35\textwidth,0.14\textwidth) {\scalebox{0.80}{$b=\frac{n}{\,\,\,\,n_{_{lte}}}$}};
      \node at (-0.17\textwidth,0.01\textwidth) {\scalebox{0.8}{log column mass}};   
    \end{tikzpicture}
    & 
    \begin{tikzpicture} 
      \node[anchor=south east, inner sep=0] (image) at (0,0) {
        \subfloat{\includegraphics[width=0.38\textwidth]{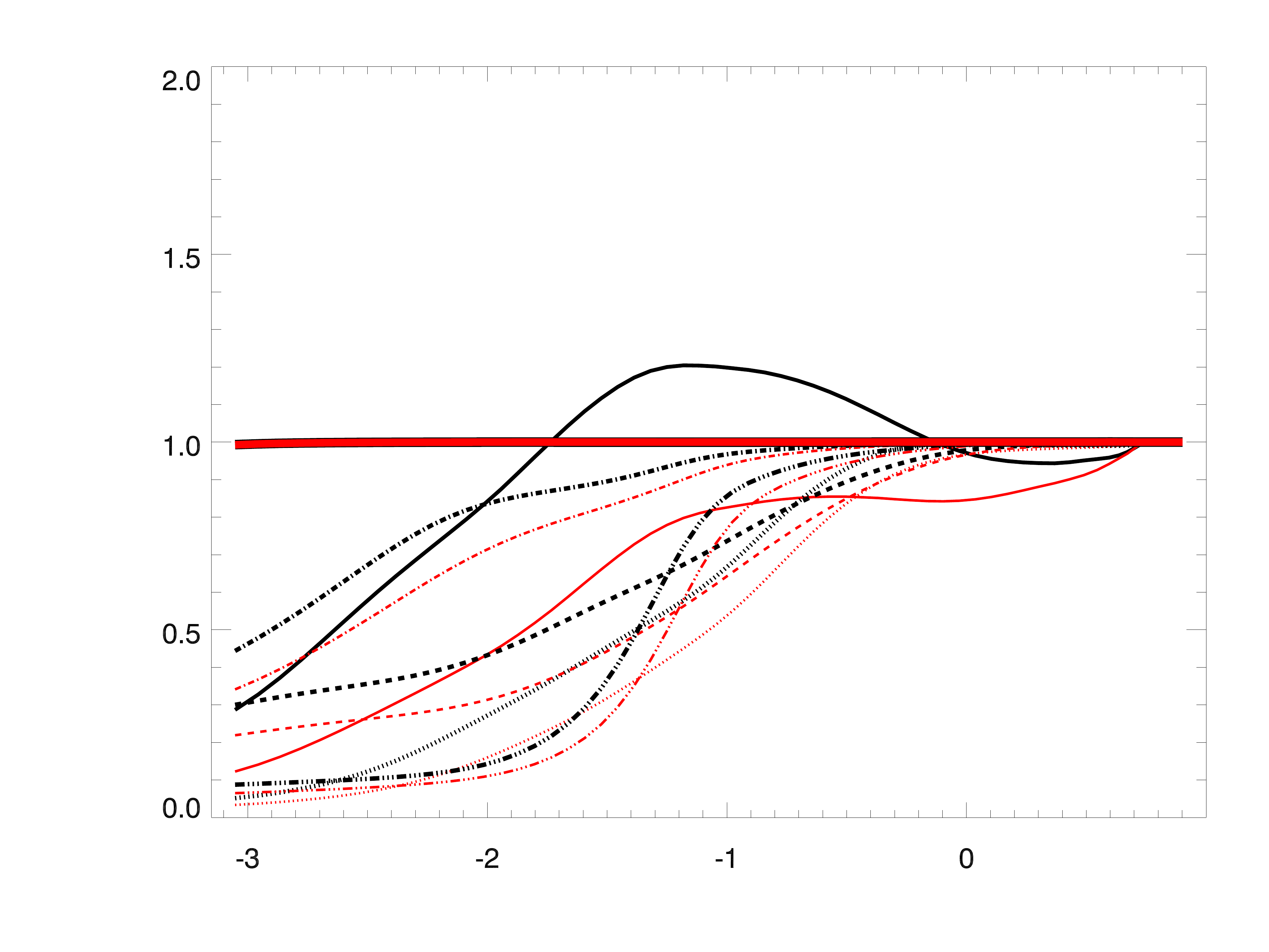}}
      };
      \node at (-0.10\textwidth,0.25\textwidth) {\scalebox{1.00}{Sun}};
      \node at (-0.17\textwidth,0.01\textwidth) {\scalebox{0.8}{log column mass}};   
       \draw[color=gray,thick] (-0.045\textwidth,0.12\textwidth) -- (-0.045\textwidth,0.20\textwidth);
       \draw[color=gray,thick] (-0.117\textwidth,0.12\textwidth) -- (-0.117\textwidth,0.20\textwidth);
    \end{tikzpicture}
    & 
        \begin{tikzpicture} 
      \node[anchor=south east, inner sep=0] (image) at (0,0) {
        \subfloat{\includegraphics[width=0.38\textwidth]{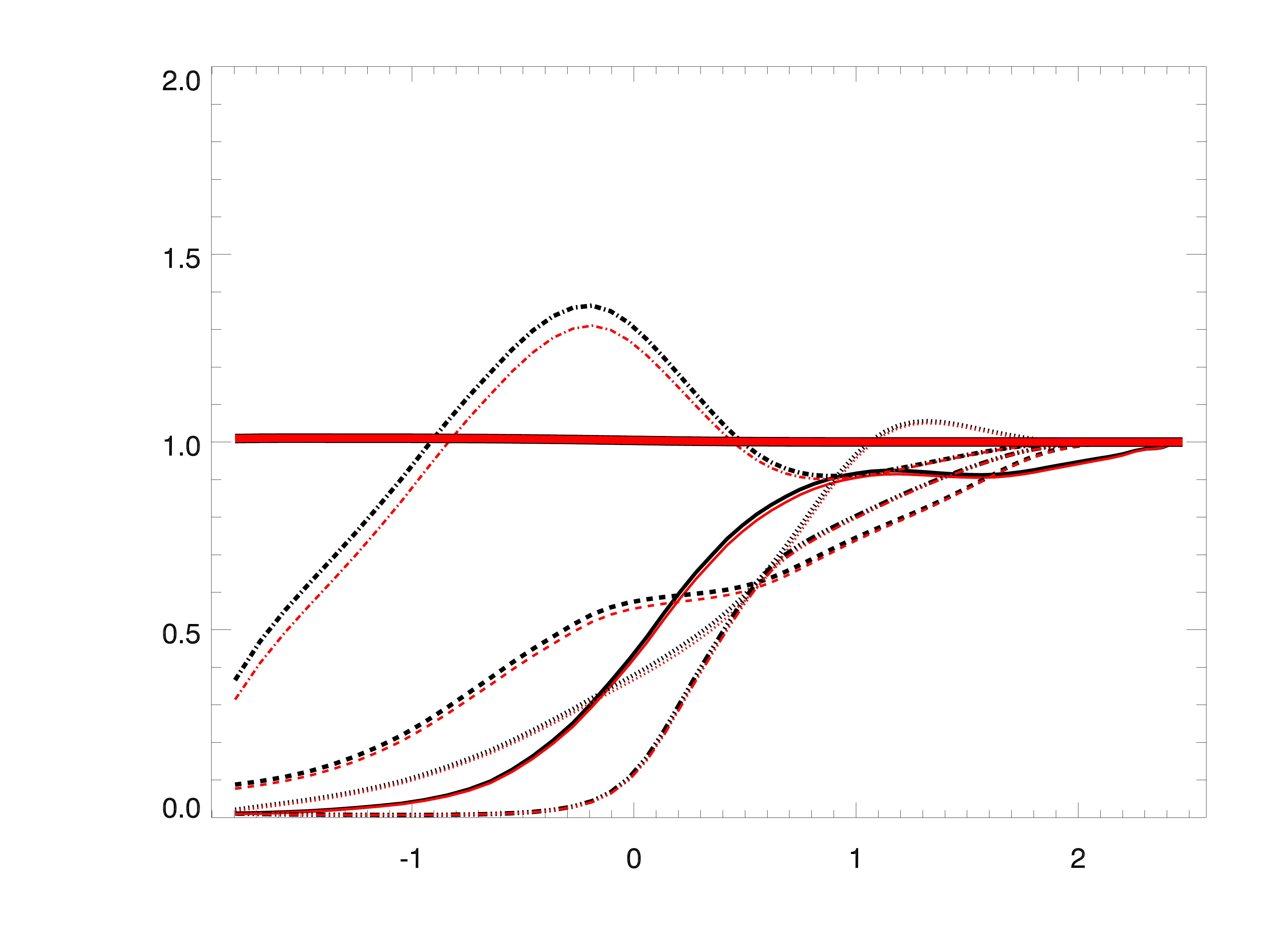}}
      };
      \node at (-0.10\textwidth,0.25\textwidth) {\scalebox{1.00}{Arcturus}};
      \node at (-0.17\textwidth,0.01\textwidth) {\scalebox{0.8}{log column mass}};   
       \draw[color=gray,thick] (-0.045\textwidth,0.20\textwidth) -- (-0.045\textwidth,0.12\textwidth);
       \draw[color=gray,thick] (-0.107\textwidth,0.20\textwidth) -- (-0.107\textwidth,0.12\textwidth);
    \end{tikzpicture}        
  \end{tabular}
\caption{Departure coefficients for some levels of \Na,\, \Mg,\, \K\ and \Ca\ obtained calculated in single-element mode (\NLTEs, in red) and in multiple-elements mode (\NLTEm, in black) for Procyon, the Sun and Arcturus. For each star, the same element abundances were used in all the LTE, \NLTEs\ and \NLTEm\ calculations.The fact that the $b$ coefficient for the \Mg{ii} ground state increases toward the surface (due to NLTE effects in the \Mg{ii} h \& k lines) in Arcturus, but stays closely to 1 for the other two stars, follows from \Mg{ii} being the dominant stage of ionisation, and its ground state the most populated state,  for Sun and Procyon, but not for the cooler star Arcturus. The two grey vertical lines in each figure mark the depths in the atmospheres where $\tau_{ross}$=0.01 (left) and 1.0 (right)}
\label{fig:bcomp}
\end{figure*}

We performed \NLTEs\ calculations for \Na,\, \Mg,\, \K\ and \Ca. We also performed \NLTEm\ calculations with all possible combination of two, three and the four elements in NLTE. As a result we found that departures from LTE in \Mg\ is affecting mainly the \Ca\ populations but does not have significant effects on \Na\ or \K. \Ca\ has a marginal effect on \Mg, especially in Arcturus, but in this star NLTE effects in \Mg\ have a diminished influence on the NLTE \Ca\ populations. \Na\ and \K\ are not affecting the populations of each other, \Mg\ or \Ca\ in any of the stars investigated in this work. 

We emphasise that all the calculations described have been performed with exactly the same atomic data: the opacity tables used for the different \NLTEs\ and \NLTEm\ calculations are constructed exactly in the same fashion with the exact same b-b and b-f data, only removing the contribution to the opacity of the elements to be calculated in NLTE. The model atoms of \Mg{i \& ii} and \Ca{i \& ii} for \multi\ have the same radiative and collisional sources than the ones used for \tlusty\ (the differences in the data implemented in the two codes are described in \S~\ref{sec:comparison}). The same \Na{i}, \Mg{i \& ii}, \K{i}\ and \Ca{i \& ii} model atoms where used in all the NLTE calculations in \tlusty and \synspec. Thus, the differences found in the \NLTEs\ and \NLTEm\ populations and spectra are due only to inter-element NLTE effects.    

Given that in all the calculations we keep fixed the atmospheric structure i.e., we are still within the trace-element approach framework, the influence of one NLTE element on the others enters trough its contribution to the opacity. A comparison of the NLTE departure coefficients $b$ between the \NLTEs\ and \NLTEm\ cases can give us information on the importance of inter-element NLTE effects (see below).

\subsection{Inter-element NLTE effects}\label{sec:inter-element} 

Our experiments show that the \Ca\ NLTE populations are sensitive to \Mg\ NLTE opacities, but the \Mg\ NLTE populations are not affected by \Ca\ NLTE opacities significantly. This opens the question of how much the effect of \ion{Fe}{} NLTE opacities will affect the NLTE effects of other elements considering that \ion{Fe}{} is the metal that contribute the most to background opacities at solar metallicities, in particular in the UV region.  We now proceed to interpret the behaviour of \NLTEm\ with respect to \NLTEs in the stars and atomic elements studied in this work.

\begin{figure*}[t]
  \hspace{-0.03\textwidth}
  \begin{tabular}{c}
    \begin{tikzpicture} 
      \node[anchor=south east, inner sep=0] (image) at (0,0) {
        \subfloat{\includegraphics[width=1.0\textwidth]{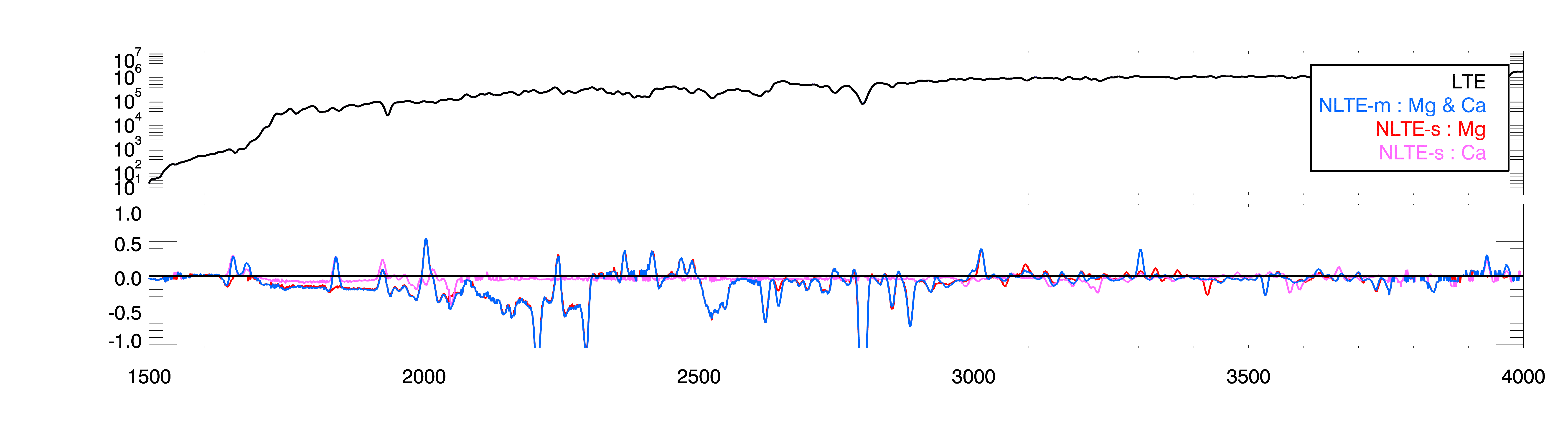}}
      };
      \node at (-0.85\textwidth,0.23\textwidth) {\scalebox{1.00}{Procyon}};
      \node[rotate=90] at (-0.95\textwidth,0.20\textwidth) {\scalebox{0.75}{$H_\lambda$}};
      \node[rotate=90] at (-0.95\textwidth,0.10\textwidth) {\scalebox{0.75}{$\Delta_{\rm{LTE}}$}(\%)};
    \end{tikzpicture}
    \\[-0.08\textwidth]
    \begin{tikzpicture} 
      \node[anchor=south east, inner sep=0] (image) at (0,0) {
        \subfloat{\includegraphics[width=1.0\textwidth]{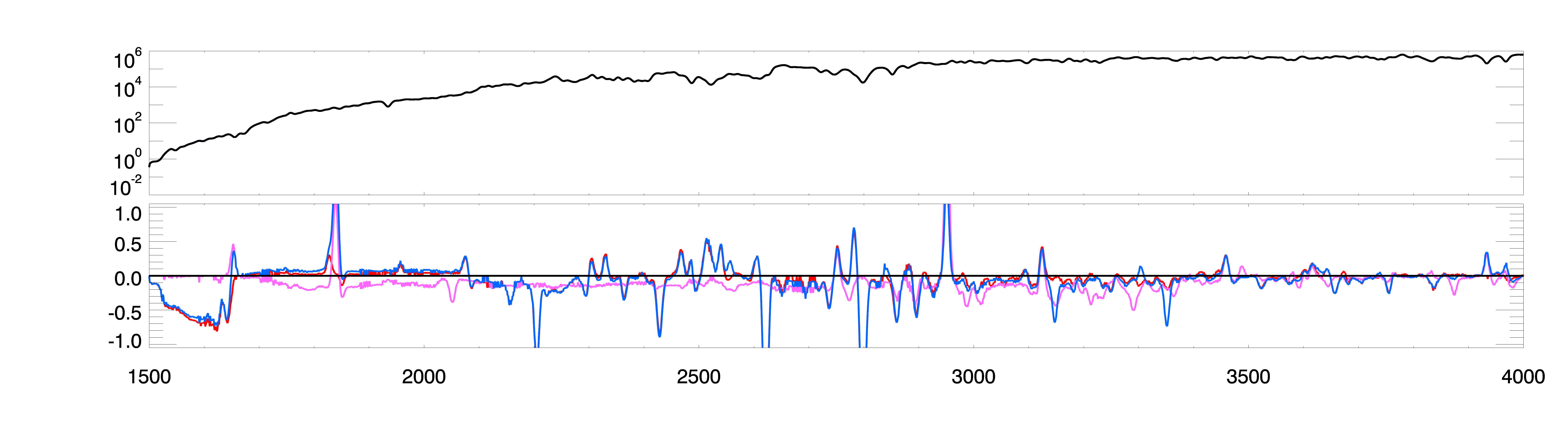}}
      };
      \node at (-0.85\textwidth,0.23\textwidth) {\scalebox{1.00}{Sun}};
      \node[rotate=90] at (-0.95\textwidth,0.20\textwidth) {\scalebox{0.75}{$H_\lambda$}};
      \node[rotate=90] at (-0.95\textwidth,0.10\textwidth) {\scalebox{0.75}{$\Delta_{\rm{LTE}}$}(\%)};
    \end{tikzpicture}
    \\[-0.08\textwidth]
        \begin{tikzpicture} 
      \node[anchor=south east, inner sep=0] (image) at (0,0) {
        \subfloat{\includegraphics[width=1.0\textwidth]{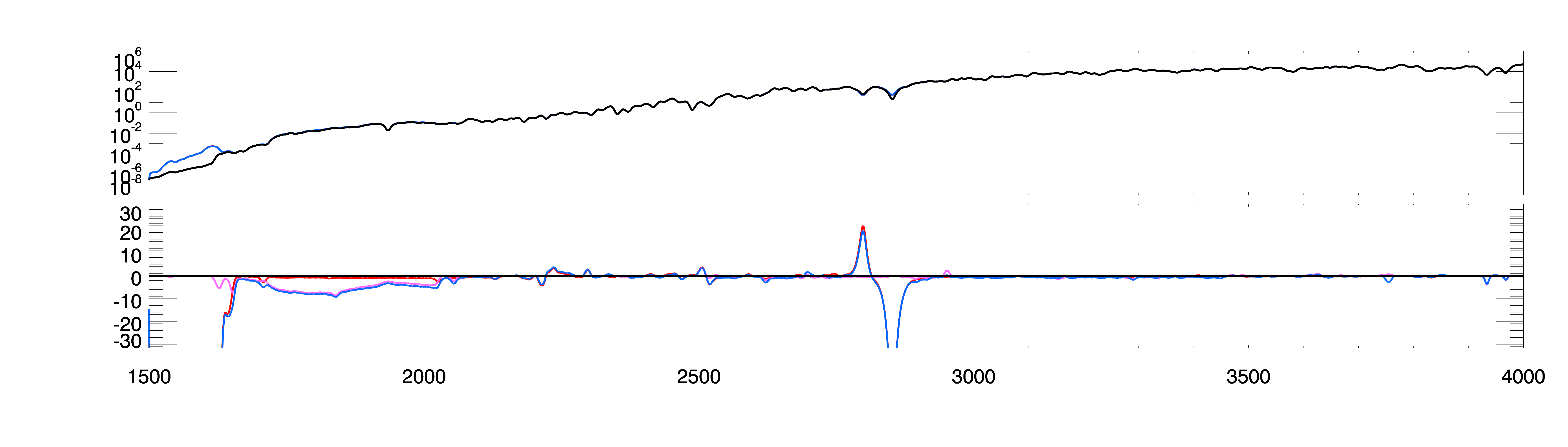}}
      };
      \node at (-0.85\textwidth,0.23\textwidth) {\scalebox{1.00}{Arcturus}};
      \node[rotate=90] at (-0.95\textwidth,0.20\textwidth) {\scalebox{0.75}{$H_\lambda$}};
      \node[rotate=90] at (-0.95\textwidth,0.10\textwidth) {\scalebox{0.75}{$\Delta_{\rm{LTE}}$}(\%)};
      \node at (-0.47\textwidth,0.024\textwidth) {\scalebox{0.75}{$\lambda$ (\AA)}};
    \end{tikzpicture}
  \end{tabular}
\caption{UV fluxes obtained for Procyon (top), the Sun (middle) and Arcturus (bottom) calculated in LTE (black), \NLTEm\ (red) \NLTEs:Mg (pink) and \NLTEs:Ca (blue). For each star, the Eddington flux (upper panel) and the \% difference between the LTE and the NLTE fluxes (lower panel) are plotted. The units of the Eddington flux $H_\lambda$ are erg cm$^{-2}$s$^{-1}$\AA$^{-1}$.}
\label{fig:uvcomp}
\end{figure*}

\begin{figure}[ht]
    \centering
    \hspace{-0.04\textwidth}
    \begin{tikzpicture}
    \node[anchor=south east, inner sep=0] (image) at (0,0) {
        \subfloat{\includegraphics[width=0.47\textwidth]{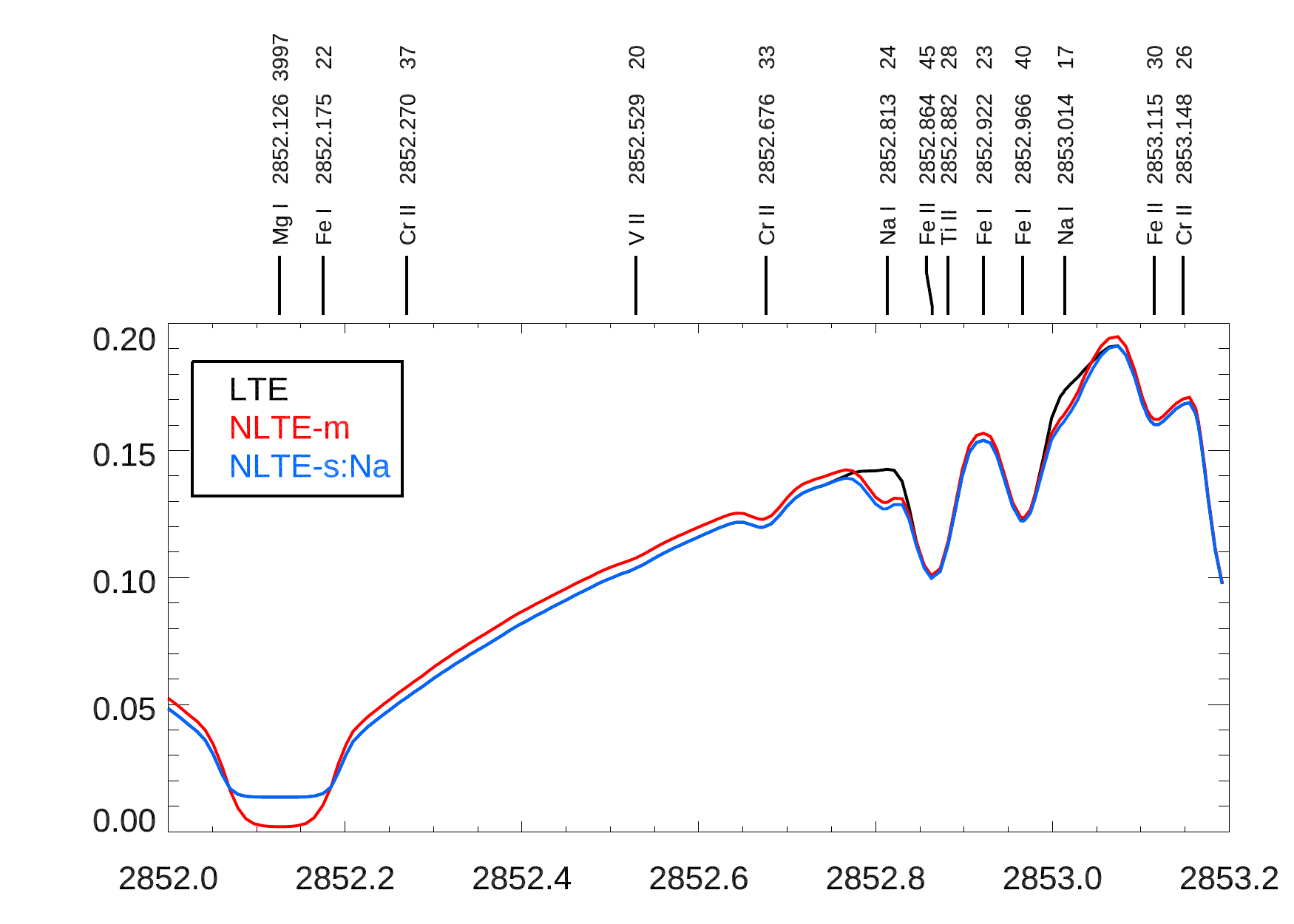}}};
        \node at (-0.23\textwidth,-0.01\textwidth) {\scalebox{1.00}{$\lambda$(\AA)}};
        \node[rotate=90] at (-0.45\textwidth,0.12\textwidth) {\scalebox{1.00}{Normalised Flux}};
    \end{tikzpicture}
    \caption{LTE (black), \NLTEm (red) and \NLTEs:Na (blue) normalised synthetic fluxes of Procyon around the red wing of the 2852~\AA\ \Mg{i} line. The atomic species of some lines are marked together with its LTE equivalent width in m\AA.}
    \label{fig:na2853}
\end{figure}

\subsubsection{Procyon}

In the case of Procyon the interplay between \Mg\ and \Ca\ leads to an increase of the NLTE flux in the $\sim$2\,000-2\,300~\AA\ region, which is the location of the photo-ionisation threshold of the ground level of \Ca{i} (2\,028~\AA). Thus, the resulting background UV opacity seen by \Ca\ is different between the \mbox{\NLTEs:Ca} and the \NLTEm\ calculations. Photon loss competes with over-ionisation, but at column-mass $\sim-1$ the overpopulation of the \Mg{i} lower levels in NLTE with respect to LTE reduces the UV flux in those layers giving advantage to photo loses of the resonance \Ca{i} lines ($\sim$2\,500~\AA\ and 4\,226.7~\AA) which results in an increase of the population of the lowest levels of \Ca\ in \NLTEm\ with respect to \mbox{\NLTEs:Ca} (i.e., when \Mg\ is treated in LTE). In contrast, \Mg\ seems to be affected very little by \Ca\ and its NLTE populations are almost identical in the \NLTEs\ and \NLTEm\ calculations. Regarding the alkali elements; given that the main NLTE mechanism driving the statistical balance of \Na\ and \K\ is photon-suction, these elements are unaffected by the changes in the UV flux due to the \Mg\ NLTE populations. As we mentioned at the introduction of this subsection, the  3s($^2$S) and 5p($^2$P$_{1/2,\,3/2}$) are the levels involved in the the 2852~\AA\ \Na\ lines. Their departure coefficients change due to differences in the background opacity given that these lines are in the red wings of the 2852~\AA\ \Mg{i} line, whose formation in NLTE is not the same as in LTE. 

\subsubsection{Sun}

As in the case of Procyon, the UV flux in the \NLTEm\ calculations is very similar to the \mbox{\NLTEs:Mg} calculations, so again \Ca\ is much more sensitive to \Mg\ than \Mg\ to \Ca. The main difference in departure coefficients between \mbox{\NLTEs:Ca} and \NLTEm\ results are on the ground level of \Ca{i} because a similar process to the one occurring in Procyon takes also place in the Sun, and in this case it is enhanced due to the broader lines that favour photon losses. The UV synthetic spectra show that the flux of the \mbox{\NLTEs:Mg} and \NLTEm\ are very similar (middle panel of Fig.\ref{fig:uvcomp}) and thus the departure coefficients of \Mg\ in the two cases are almost identical. \Na\ and \K\ have NLTE effects driven by photo-suction making their departure coefficients of the \NLTEs\ and \NLTEm\ calculations almost identical.

\begin{figure}[h]
    \hspace{-0.02\textwidth}
    \begin{tikzpicture}
    \node[anchor=south east, inner sep=0] (image) at (0,0) {
        \subfloat{\includegraphics[width=0.45\textwidth]{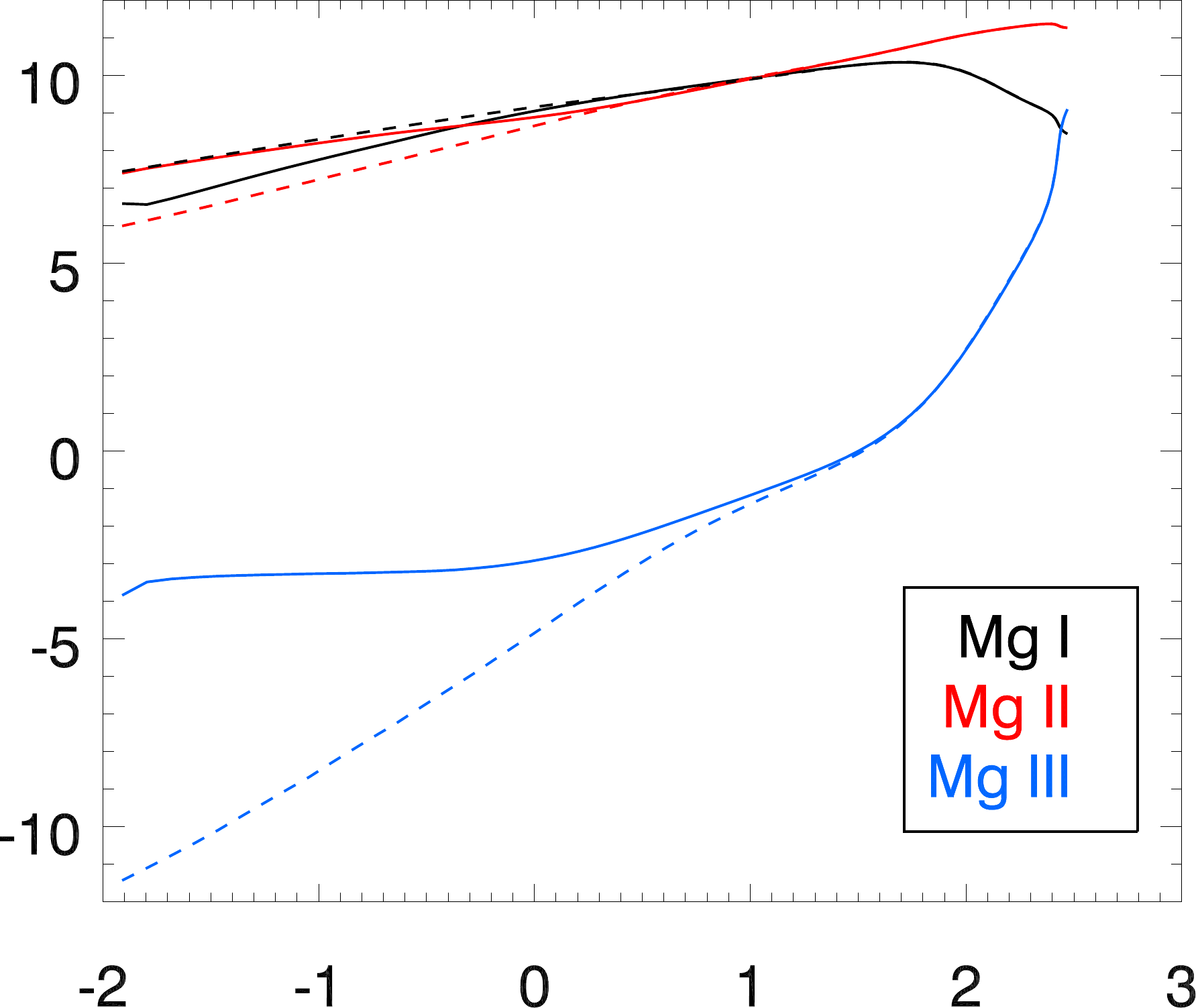}}};
      \node at (-0.2\textwidth,-0.02\textwidth) {\scalebox{1.1}{log column mass}}; 
      \node[rotate=90] at (-0.46\textwidth,0.2\textwidth) {\scalebox{1.1}{log number density}}; 
    \end{tikzpicture}
    \caption{\Mg\ number densities of different ionisation stages in Arcturus. The solid lines represent the NLTE populations while the dashed lines represent the LTE populations.}
    \label{fig:Mgpop}
\end{figure}

\subsubsection{Arcturus}

\Mg{i} is not a minority species anymore in the atmosphere of Arcturus; in fact, at column-mass $\lesssim 1$ the population of \Mg{i} is larger than the population of \Mg{ii}, as illustrated in \fig{fig:Mgpop}. Since there is not a \Mg{ii} reservoir in Arcturus any more, photo-ionisation of the 3p(1P$^*$) level is not efficient compared with the photon pumping that occurs in the 3p(1P)-3s(1S) \Mg{i} line at 2\,850~\AA, making this level to have $b>1$ along the atmosphere (see Mg in Arcturus panel in \fig{fig:bcomp}).  

As seen in \fig{fig:uvcomp}, the \NLTEm\ UV fluxes are now more influenced by \Ca\ than by \Mg\ in some regions (like the 1\,700 - 2\,000~\AA\ ). But the UV flux in Arcturus is three to four orders of magnitude weaker than in the Sun, thus the absolute changes in the UV fluxes due to the NLTE populations of \Ca\ and \Mg\ are not large enough to affect the \NLTEm\ populations of the four atomic elements significantly with respect to their \NLTEs\ populations. On the other hand, due to the lower gravity of Arcturus, metallic lines broaden, so lines with different NLTE / LTE behaviour in the wings (mostly resonance lines) will also affect a larger part of the spectrum than in case of dwarfs. This explains why the departure $b$ coefficient of the ground level of \K\ has an small but noticeable \NLTEm/\NLTEs\ difference in Arcturus and no difference in the Sun and Procyon.

\Ca{i}, on the other hand, is still a minority species in Arcturus and thus UV over-ionisation still dominates, but since the UV is much weaker than in the other two stars, the effects of NLTE \Mg\ opacities on the NLTE \Ca\ populations is not as strong as in the solar and Procyon cases (see \fig{fig:bcomp}).

\subsection{Physics of the inter-element effects}

\Na\ and \K\ are not affected by any of the other elements. We note though, that the \Na{i} levels 3s($^2$S) and 5p($^2$P) have slightly different departure $b$ coefficients between the \NLTEs\ and \NLTEm\ calculations, in particular for Procyon. The transition between those levels produces the  2\,852.8  and 2\,853.0~\AA\ lines, which lie in the red wing of the \Mg{i} 2\,852~\AA\ line. This is an illustration of how the NLTE opacities of some elements can affect the statistical balance of other elements: the background opacity at around 2\,853 ~\AA\ is dominated by the red wing of the \Mg{i} 2\,852 line, which has different strength along the stellar atmosphere when \Mg\ is treated in NLTE and in LTE (see \fig{fig:na2853}).
The differences observed between the red (\NLTEs) and black lines (\NLTEm) in \fig{fig:bcomp} are the result of the combination of the overlap between b-b and b-f transitions of the different elements treated in NLTE, like the one just mentioned.

Given that the main NLTE mechanism affecting the \Ca\ and \Mg\ populations is UV over-ionisation, the differences between the \NLTEs\ and \NLTEm\ populations for both \Ca\ and \Mg\ come mostly from their effects on the radiation field of that region. Figure \ref{fig:uvcomp} helps to understand the differences between the departure coefficients $b$ shown in \fig{fig:bcomp}. The figure shows three panels (one for each star) with the absolute flux obtained in LTE, \NLTEm, and \NLTEs\ for \Mg\ and \Ca\ at the top and the percentage difference between the LTE flux and the three NLTE fluxes (single and multi element) at the bottom of each panel. Broad lines with wings affected by NLTE will affect lines of other elements with lines formed the wings because their background opacity will have a contribution of the NLTE wings. For the Sun and hotter stars, the UV flux is strong enough to affect species sensitive to over-ionisation in \NLTEm\ if there are other species in the calculations that affect the continuum fluxes due to NLTE effects. Our results show also that, NLTE effects on broad lines will result in different background opacities perceived by lines of other elements in the same region, potentially affecting the NLTE results of those other elements.

The fundamental effect of NLTE calculations is the redistribution of the population of different levels in the different ions of a given element, with respect to LTE. As a consequence for an element X, spectral lines and the continuum contribution is different in LTE and NLTE for an atmosphere with the same X abundance. That is the reason why even though the derived LTE and NLTE \Mg\ solar abundance is the same, the derived \NLTEs\ and \NLTEm\ abundances of \Ca\ are different. Figure \ref{fig:na2853} illustrate this; the \Mg\ abundance used in the three profiles shown is the same (A(Mg)=7.42~dex) and the LTE and \NLTEs:Na calculations provide the same profile of the broad 2852~\AA\ \Mg{i} line, the NLTE core and wings of this line differs in the \NLTEm\ Calculations since \Mg\ is now in NLTE with the same A(\Mg). The reason for the different profiles in LTE and NLTE is due to the redistribution of the population of the levels involving that transition (3p $^1$P - 3s $^1$S) that leads to a deeper core and weaker wings of the \Mg\ NLTE profile with respect to the \Mg\ LTE line profile. 


\section{Comparison with observations}

Finally we derived \Na, \Mg, \K\ and \Ca\ abundances for Procyon, the Sun and Arcturus in LTE, \NLTEs\ and \NLTEm. The derived abundances and macro-turbulent velocities are presented in Table \ref{tab:results}.

\begin{table*}[t]
  \caption{\Na, \Mg, \K\ and \Ca\ abundances derived for the LTE, \NLTEs\ and \NLTEm\ calculations. Macro-turbulent velocity was also a free parameter in each calculation. The second column shows the number of points used to perform the best fit calculation. }\label{tab:results}
  \centering
  \hspace{0.02\textwidth}
  \begin{tabular}{l r c c r c c r c c r}\hline\hline
    Star &  & \multicolumn{3}{c}{LTE}  &  \multicolumn{3}{c}{\NLTEs}  &  \multicolumn{3}{c}{\NLTEm}  \\
         &  N &  A(X) & V$_{mac}$ & $\chi^2$  &     A(X) & V$_{mac}$ & $\chi^2$  &     A(X) & V$_{mac}$ & $\chi^2$ \\\hline
    \\[-0.3cm]
    \multicolumn{10}{c}{Na} \\
    Procyon  & 935  & 6.36$\pm$0.09 &  2.87$\pm$1.44 &  0.92 & 6.14$\pm$0.05 &  6.68$\pm$0.85 &  0.25 & 6.13$\pm$0.05 &  6.69$\pm$0.86 &  0.25  \\
    Sun      & 2358 & 6.29$\pm$0.05 &  1.45$\pm$4.19 &  8.09 & 6.15$\pm$0.04 &  1.25$\pm$1.28 &  4.35 & 6.15$\pm$0.04 &  1.25$\pm$1.29 &  4.35   \\
    Arcturus & 501  & 5.91$\pm$0.06 &  2.18$\pm$2.30 &  0.65 & 5.80$\pm$0.05 &  3.82$\pm$1.33 &  0.62 & 5.80$\pm$0.05 &  3.81$\pm$1.32 &  0.62   \\\hline\\[-0.3cm]
    \multicolumn{10}{c}{Mg} \\
    Procyon  & 2504 & 7.42$\pm$0.06 &  6.52$\pm$1.77 &  1.14 & 7.42$\pm$0.05 &  7.10$\pm$1.45 &  1.03 & 7.42$\pm$0.05 &  7.12$\pm$1.46 &  1.02   \\
    Sun      & 2729 & 7.56$\pm$0.02 &  2.32$\pm$2.28 &  1.86 & 7.56$\pm$0.02 &  3.36$\pm$1.55 &  1.40 & 7.56$\pm$0.02 &  3.38$\pm$1.55 &  1.40    \\
    Arcturus & 1192 & 7.40$\pm$0.05 &  3.93$\pm$1.68 &  3.26 & 7.39$\pm$0.05 &  5.13$\pm$1.19 &  3.02 & 7.39$\pm$0.05 &  5.16$\pm$1.20 &  3.03    \\\hline\\[-0.3cm]
    \multicolumn{10}{c}{K} \\
    \\[-0.3cm]
    Procyon  & 120  & 5.71$\pm$3.06 &  2.32$\pm$1.56 &  0.06 & 5.05$\pm$0.04 &  5.60$\pm$0.26 &  0.01 & 5.05$\pm$0.04 &  5.60$\pm$0.26 &  0.01  \\
    Sun      & 182  & 5.36$\pm$0.07 &  0.99$\pm$1.03 &  0.04 & 5.09$\pm$0.04 &  1.30$\pm$1.10 &  0.01 & 5.09$\pm$0.04 &  1.30$\pm$1.10 &  0.01   \\
    Arcturus & 182  & 4.98$\pm$0.05 &  3.95$\pm$0.30 &  0.02 & 4.54$\pm$0.04 &  5.25$\pm$0.20 &  0.02 & 4.55$\pm$0.05 &  5.23$\pm$0.20 &  0.02  \\\hline\\[-0.3cm]
    \multicolumn{10}{c}{Ca} \\
    Procyon  & 1410 & 6.17$\pm$0.07 &  3.68$\pm$1.12 &  2.78 & 6.13$\pm$0.06 &  5.01$\pm$0.81 &  2.23 & 6.10$\pm$0.06 &  4.95$\pm$0.74 &  2.27  \\
    Sun      & 6116 & 6.35$\pm$0.05 &  0.77$\pm$1.77 & 5.92 & 6.37$\pm$0.05 &  0.71$\pm$1.76 &  4.76 & 6.30$\pm$0.04 &  0.69$\pm$1.57 & 3.93    \\
    Arcturus & 2026 & 5.97$\pm$0.08 &  3.28$\pm$1.36 &  7.88 & 5.99$\pm$0.08 &  3.60$\pm$1.20 &  8.81 & 5.98$\pm$0.08 &  3.61$\pm$1.10 & 8.80    \\\hline
  \end{tabular}
\end{table*}

We used the all-lines method described in \cite{CaPaperI}, where all the lines of the spectrum were analysed simultaneously; therefore a single abundance and macro-turbulent velocity was used for all the lines of a given element simultaneously. In order to determine the error associated with the derived abundances and macro-turbulent velocities, we used the method described in section 5 of \cite{2017A&A...597A..16P} where the model errors are prioritised. For all the three stars and the four elements studied in this work, the residuals ($\chi^2$) of the best fit in NLTE are reduced relative to the LTE analysis, except for \Ca\ in Arcturus (we found the same anomaly in \citeauthor{CaPaperI}). The $\chi^2$ of the \NLTEm\ calculations is either lower or the same as the $\chi^2$ of \NLTEs\ calculation. The case of \Ca\ in the Sun is particularly interesting: while its \NLTEs\ abundance correction is marginally positive (0.02), the \NLTEm\ abundance correction is negative (-0.05) and the \NLTEm\ abundance becomes in excellent agreement with the meteoritic value. A more detailed description is given below.

The closest comparison of our results and previous works are for Na \cite{2011A&A...528A.103L}, for Mg \cite{2015A&amp;A...579A..53O}, for K \cite{2019A&A...627A.177R} and for Ca \cite{CaPaperI}. The LTE / \NLTEs\ results are generally in very good agreement with those studies considering the differences in methodology (equivalent with vs line profile fitting, line-by-line abundance determination vs all-lines at once, etc) and the input data used (different model atmospheres, background opacities, synthetic spectra calculations, radiative transfer codes, etc.)

\subsection{Na}
\begin{figure}[h]
    \hspace{0.0\textwidth}
    \begin{tikzpicture}
    \node[anchor=south east, inner sep=0] (image) at (0,0) {
        \subfloat{\includegraphics[width=0.45\textwidth]{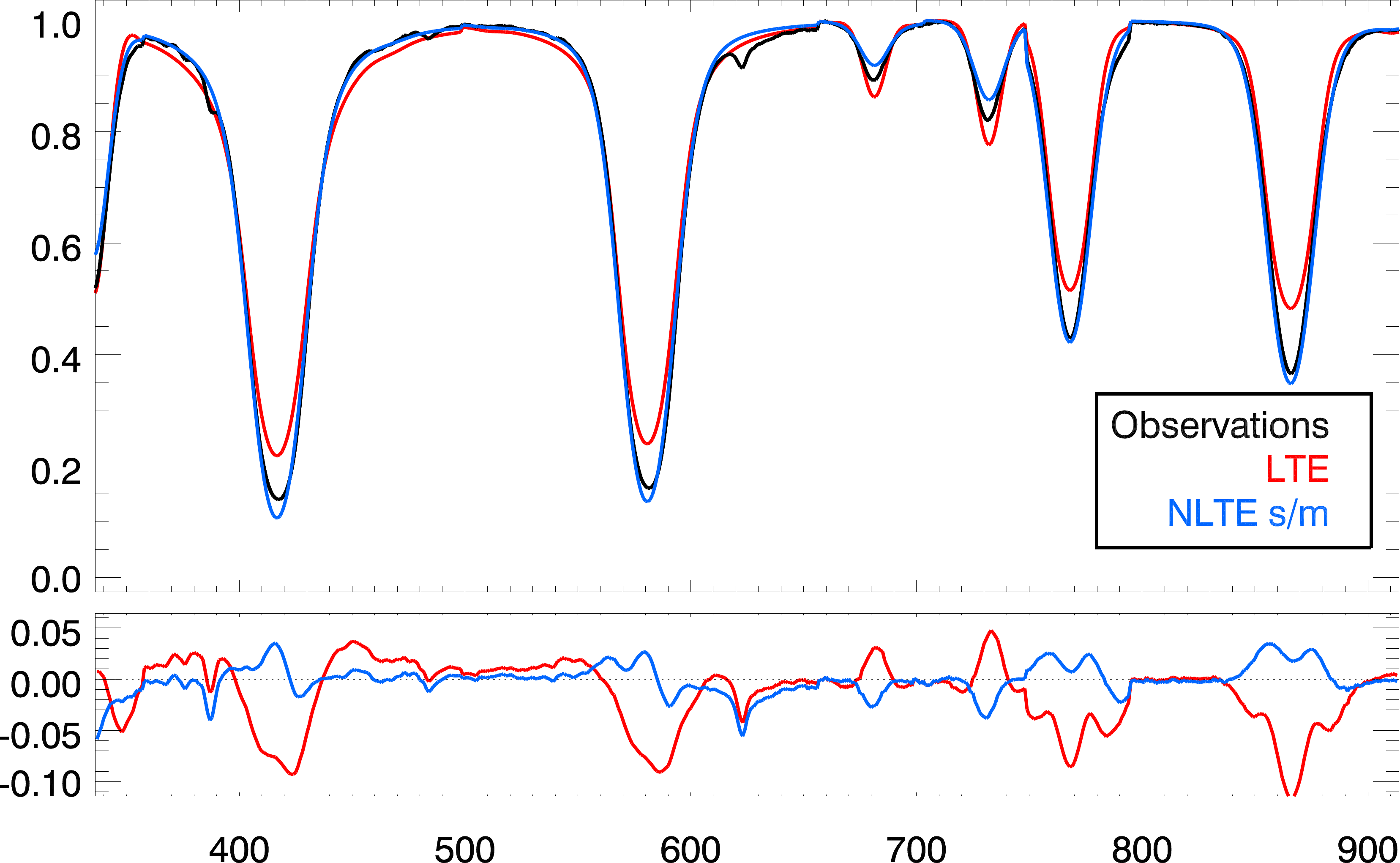}}};
      \node at (-0.22\textwidth,-0.01\textwidth) {\scalebox{0.8}{mask index}}; 
      \node[rotate=90] at (-0.46\textwidth,0.2\textwidth) {\scalebox{0.8}{Relative Flux }}; 
      \node[rotate=90] at (-0.46\textwidth,0.05\textwidth) {\scalebox{0.8}{$\Delta_{syn-obs}$}}; 
    \end{tikzpicture}
    \caption{\emph{Top.} Selected lines of \Na\ observed in Procyon (black) and the best fit for the LTE (red) and NLTE(blue) calculations with the parameters in derived in Table \ref{tab:results}. \emph{Bottom.} Difference between observations and the synthetic spectrum in LTE (red) and in NLTE (blue).}
    \label{fig:Proc_Na_spec}
\end{figure}
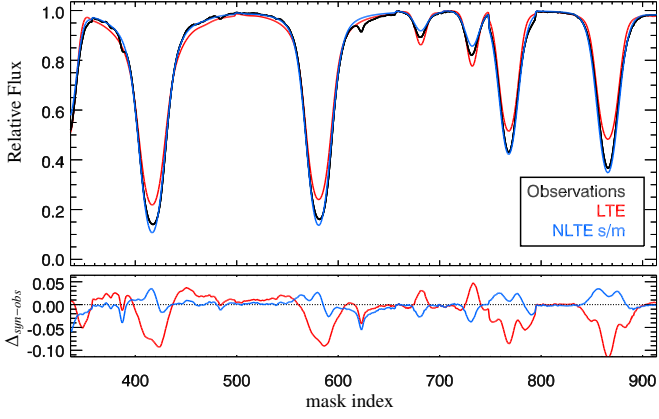

\cite{2011A&A...528A.103L} used the \multi\ code to calculate LTE and \NLTEs\ equivalent widths in order to derive \Na\ abundance corrections for a grid of MARCS model atmospheres. Using their Figure 4, their abundance corrections for Procyon, the Sun and Arcturus are $\sim -0.15, -0.1$  and $-0.2$ respectively while our corrections are $-0.22(9)$, $-0.14(5)$ and $-0.11(5)$ for the three stars. 
As expected from the departure coefficients in \fig{fig:bcomp} we find the derived \NLTEs\ to be the same as the \NLTEm\ abundances. Figure \ref{fig:Proc_Na_spec} shows selected lines of \Na\ in a comparison between observations of Procyon and synthetic spectra in LTE and NLTE. since the \NLTEs\ and \NLTEm\ synthetic spectra are indistinguishable, only one of those were plotted. The residuals between observations and the best fit of the synthetic spectra is reduced in NLTE with respect to LTE. We see that for Procyon the best fit in NLTE is 73\% better than the best fit in LTE, while for the Sun and Arcturus is 46\% and only 5\% better respectively.  

Our derived solar abundance becomes slightly lower than the value found in CI carbonateous chondrites, after scaling through several well-determined refractory elements (Asplund 2009; Lodders 2019).

\subsection{Mg}

\cite{2015A&amp;A...579A..53O} also used the \multi\ code for the determination of LTE and \NLTEs\ \Mg\ abundances. There is excellent agreement between our absolute LTE and \NLTEs\ abundances. The NLTE best fit are 10\%, 24\% and 7\% better than the LTE best fit for Procyon, the Sun and Arcturus respectively. As in the case of \Na, The effects of the other elements calculated in NLTE are negligible and therefore the \NLTEs\ and \NLTEm\ results for \Mg\ are almost identical. 

In this case our derived solar abundance is in better agreement with the meteoritic value, solving a 2$\sigma$ inconsistency shown in previous determinations (Asplund 2009).

\subsection{K}

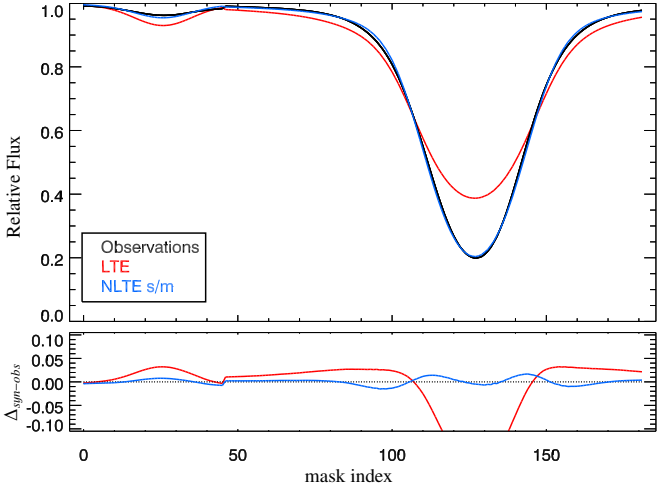
\begin{figure}[h]
    \hspace{-0.02\textwidth}
    \begin{tikzpicture}
    \node[anchor=south east, inner sep=0] (image) at (0,0) {
        \subfloat{\includegraphics[width=0.45\textwidth]{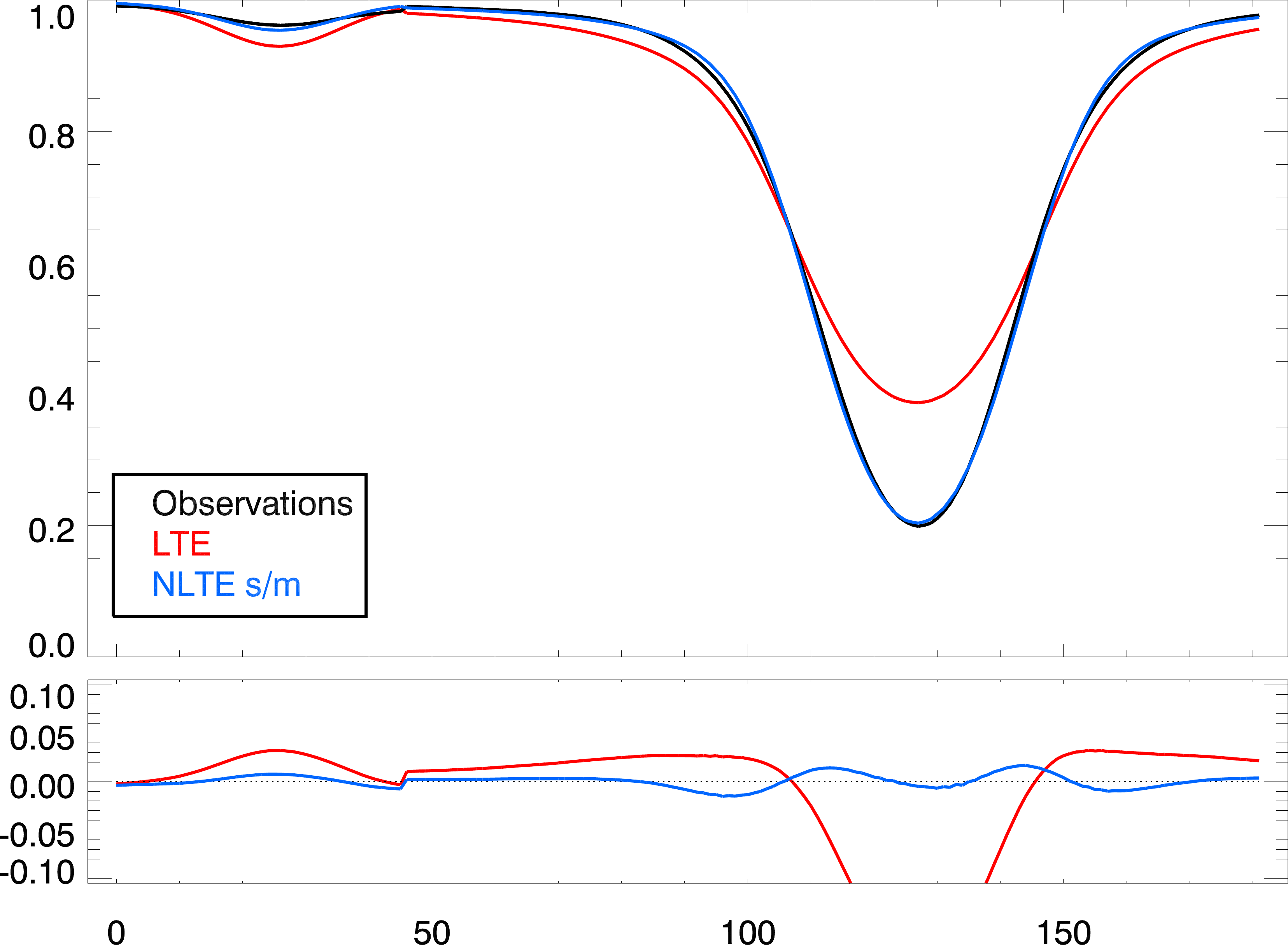}}};
      \node at (-0.22\textwidth,-0.01\textwidth) {\scalebox{0.8}{mask index}}; 
      \node[rotate=90] at (-0.46\textwidth,0.2\textwidth) {\scalebox{0.8}{Relative Flux }}; 
      \node[rotate=90] at (-0.46\textwidth,0.05\textwidth) {\scalebox{0.8}{$\Delta_{syn-obs}$}}; 
    \end{tikzpicture}
    \caption{\emph{Top.} lines of \K\ Used in the solar observations (black) compared with the best fit of the LTE (red) and NLTE(blue) calculations with the parameters in derived in Table \ref{tab:results}. \emph{Bottom.} Difference between observations and the synthetic spectrum in LTE (red) and in NLTE (blue).}
    \label{fig:Sun_K_spec}
\end{figure}

\cite{2019A&A...627A.177R} presented LTE and NLTE abundances for the Sun, Procyon and other stars by fitting the synthetic equivalent widths with the observed ones. The NLTE abundance derived for the Sun is in excellent agreement with ours. In the case of Procyon although the abundance correction is the same (0.68 by \citeauthor{2019A&A...627A.177R} and 0.66 by us) our LTE and NLTE abundances are higher by 0.19~dex. In Arcturus the LTE and NLTE residuals of the best fit are same: we found an abundance correction of 0.44~dex.

\subsection{Ca}

The most interesting case is \Ca, whose NLTE populations are affected the most by the NLTE populations of the other elements, specifically by  \Mg. \cite{CaPaperI} presents NLTE-s abundance corrections of \Ca\ for the three stars studied in this work. In order to make a cleaner comparison, we use the same masks as the ones used in the all-lines calculation in \cite{CaPaperI}. 

\begin{figure*}[!ht]
    \hspace{-0.02\textwidth}
    \\[-0.05\textwidth]
\begin{tabular}{c}
    \begin{tikzpicture}
    \node[anchor=south east, inner sep=0] (image) at (0,0) {
        \subfloat{\includegraphics[width=1.00\textwidth]{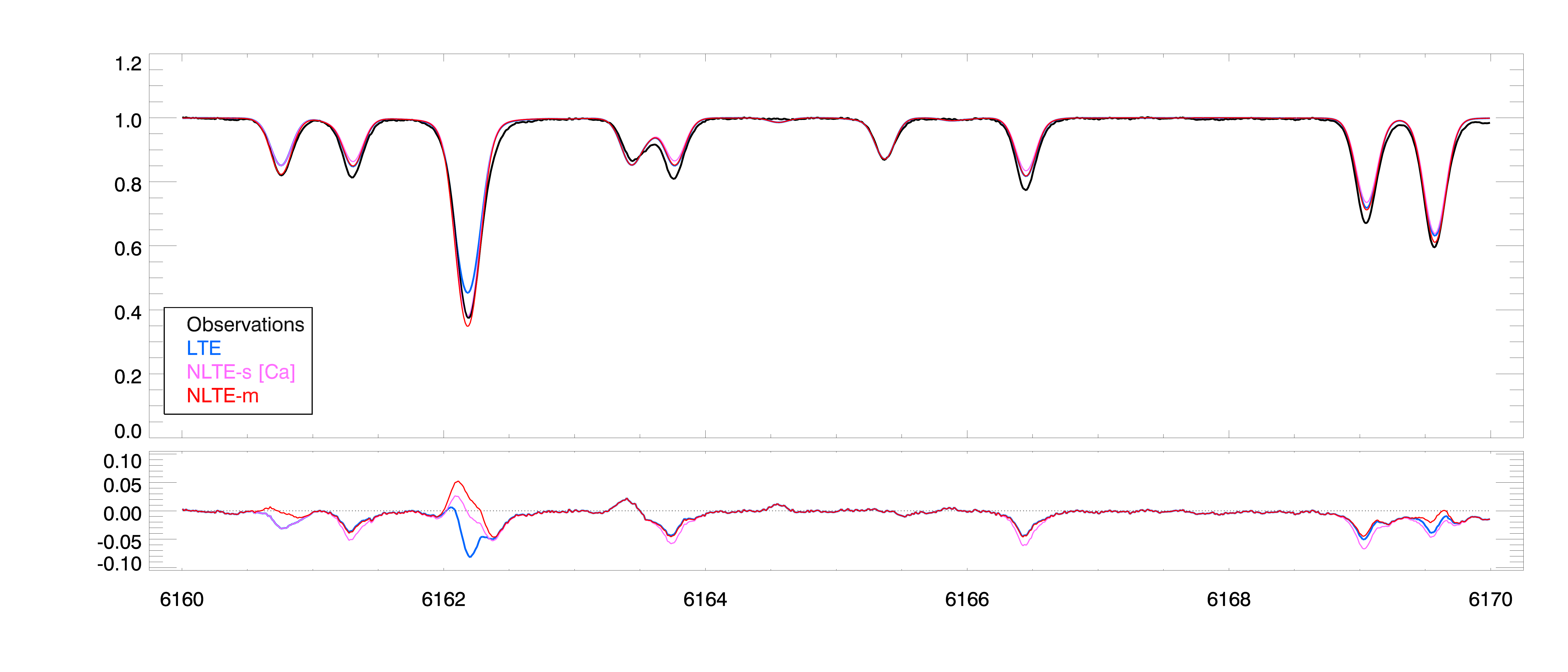}}};
      \node at (-0.47\textwidth,0.03\textwidth) {\scalebox{0.8}{$\lambda$(\AA)}}; 
      \node at (-0.15\textwidth,0.17\textwidth) {\scalebox{1.5}{Procyon}}; 
      \node[rotate=90] at (-0.96\textwidth,0.30\textwidth) {\scalebox{0.8}{Relative Flux }}; 
      \node at (-0.82\textwidth,0.37\textwidth) {\scalebox{0.8}{Na}}; 
      \node at (-0.77\textwidth,0.37\textwidth) {\scalebox{0.8}{Ca}}; 
      \node at (-0.70\textwidth,0.37\textwidth) {\scalebox{0.8}{Ca}}; 
      \node at (-0.595\textwidth,0.37\textwidth) {\scalebox{0.8}{Ni}}; 
      \node at (-0.57\textwidth,0.37\textwidth) {\scalebox{0.8}{Ca}}; 
      \node at (-0.435\textwidth,0.37\textwidth) {\scalebox{0.8}{Fe}}; 
      \node at (-0.34\textwidth,0.37\textwidth) {\scalebox{0.8}{Ca}}; 
      \node at (-0.13\textwidth,0.37\textwidth) {\scalebox{0.8}{Ca}}; 
      \node at (-0.085\textwidth,0.37\textwidth) {\scalebox{0.8}{Ca}}; 
      \node[rotate=90] at (-0.96\textwidth,0.10\textwidth) {\scalebox{0.8}{$\Delta_{syn-obs}$}}; 
%
    \end{tikzpicture}
    \\[-0.07\textwidth]
     \begin{tikzpicture}
    \node[anchor=south east, inner sep=0] (image) at (0,0) {
        \subfloat{\includegraphics[width=1.00\textwidth]{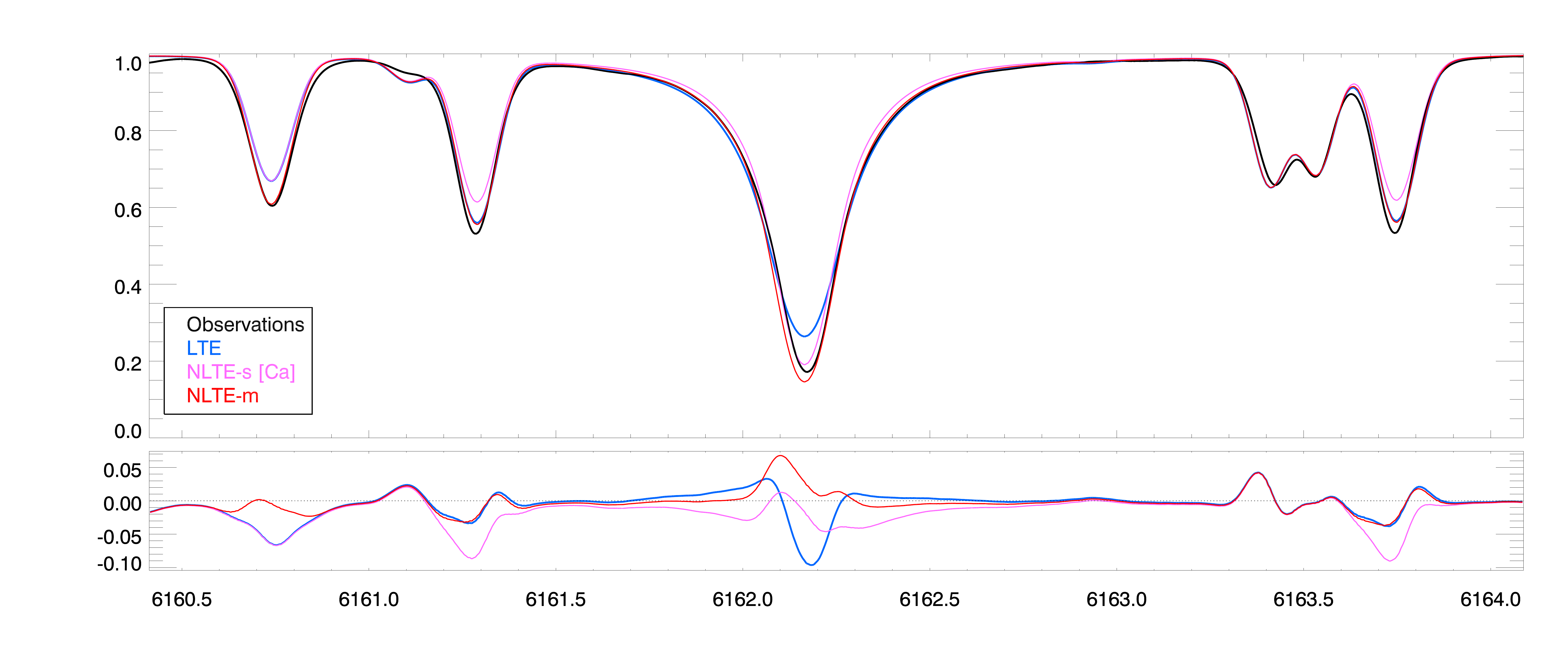}}};
      \node at (-0.47\textwidth,0.03\textwidth) {\scalebox{0.8}{$\lambda$(\AA)}}; 
      \node at (-0.15\textwidth,0.17\textwidth) {\scalebox{1.5}{Sun}}; 
      \node[rotate=90] at (-0.96\textwidth,0.30\textwidth) {\scalebox{0.8}{Relative Flux }}; 
      \node at (-0.83\textwidth,0.37\textwidth) {\scalebox{0.8}{Na}}; 
      \node at (-0.70\textwidth,0.37\textwidth) {\scalebox{0.8}{Ca}}; 
      \node at (-0.49\textwidth,0.37\textwidth) {\scalebox{0.8}{Ca}}; 
      \node at (-0.19\textwidth,0.37\textwidth) {\scalebox{0.8}{Ni}}; 
      \node at (-0.16\textwidth,0.37\textwidth) {\scalebox{0.8}{Fe}}; 
      \node at (-0.11\textwidth,0.37\textwidth) {\scalebox{0.8}{Ca}}; 
      \node[rotate=90] at (-0.96\textwidth,0.10\textwidth) {\scalebox{0.8}{$\Delta_{syn-obs}$}}; 
%
    \end{tikzpicture}   
    \\[-0.07\textwidth]
    \begin{tikzpicture}
    \node[anchor=south east, inner sep=0] (image) at (0,0) {
        \subfloat{\includegraphics[width=1.00\textwidth]{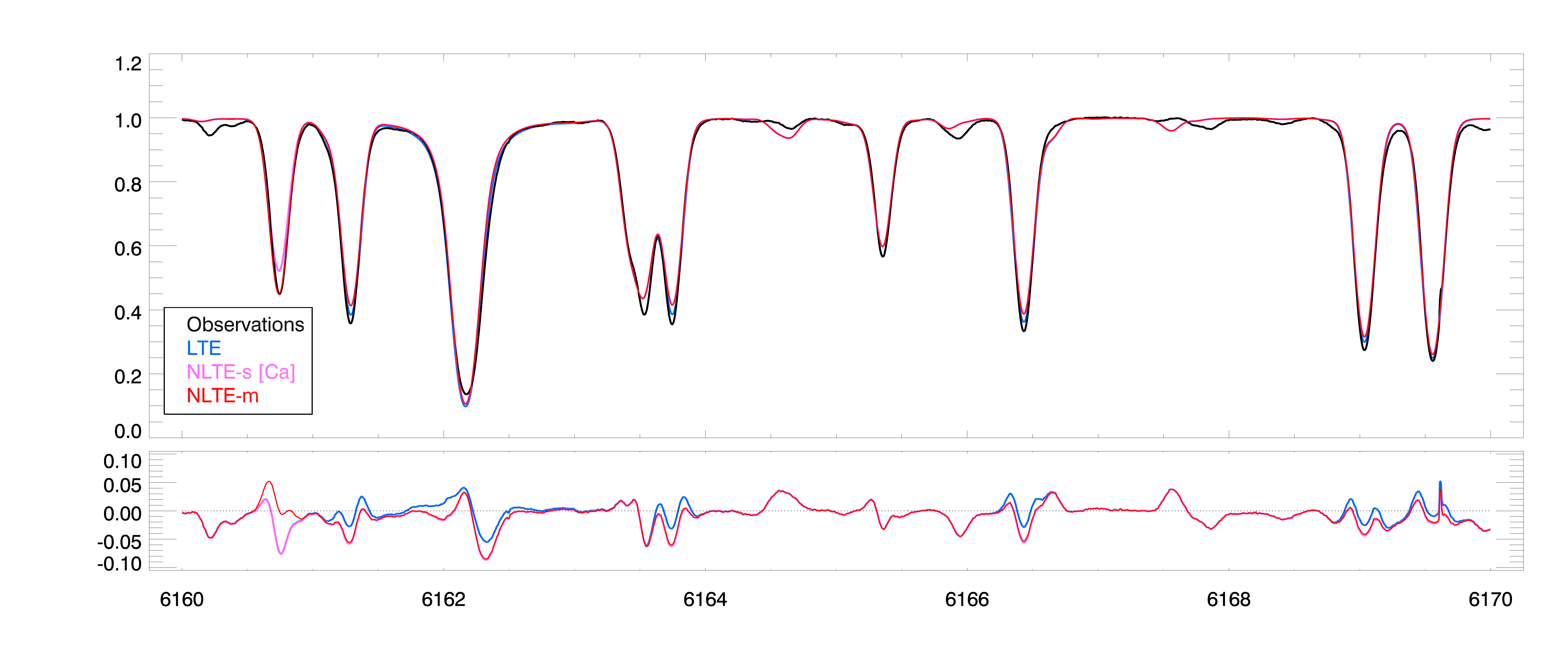}}};
      \node at (-0.47\textwidth,0.03\textwidth) {\scalebox{0.8}{$\lambda$(\AA)}}; 
      \node at (-0.15\textwidth,0.17\textwidth) {\scalebox{1.5}{Arcturus}}; 
      \node[rotate=90] at (-0.96\textwidth,0.30\textwidth) {\scalebox{0.8}{Relative Flux }}; 
      \node at (-0.82\textwidth,0.37\textwidth) {\scalebox{0.8}{Na}}; 
      \node at (-0.77\textwidth,0.37\textwidth) {\scalebox{0.8}{Ca}}; 
      \node at (-0.70\textwidth,0.37\textwidth) {\scalebox{0.8}{Ca}}; 
      \node at (-0.60\textwidth,0.371\textwidth) {\scalebox{0.8}{Ni}}; 
      \node at (-0.59\textwidth,0.355\textwidth) {\scalebox{0.8}{Fe}}; 
      \node at (-0.57\textwidth,0.37\textwidth) {\scalebox{0.8}{Ca}}; 
      \node at (-0.435\textwidth,0.37\textwidth) {\scalebox{0.8}{Fe}}; 
      \node at (-0.34\textwidth,0.37\textwidth) {\scalebox{0.8}{Ca}}; 
      \node at (-0.13\textwidth,0.37\textwidth) {\scalebox{0.8}{Ca}}; 
      \node at (-0.085\textwidth,0.37\textwidth) {\scalebox{0.8}{Ca}}; 
      \node[rotate=90] at (-0.96\textwidth,0.10\textwidth) {\scalebox{0.8}{$\Delta_{syn-obs}$}}; 
%
    \end{tikzpicture} \\[-0.045\textwidth]  
\end{tabular}   
    \caption{Comparison between synthetic spectra and observations of the section around 6162~\AA\ for Procyon (top), the Sun (middle) and Arcturus (bottom) where several \Ca\ lines can be observed. The black lines represent the observations; blue, pink and red are the synthetic spectra in LTE, NLTE-s[Ca] and NLTE-m respectively. Bellow each spectra the residuals (synthetic-observed) are plotted. The \Ca\ abundance and V$_{mac}$ adopted for the LTE, NLTE-s~[Ca] and NLTE-m synthetic spectra are [A(Ca), V$_{mac}$(Km/s)] = [6.10,~5.1] for Procyon; [6.30,~1.0] for the Sun and [5.99,~3.6] for Arcturus.}
    \label{fig:Ca_spec}
\end{figure*}
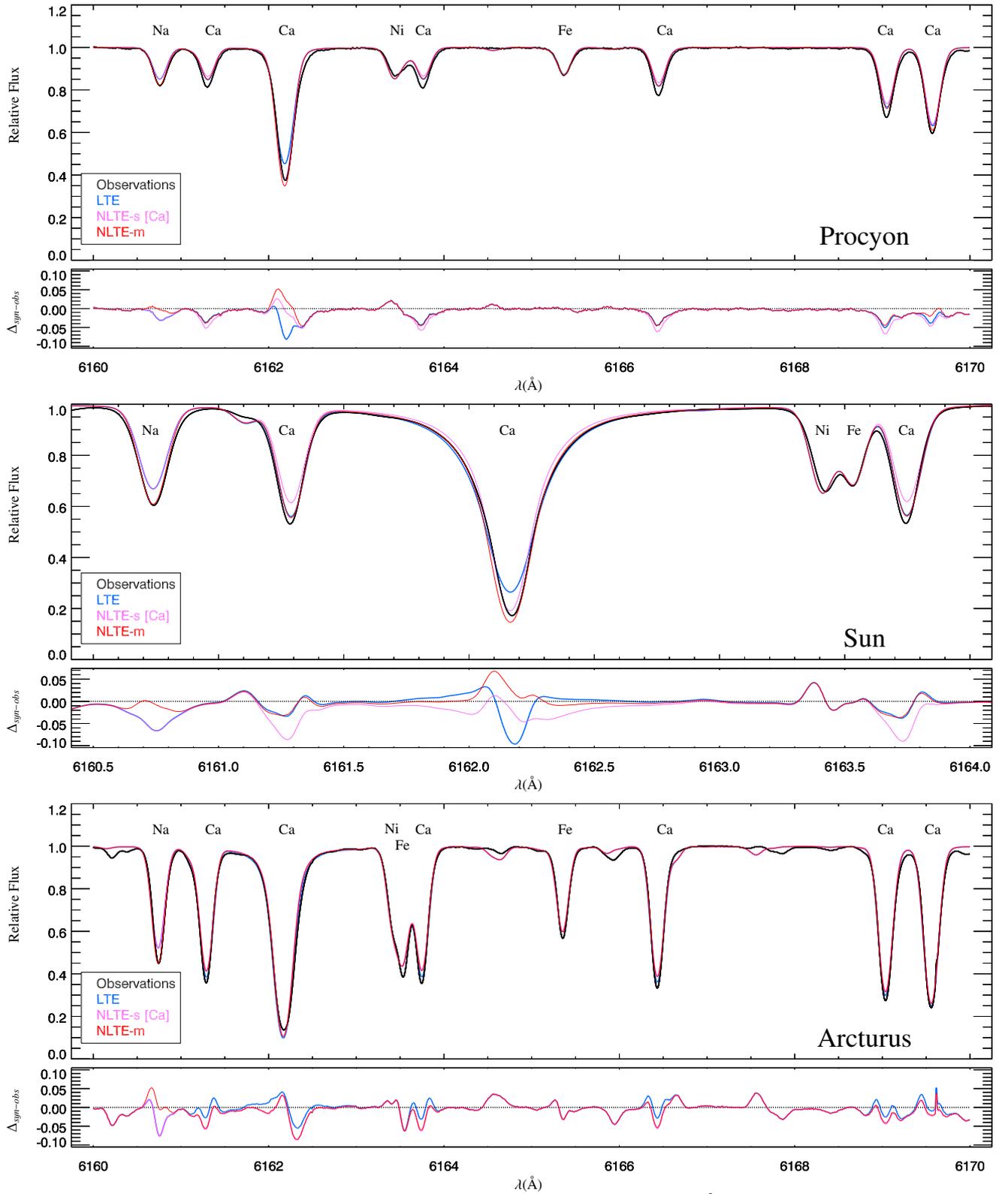

For Procyon, \citeauthor{CaPaperI} found an abundance correction of $\Delta_{nlte,s}(Ca)=-0.08$ dex while in this work we found $\Delta_{nlte,s}(Ca)=-0.05$ dex although our new LTE \Ca\ abundance is higher than in \citeauthor{CaPaperI}. A possible explanation of the discrepancy is that \multi\ calculates line profiles only of the element under study, while \synspec\ computes synthetic spectra including the contribution from several atomic and molecular species. The effects in Procyon of having a more realistic synthetic spectrum is probably affecting more the derived abundances than in the Sun or in Arcturus. A similar phenomenon happens with \K\ in Procyon when compared with \cite{2019A&A...627A.177R}, which also used \multi\ in the calculation of the synthetic spectral lines. Our multi-element calculation decreases the derived \NLTEm\ abundance by 0.03 dex with respect to the \NLTEs\ results. The top panel of \fig{fig:Ca_spec} shows the region around 6165~\AA\ where there are several lines of \Ca. It is clear that the core of the 6162~\AA\ line is the weakest in LTE (blue line), then in \NLTEs\ the core of this line match the observations quite well and the \NLTEm\ calculations strengthen the core a bit more. If only the 6162~\AA\ line is considered, the fit with observations is better in \NLTEs\ than in \NLTEm, but by checking the other lines in the window, we note that the other \NLTEs\ \Ca\ lines have weaker cores than the LTE profiles and the observations. The NLTE-m spectra have similar line profiles to the LTE for four of the \Ca\ lines (6161.4, 6163.7, 6166.5 and 6169.1). The 6169.6~\AA\ line has a NLTE-m profile that match better the observations than the LTE and NLTE-s cases. We found that our NLTE-s best fit is 20\% and our NLTE-m best fit is 18\% better than the LTE best fit.

For the Sun, \cite{CaPaperI} derived an LTE \Ca\ abundance in agreement to ours, but we found a negligible abundance correction of 0.02~dex while the NLTE-m results show a correction of $-0.05$~dex. The central panel in \fig{fig:Ca_spec} shows solar observations and the synthetic spectra in the region around 6162~\AA\ where three \Ca\ lines are clearly visible. There is also a \Na\ which has a NLTE line profile in the NLTE-m spectra. The NLTE calculations show that while the core of the 6162~\AA\ line matches better the observations in the NLTE-s and NLTE-m cases than in the LTE one, the wings of the NLTE-s and the other two \Ca\ lines seen in the figure, suggest that the NLTE-s abundance must increase in order to fit the observations (as the results in Table \ref{tab:results} show). The LTE spectrum is unable to reproduce the core and the wings of the 6162~\AA\ line. While the 6161.3 and 6163.7 \Ca\ lines are reproduced better in LTE than in NLTE-s. The NLTE-m  6161.3 and 6163.7~\AA\ lines are similar to the LTE lines and with the same abundance the core and the profile of the 6162~\AA\ can be reproduced. The results of this sort of balance shows that our NLTE-s best fit is 20\% better than the best fit in LTE and our NLTE-m best fit is 34\% better than the best fit in LTE. We should also point out that the NLTE-m \Ca\ abundance we found is in excellent agreement with the meteoritic value.

Like the other atomic elements studied in this work, \Ca\ in Arcturus seems to be unaffected by NLTE multi-element effects. The LTE and NLTE-s abundances found by \cite{CaPaperI} agree with ours and the small NLTE corrections are in the same direction ($-0.02$ in this work and $-0.01$ ~dex in \citeauthor{CaPaperI}). As in the work from \citeauthor{CaPaperI}, the NLTE-s best fit is worse than the LTE best fit (32\% in that work and 12\% in this work). Will the addition of iron in NLTE-m calculations improve the situation of \Ca\ in NLTE for Arcturus?, is the NLTE ionisation balance of the important electron donors affecting the total electron populations and therefore the atmospheric structure in Arcturus in a sensible manner? Further investigations will answer those questions.

\section{The APOGEE spectral window}

The only difference between the analysis of the observations of the Sun and Arcturus in the APOGEE window, and the previous comparison in the optical are the {b-b data adopted for \synspec: We adopted the ASPCAP atomic line-list \citep{2015ApJS..221...24S} formatted for \synspec. The molecular b-b data use  isotopic ratios suited for the Sun and Arcturus}.

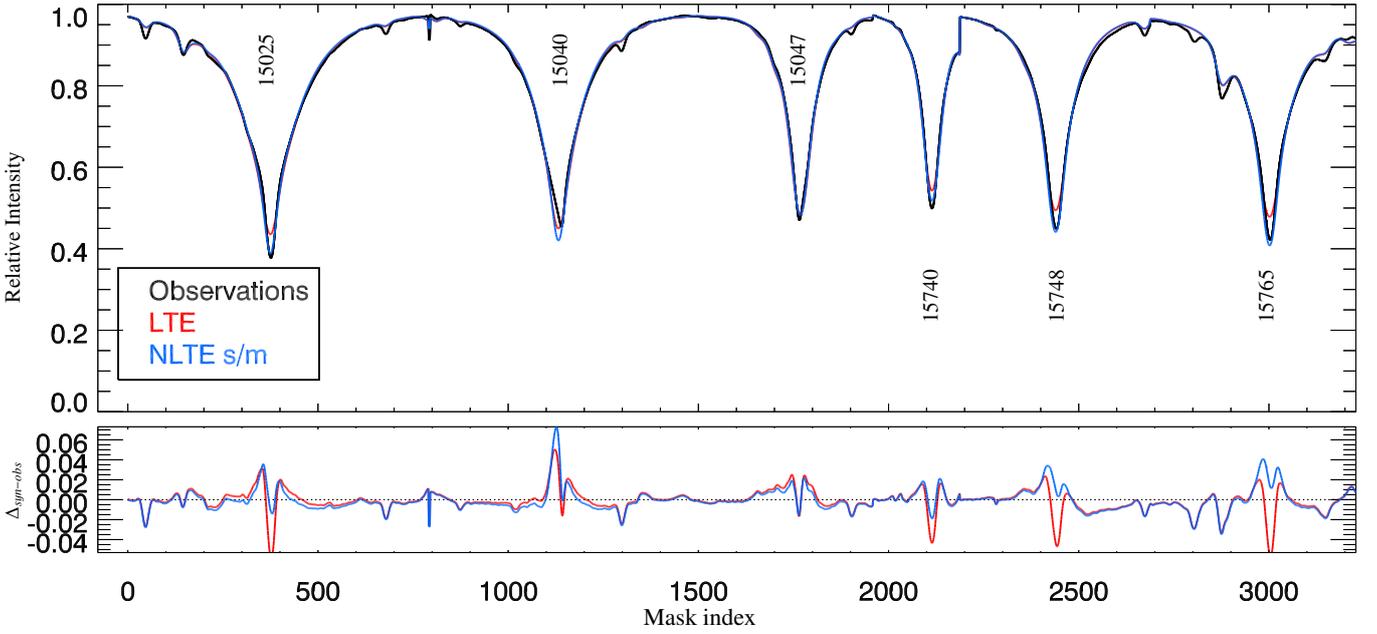
\begin{figure*}[!ht]
     \begin{tikzpicture}
    \node[anchor=south east, inner sep=0] (image) at (0,0) {
        \subfloat{\includegraphics[width=0.95\textwidth]{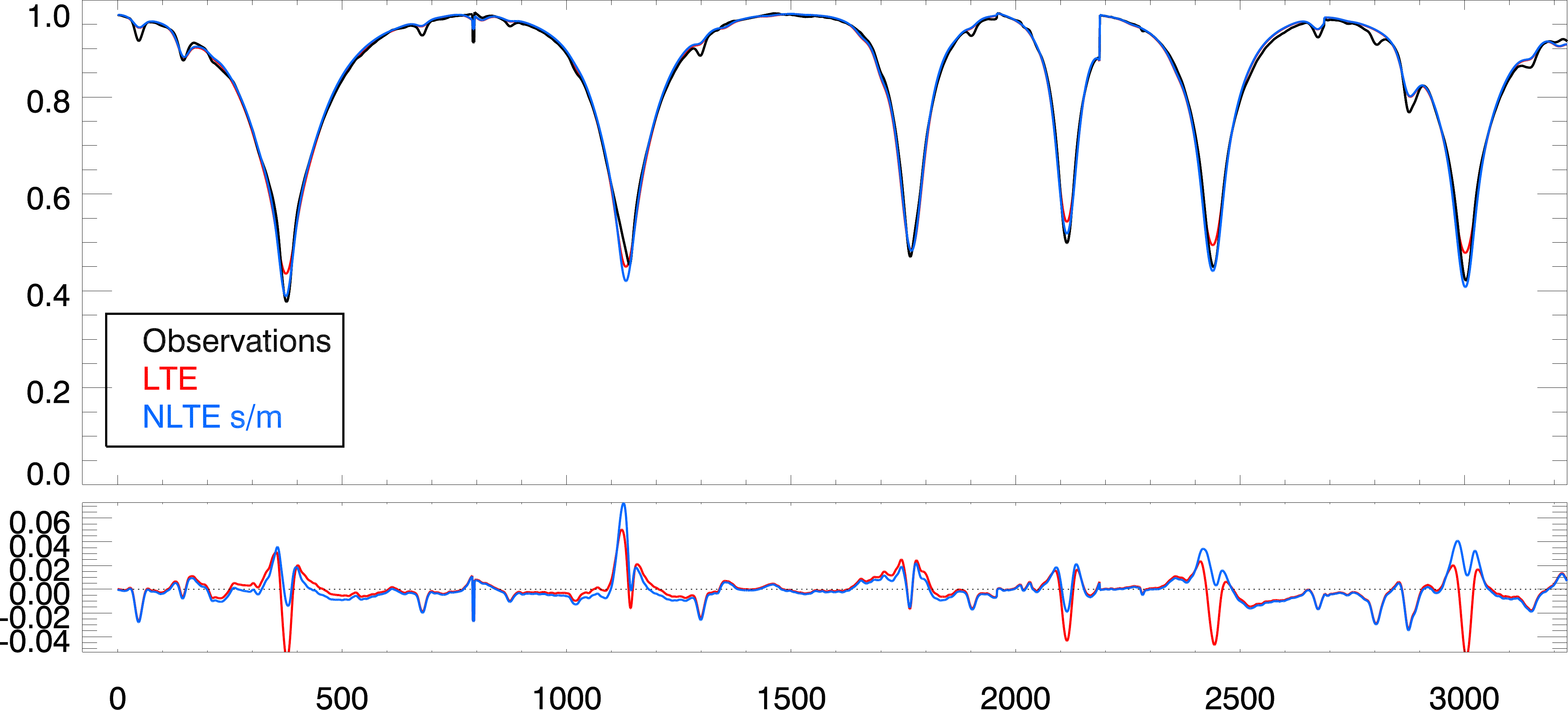}}};
      \node[rotate=90] at (-0.78\textwidth,0.39\textwidth) {\scalebox{0.9}{15025}}; 
      \node[rotate=90] at (-0.57\textwidth,0.39\textwidth) {\scalebox{0.9}{15040}}; 
      \node[rotate=90] at (-0.40\textwidth,0.39\textwidth) {\scalebox{0.9}{15047}}; 
      \node[rotate=90] at (-0.305\textwidth,0.22\textwidth) {\scalebox{0.9}{15740}}; 
      \node[rotate=90] at (-0.215\textwidth,0.22\textwidth) {\scalebox{0.9}{15748}}; 
      \node[rotate=90] at (-0.065\textwidth,0.22\textwidth) {\scalebox{0.9}{15765}}; 
      \node[rotate=90] at (-0.96\textwidth,0.27\textwidth) {\scalebox{0.9}{Relative Intensity}}; 
      \node[rotate=90] at (-0.96\textwidth,0.08\textwidth) {\scalebox{0.8}{$\Delta_{syn-obs}$}}; 
      \node at (-0.47\textwidth,-0.011\textwidth) {\scalebox{1.0}{Mask index}};    
    \end{tikzpicture}   
   \caption{Comparison between synthetic spectra and centre of disc solar intensity observations of \Mg lines in the APOGEE window. The black lines represent the observations; red and blue are the synthetic spectra in LTE and NLTE respectively. Bellow each spectra the residuals (synthetic-observed) are plotted. The \Mg\ abundance adopted for both synthetic spectra is A(Mg) = 7.55 dex. The label of each line represent the wavelength (in \AA).}
    \label{fig:Mg_apospec}
\end{figure*}

The solar observations in the $H$-band we adopt correspond to the centre of disc intensity spectrum from \cite{sun_ir}. For \Na, \K\ and \Ca\ there are no significant NLTE effects in the line profiles and therefore the derived abundances in LTE and NLTE are the same. There are only two useful lines of \Na\ in the $H$-band solar spectrum. The 15992~\AA\ line can be reproduced with A(Na)=6.12~dex, in good agreement with the NLTE \Na\ solar abundance we derived in the optical. The \Na\ 16388.8~\AA\ line is weaker than the observations using the same abundance, but there are issues to determine the continuum location around this line: it lies in the blue wing of an H-Brackett line ([n$_l$,n$_u$]=[4,12] at 1.6398~\AA) and has Fe, Ni and Ti blends. The \Mg\ abundance found using the spectra in the APOGEE region is A(Mg)=7.55~dex, in excellent agreement with our derived abundance from the optical transitions. Like in the optical, we found that the \Mg\ abundances that correspond to the best fittings in NLTE and LTE are the same, but as seen in \fig{fig:Mg_apospec} the LTE line profile is unable to reproduce the core of the strong \Mg\ lines. The observed \K\ lines in the $H$-band are insensitive to NLTE effects. The derived abundance in the sun is A(K)=5.05~dex,  in good agreement as well with our NLTE results for the optical lines. The derived abundance of \Ca\ found from the 5 lines deemed as useful in the $H$-band is A(Ca)=6.30~dex, in excellent agreement with our NLTE results for the optical. As mentioned before, the LTE and NLTE line profiles for the transitions observed in the $H$-band do not differ significantly, and therefore the derived abundance is the same.

\begin{figure*}[!ht]
     \begin{tikzpicture}
    \node[anchor=south east, inner sep=0] (image) at (0,0) {
        \subfloat{\includegraphics[width=0.95\textwidth]{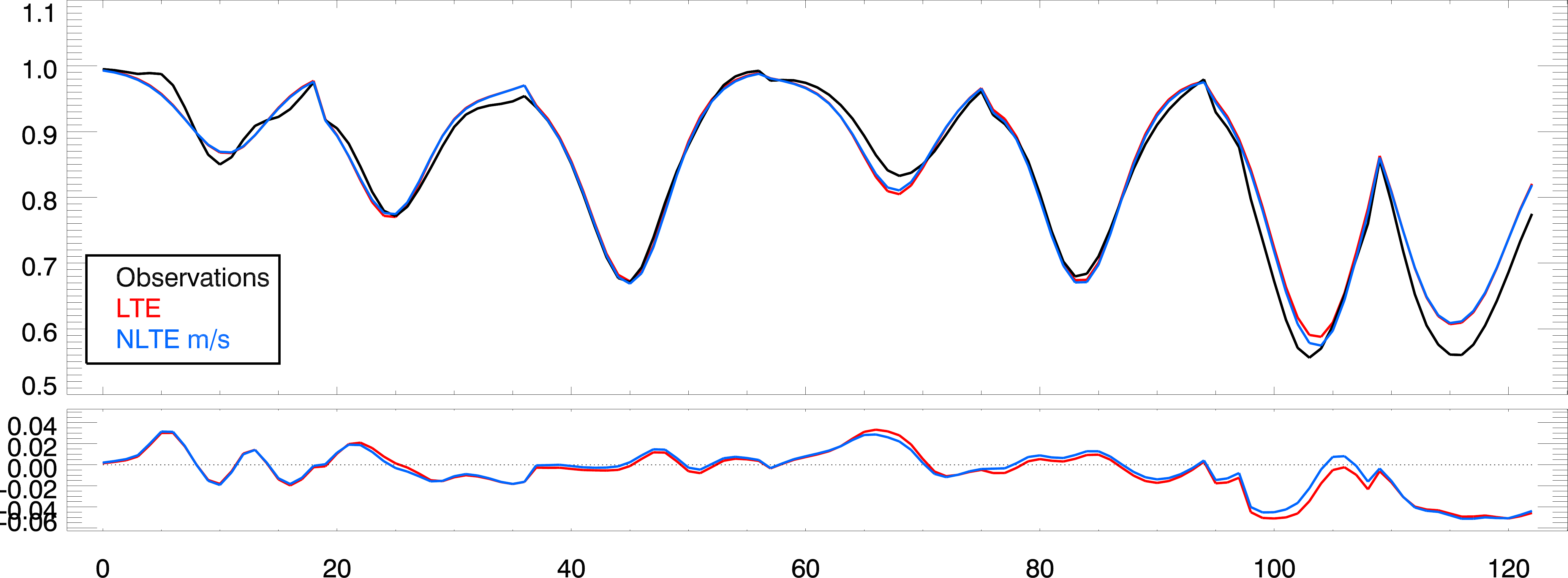}}};
      \node[rotate=90] at (-0.96\textwidth,0.25\textwidth) {\scalebox{0.9}{Relative Flux}}; 
      \node[rotate=90] at (-0.96\textwidth,0.08\textwidth) {\scalebox{0.8}{$\Delta_{syn-obs}$}}; 
      \node at (-0.47\textwidth,-0.011\textwidth) {\scalebox{1.0}{Mask index}};    
    \end{tikzpicture}   
   \caption{Comparison of \Ca\ lines in the apogee window between synthetic spectra and the observations of Arcturus. The black lines represent the observations; red and blue are the best fit synthetic spectra in LTE and NLTE respectively. The lower panel shows the residuals (synthetic-observed).}
    \label{fig:Ca_apospec}
\end{figure*}
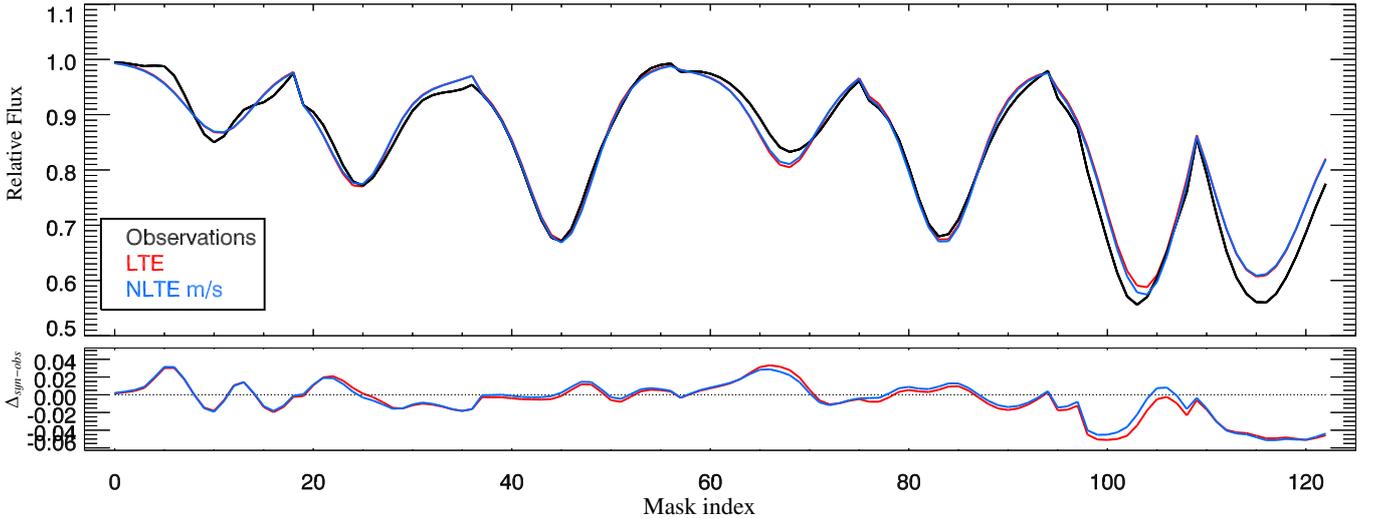

In the case of Arcturus we adopted the observations from \cite{1995PASP..107.1042H}. For \Na, the 15992~\AA\ line has an strong blend with two CN lines and the other visible line (16389~\AA) is rather weak but has NLTE effects thus the derived abundances are A(Na$_{nlte}$)=5.82 and A(Na$_{lte}$)=5.87~dex; both results consistent with our results in the optical. For \Mg, abundances of the best fit in LTE and NLTE are A(Mg)=7.55 and 7.40~dex; like in the sun, the core of the strong lines can not be reproduced in LTE. \K\ has two relatively strong lines at 15170~\AA. We found that there are no significant NLTE effects for these two lines and the derived abundance is A(K)=4.80~dex, an intermediate value between the LTE and the NLTE abundances we found in the optical. There are some changes in the \Ca\ line profiles observed in Arcturus. The derived abundances were A(Ca$_{nlte}$)=5.96 and A(Ca$_{lte}$)=6.00~dex, both in good agreement with the derived abundance from optical transitions. However, in this case our best fit in NLTE is 12\% better than the best fit in LTE. Figure \ref{fig:Ca_apospec} shows the results of the best fit under LTE and NLTE  compared with the observations for the selected \Ca\ lines. The redistribution of the \Ca\ populations in NLTE shows that for some lines core of the NLTE profile is deeper than the core of the LTE one, while for other lines the opposite is true. Since the NLTE profile is closer to the observations, the line by line analysis will present higher dispersion in the abundances derived in LTE than in NLTE.

%
%
%

Regarding the \Mg{i} triplet in the $H$-band, we should stress that the impact of the inability of the LTE calculations to match the observed cores is minimised when the abundance determination is performed fitting line profiles at high resolution, as we are doing here. The cores of the lines span only a few frequencies, the rest of the line profile is reproduced well, and an abundance increase to improve the fitting in the core will necessarily ruin the match in the wings of the line profile. Therefore the best-fitting abundance is fairly insensitive to the NLTE changes in the line core. If the analysis is carried out fitting line profiles at lower resolution, or using equivalent widths, the NLTE abundance corrections will be enhanced. This implies that the abundance corrections expected for these lines in the APOGEE spectra will be significantly augmented compared to our estimates derived for very high-resolution data. We used the NLTE line profile of the 15\,748~\AA\ \Mg{i} line as a test case; in both stars the the NLTE cores are stronger than the LTE ones, but in the Sun this line has broad wings and in Arcturus its wings are narrow. We fixed the NLTE abundance and determined the LTE value that minimises the residuals between the two profiles. This experiment was performed at different spectral resolutions, by performing a Gaussian convolution with increasing FWHM in both LTE and NLTE spectra. As seen in Table \ref{tab:acorr-fwhm}, we find that at increasing resolution the LTE and NLTE profiles become more similar and at the same time the abundance correction increases.

\begin{table}
\centering
\caption{NLTE abundance corrections at different resolutions for our test case: The \Mg{i} 15\,748~\AA line in the Sun and Arcturus}\label{tab:acorr-fwhm}
\tiny
\begin{tabular}{c c c c c}\hline\hline
    & \multicolumn{2}{c}{Sun} & \multicolumn{2}{c}{Arcturus} \\\hline
 FWHM [\AA] &  $\Delta$A(Mg) &  RMS$_{LTE-NLTE}$ & $\Delta$A(Mg) &  RMS$_{LTE-NLTE}$  \\[3pt]\hline
 0.00  & -0.02 & 7.25e-04 &  -0.04   & 1.02e-03 \\
 0.25  & -0.03 & 1.92e-04 &  -0.09   & 6.71e-04 \\
 0.50  & -0.04 & 1.12e-04 &  -0.12   & 3.14e-04 \\
 0.75  & -0.05 & 6.96e-05 &  -0.14   & 1.47e-04 \\
 1.00  & -0.05 & 4.49e-05 &  -0.14   & 7.49e-05 \\
 1.50  & -0.05 & 2.23e-05 &  -0.14   & 1.92e-05 \\\hline
\end{tabular}
\tablefoot{The abundance correction is defined as \mbox{$\Delta$A(Mg) = A$_{NLTE}$(Mg)$-$A$_{LTE}$(Mg).} For the sun and Arcturus we use A$_{NLTE}$(Mg) = 7.53 and 7.43 dex respectively.}
\end{table}

\subsection{Other NLTE work in the APOGEE window}

Other NLTE studies in the $H$-band spectra for the atoms studied in this work are the ones from \cite{2017ApJ...835...90Z} for \Mg\,and \cite{2019ApJ...881...77Z} for \Ca. 

In \cite{2017ApJ...835...90Z}, they adopted for the Sun A(Mg)=7.53~dex and calibrated the oscillator strength of the 15750~\AA\ lines in order to fit the observations. Their figure 2 does not show a deeper NLTE core compared with the LTE line profile, a feature we see in our synthetic spectra (see  \fig{fig:Mg_apospec}). We believe the reason for this particular discrepancy is due to the collisional data adopted for the high-lying levels. Hydrogen collisional excitation were ignored for the high-lying levels of \Mg{i} while we used the formula from \cite{Kaulakys:1986tl}. Electron collisional ionisation were taken from \cite{1962amp..conf..375S}, while we used the formula from \citep{Vrinceanu:2005em} and electron collisional excitation for the high-lying levels were taken from \cite{1962ApJ...136..906V}, while we used the BSR results of \cite{2017A&A...606A..11B}. The derived NLTE populations of the high-lying levels depend directly on the collisional rates that involve those levels.
They also derived LTE and NLTE abundances for Arcturus and found an abundance correction of $\Delta$A(Mg)$_{nlte-lte}=-0.03$~dex which is consistent with our results. 

The work on \Ca\ presented in \cite{2019ApJ...881...77Z} study both, the optical and the $H$-band spectra. The derived LTE abundances agree with ours for the Sun, but for Arcturus they are $\sim$0.1~dex lower than ours. The reason for this discrepancy can be due to differences in the abundance adopted and the radiative data (they used astrophysical f-values by setting the solar A(Ca)=6.31~dex while we adopted f-values based on \citeauthor{YU2018263} 2018, see Table \ref{tab:ca_gfs}).

\begin{table}
    \centering
    \caption{log(gf) values adopted for the \Ca{i} lines visible in the H-band in \cite{2019ApJ...881...77Z} and in this work \citep[taken from ][]{YU2018263}.}\label{tab:ca_gfs}
    \begin{tabular}{c c c} \\\hline\hline
    $\lambda$ (\AA)  & \cite{2019ApJ...881...77Z}   &  This work\\\hline
        16\,136.823  & $-0.525$  &  $-0.973$ \\
        16\,150.763  & $-0.215$  &  $-0.369$ \\
        16\,155.236  & $-0.685$  &  $-0.624$ \\
        16\,157.364  & $-0.165$  &  $-0.411$ \\
        16\,197.075  & \,\,\,$0.098$  &  $-0.016$ \\\hline
    \end{tabular}
\end{table}

For Arcturus they found in the optical an abundance correction of 0.01~dex, which is in excellent agreement with our NLTE-s results. For the $H$-band, \citeauthor{2019ApJ...881...77Z} found no abundance correction for Arcturus, while we found -0.04~dex. We think this discrepancy is again due to differences in the collisional data adopted for the high-lying levels.  In this work, hydrogen collisional excitation for the high-lying levels of \Ca{i} were ignored, while we used the formula from \cite{Kaulakys:1986tl}. For electron collisional ionisation \citeauthor{2019ApJ...881...77Z} used the formula from \cite{1962amp..conf..375S},  while we adopted the formula from \cite{Vrinceanu:2005em}; finally, electron collisional data involving the levels producing the \Ca{i} lines visible in the H-band are from \cite{1962amp..conf..375S}, while we used an extension of the data presented in \cite{PhysRevA.99.012706}.

\section{Conclusions}

Traditional (1D and 3D) NLTE calculations in cool stars adopt the trace-element approximation, where all the atmospheric parameters are kept fixed while the populations of a single atomic element are allowed to change. With the aim of removing such approximation, we performed for the first time NLTE radiative transfer calculations in cool stars including \emph{several} atomic elements simultaneously, still within the trace-element approximation but including the inter-element effects through the background opacities.   

Atomic elements in ionisation stages where over-ionisation is an important NLTE mechanism are most likely affected by \Mg\ NLTE effects. For late-type stars around solar temperature and hotter, there is enough change in the UV flux due to \Mg-NLTE effects that it can affect the statistical balance of \Ca, but not \K\ or \Na. 

Due to the sensitivity of the NLTE results for a particular radiative transition on the collisional data of the levels associated to that transition, special care must be taken when performing NLTE calculations of spectral lines involving high-lying level, for which usually accurate collisional data are missing.  

As found in previous studies \citep{CaPaperI}, the best fit for Ca in Arcturus in LTE is better than the best fit for Ca in NLTE-s[Ca] and NLTE-m. Perhaps the calcium NLTE populations in Arcturus are sensitive to other elements than the ones studied here -- iron is the obvious candidate. Future calculations expanding the elements to calculate in the NLTE-m mode will help to clarify this issue.

Observations at lower resolution will cause derived LTE abundances to increase due to the inability of the LTE line profiles to match the cores of strong lines (like the \Mg{i} triplet in the $H$-band). 

The derived NLTE abundance corrections in the optical and in the H-band differ, but the NLTE abundances derived are consistent between the two spectral regions. 

The goal of this effort is to update the synthetic spectral library that will be used by ASPCAP in DR17 which will include inter-element NLTE effects for \Na, \Mg, \K\ and \Ca\ in the analysis of APOGEE spectra.   

This work demonstrates that in cool stars inter-element NLTE effects (via background opacities) can have an impact on the derived abundances of the same order as "traditional" NLTE effects. We expect the inter-element NLTE effects to increase with temperature (due to increased radiative fluxes) and higher surface gravities (due to wider lines). This an step towards \emph{full} NLTE stellar atmospheric modelling of cool stars. Our next step in this line will be to include more elements that contribute opacity in the next \NLTEm\ calculations and study its effects on the NLTE results derived for other species. 


In Section 4 of \cite{Asplund:2005bp}, the author points out the lack of studies regarding the effects of NLTE and 3D on the structure of cool stars, and the possible implications on parameters determined via stellar spectra e.g. \teff\ and \logg\ derived from ionisation balance / line broadening. More recently, \citeauthor{Asplund:2009eu} (2009, Section 2) points out that even though we have had great progress in the quality and quantity of available atomic data we still require to go beyond LTE in order to attain the desired \%1 level of precision in derived solar abundances that are fundamental for the whole astrophysics community. Works regarding the effects of the trace element approach on the NLTE results in cool stars is still an area to explore and this is our first effort on this front.

\begin{acknowledgements}

We want to thank the referee for the careful reading of the manuscript and for giving such constructive comments and suggestions which helped improving the quality and readability of the paper.  

We thank P. S. Barklem, M. Bautista, S. Nahar, J. Holtzman and O. Zatsarinny for providing assistance. I. Hubeny is thankful for funding for his visit to the IAC by the Severo Ochoa program, awarded by the Government of Spain to the IAC to recognise, reward and promote outstanding scientific research in Spanish centres and units with a high level of excellence in the international arena. The research of Y. Osorio and C. Allende Prieto is partially funded by the Spanish MINECO under grant AYA2014-56359-P. This research has made use of NASA's Astrophysics Data System Bibliographic Services, TOPBASE,  the NORAD-Atomic-Data web-page, and the VALD database, operated at Uppsala University, the Institute of Astronomy RAS in Moscow, and the University of Vienna. We build on software and data written, collected, maintained, and made publicly available by R. L. Kurucz and F. Castelli.

SzM has been supported by the J{\'a}nos Bolyai Research Scholarship of the Hungarian Academy of Sciences, by the Hungarian NKFI Grants K-119517 and GINOP-2.3.2-15-2016-00003 of the Hungarian National Research, Development and Innovation Office.

Funding for the Sloan Digital Sky Survey IV has been provided by the Alfred P. Sloan Foundation, the U.S. Department of Energy Office of Science, and the Participating Institutions. SDSS acknowledges support and resources from the Center for High-Performance Computing at the University of Utah. The SDSS web site is www.sdss.org.

SDSS is managed by the Astrophysical Research Consortium for the Participating Institutions of the SDSS Collaboration including the Brazilian Participation Group, the Carnegie Institution for Science, Carnegie Mellon University, the Chilean Participation Group, the French Participation Group, Harvard-Smithsonian Center for Astrophysics, Instituto de Astrof\'isica de Canarias, The Johns Hopkins University, Kavli Institute for the Physics and Mathematics of the Universe (IPMU) / University of Tokyo, the Korean Participation Group, Lawrence Berkeley National Laboratory, Leibniz Institut f\"ur Astrophysik Potsdam (AIP), Max-Planck-Institut f\"ur Astronomie (MPIA Heidelberg), Max-Planck-Institut f\"ur Astrophysik (MPA Garching), Max-Planck-Institut f\"ur Extraterrestrische Physik (MPE), National Astronomical Observatories of China, New Mexico State University, New York University, University of Notre Dame, Observat\'orio Nacional / MCTI, The Ohio State University, Pennsylvania State University, Shanghai Astronomical Observatory, United Kingdom Participation Group, Universidad Nacional Aut\'onoma de M\'exico, University of Arizona, University of Colorado Boulder, University of Oxford, University of Portsmouth, University of Utah, University of Virginia, University of Washington, University of Wisconsin, Vanderbilt University, and Yale University.

\end{acknowledgements}

\bibliographystyle{aa}
\bibliography{papers}

\end{document}